\documentclass[aps,prd,longbibliography,onecolumn,showpacs,amsmath,amssymb,superscriptaddress,nofootinbib,floatfix,10pt,preprintnumbers]{revtex4-1}

\usepackage[usenames,dvipsnames]{color}
\include{jdefs}
\usepackage{graphicx}
\usepackage{multirow}
\usepackage[utf8]{inputenc}
\usepackage{url}
\usepackage{xspace}
\usepackage{braket}
\usepackage{tabularx}
\definecolor{darkblue}{rgb}{0,0,0.5}
\definecolor{darkgreen}{rgb}{0.1,0,0.3}
\definecolor{darkred}{rgb}{0.6,0,0}
\usepackage[bookmarks=true, bookmarksnumbered=true, colorlinks=true,
  urlcolor=darkblue, citecolor=darkgreen, linkcolor=darkred,
  breaklinks=true]{hyperref} 
\usepackage[normalem]{ulem}

\usepackage{hyperref}

\usepackage{amsmath}
\usepackage{amssymb,amsfonts,amsthm}

\usepackage{graphicx}

\usepackage{color}

\usepackage{siunitx}

\newcommand{\alphEM}{\alpha_{\mathrm{EM}}(\mZ)^{\overline{MS}}}
\newcommand{\invalphEM}{\left[\alpha_{\mathrm{EM}}(\mZ)^{\overline{MS}}\right]^{-1}}
\newcommand{\alphs}{\alpha_{\mathrm{S}}(\mZ)^{\overline{MS}}}

\newcommand{\mc}{m_c}
\newcommand{\mb}{m_b}
\newcommand{\mt}{m_t}
\newcommand{\mZ}{M_Z}
\newcommand{\mW}{M_W}
\newcommand{\GamZ}{\Gamma_Z}
\newcommand{\GamZinv}{\Gamma_Z^{\mathrm{inv}}}
\newcommand{\swsqeff}{\sin^2\theta_{\mathrm{eff}}^{\mathrm{lept}}}

\newcommand{\brbsgamma}{BR(B\rightarrow X_s\gamma)}
\newcommand{\brbsmumu}{BR(B_s^0\to\mu^+\mu^-)}
\newcommand{\RBtaunu}{\frac{BR(B_u \to \tau \nu)}{BR(B_u \to \tau \nu)_\text{SM}}}
\newcommand{\amu}{\delta a_\mu^{\mathrm{SUSY}}}

\newcommand{\mcha}[1]{m_{\tilde{\chi}^\pm_{#1}}}
\newcommand{\mneu}[1]{m_{\tilde{\chi}^0_{#1}}}

\newcommand{\mglu}{m_{\tilde{g}}}

\newcommand{\mstop}[1]{m_{\tilde{t}_{#1}}}
\newcommand{\msbot}[1]{m_{\tilde{b}_{#1}}}
\newcommand{\msup}[1]{m_{\tilde{u}_{#1}}}
\newcommand{\msdw}[1]{m_{\tilde{d}_{#1}}}
\newcommand{\msc}[1]{m_{\tilde{c}_{#1}}}
\newcommand{\mss}[1]{m_{\tilde{s}_{#1}}}

\newcommand{\msel}[1]{m_{\tilde{e}_{#1}}}
\newcommand{\msmu}[1]{m_{\tilde{\mu}_{#1}}}
\newcommand{\mstau}[1]{m_{\tilde{\tau}_{#1}}}
\newcommand{\msnu}{m_{\tilde{\nu}}}
\newcommand{\msneu}[1]{m_{\tilde{\nu}_{#1}}}

\newcommand{\GeV}{\mathrm{GeV}}
\newcommand{\TeV}{\mathrm{TeV}}

\newcommand{\tanb}{\tan\beta}

\newcommand{\relic}{\Omega_\mathrm{DM} h^2}
\newcommand{\relicneu}{\Omega_{\tilde{\chi}^0_1} h^2}

\newcommand{\mh}{m_{h^0}}
\newcommand{\mH}{m_{H^0}}
\newcommand{\mA}{m_{A^0}}
\newcommand{\mCH}{m_{H^\pm}}

\newcommand{\mHu}{m_{H_u}}
\newcommand{\mHd}{m_{H_d}}
\newcommand{\At}{A_t}
\newcommand{\Ab}{A_b}
\newcommand{\Atau}{A_\tau}
\newcommand{\msq}[1]{m_{\tilde{Q}_{#1}}}
\newcommand{\msu}[1]{m_{\tilde{U}_{#1}}}
\newcommand{\msd}[1]{m_{\tilde{D}_{#1}}}
\newcommand{\msl}[1]{m_{\tilde{L}_{#1}}}
\newcommand{\mse}[1]{m_{\tilde{E}_{#1}}}

\newcommand{\like}{{\cal L}}
\newcommand{\lnlike}{\ln {\cal L}}

\newcommand{\npop}{N_{\rm pop}}
\newcommand{\ngen}{N_{\rm gen}}

\newcommand{\pikaiav}{{\tt PIKAIA 1.2}}
\newcommand{\Zfitv}{{\tt ZFITTER 6.42}}
\newcommand{\SSUSYv}{{\tt SOFTSUSY 4.1.0}}
\newcommand{\FHv}{{\tt FeynHiggs 2.13.0}}
\newcommand{\HBv}{{\tt HiggsBounds 4.3.1}}
\newcommand{\HSv}{{\tt HiggsSignals 1.4.0}}
\newcommand{\microv}{{\tt micrOMEGAs 4.3.2}}
\newcommand{\smodelsv}{{\tt SModelS 1.1.1}}
\newcommand{\pythiav}{{\tt PYTHIA 8.2}}
\newcommand{\nllfastv}{{\tt NLL-fast}}
\newcommand{\fastlimv}{{\tt Fastlim 1.0}}

\newcommand{\Zfit}{{\tt ZFITTER}}
\newcommand{\FH}{{\tt FeynHiggs}}
\newcommand{\pikaia}{{\tt PIKAIA}}
\newcommand{\SSUSY}{{\tt SOFTSUSY}}
\newcommand{\micro}{{\tt micrOMEGAs}}
\newcommand{\smodels}{{\tt SModelS}}

\newcommand{\chisq}{\chi^2}
\newcommand{\chisqmin}{\chi^2_{\rm min}}
\newcommand{\chisqGCE}{\chi^2_{\rm GCE}}

\newcommand{\sip}{\sigma^{SI}_{\tilde{\chi}^0_1 p}}
\newcommand{\sineu}{\sigma^{SI}_{\tilde{\chi}^0_1 n}}
\newcommand{\sdp}{\sigma^{SD}_{\tilde{\chi}^0_1 p}}
\newcommand{\sdn}{\sigma^{SD}_{\tilde{\chi}^0_1 n}}
\newcommand{\sigmav}{\langle \sigma v \rangle_0}

\begin{document}

\preprint{IPPP/18/32}
%\preprint{CPT/18/64}

\title{\mbox{ }\vspace{0.5cm}\\
The Power of Genetic Algorithms:\\ what remains of the pMSSM?\vspace{0.5cm}}

\author{Steven  Abel}
 \email{s.a.abel@durham.ac.uk}
 \affiliation{Institute for Particle Physics Phenomenology, Durham University, Durham DH1 3LE, United Kingdom} 
\author{David G. Cerde\~no}
 \email{d.g.cerdeno@durham.ac.uk}
 \affiliation{Institute for Particle Physics Phenomenology, Durham University, Durham DH1 3LE, United Kingdom} 
\author{Sandra Robles \vspace{0.3cm}} 
 \email{sandra.robles@unimelb.edu.au}
 \affiliation{ARC Centre of Excellence for Particle Physics at the Terascale, School of Physics, The University of Melbourne, Victoria 3010, Australia} 

 \begin{abstract}
 \vspace{0.5cm}
\noindent Genetic Algorithms (GAs) are explored as a tool for probing new physics with  high dimensionality. We study the 19-dimensional pMSSM, including experimental constraints from all sources and assessing the consistency of potential signals of new physics.  
We show that GAs excel at making a fast and accurate diagnosis of the cross-compatibility of a set of experimental constraints in such high dimensional models. In the case of the pMSSM, it is found that only ${\cal O}(10^4)$ model evaluations are required to obtain a best fit point in agreement with much more costly MCMC scans. This efficiency allows higher dimensional models to be falsified, and patterns in the spectrum identified,  orders of magnitude more quickly. As examples of falsification, we consider the muon anomalous magnetic moment, and the Galactic Centre gamma-ray excess observed by Fermi-LAT, which could in principle be explained in terms of neutralino dark matter. We show that both observables cannot be explained within the pMSSM, and that they provide the leading contribution to the total goodness of the fit, with $\chisq_{\amu}\approx12$ and $\chisq_{\rm GCE}\approx 155$, respectively.

\end{abstract}

\maketitle

\section{Introduction}

Experimental constraints on supersymmetry continue to make the simplest realisations of the Minimal Supersymmetric Standard Model (MSSM) less credible. One is forced to consider less constrained alternatives such as the pMSSM \cite{Djouadi:1998di}. 
This is the most general version of the $R$-parity conserving MSSM under the assumption of CP conservation, Minimal Flavour Violation, and degenerate first and second generation sfermion masses\footnote{Usually the pMSSM is defined with these last two conditions imposed at the electroweak scale, but for this study it will make little difference if we impose them at the Grand Unified Theory (GUT) scale.}.
It has a multi-dimensional parameter space -- 23 in total, consisting of 19 fundamental parameters and 4 nuisance parameters. 

Analysis of such high dimensionality models becomes very difficult. The traditional technique of ``slice-and-scan'' that suffices for the Constrained MSSM (CMSSM) for example, is entirely infeasible. Typically one uses Monte-Carlo and nested sampling approaches as in Refs.~\cite{Berger:2008cq,AbdusSalam:2009qd,Cahill-Rowley:2013yla,Cahill-Rowley:2014twa}. It is probably fair to say that, even if analysis can be made feasible by these methods, it is not always clear what one should conclude from the results. Suppose for instance that upon scanning a 23D cube of the parameter space of the pMSSM one found that in every 2 dimensional slice the allowed region occupies the inside of a circle that just touches the edges of the cube. 
This ``allowed ball'' would appear to almost fill the 23D cube inside which it just fits, and yet it would actually occupy only 0.4\% of the volume. 
This is the infamous  ``large dimensionality problem'': taking slices of a high dimensional object inevitably gives a very misleading impression of its structure. 
On a more practical level, how can one attempt to falsify a model such as the pMSSM, when superficially it seems that virtually any set of observables could be accommodated somewhere in the parameter-space? And compounding the problem associated with the multi-modality of variables, is the multi-modality of observables. If several suitable areas of parameter space are discovered, do they represent a single cluster or several disjoint favoured regions? Do they give a prediction for the spectrum? Which observables have most influence over the favoured regions? 

All of these issues suggest the use of heuristic search and visualisation techniques. In this paper we consider the effectiveness of Genetic Algorithms (GAs), in assessing and analysing the pMSSM. GAs seek optimal solutions by evolving a population of models in the search-space which, by means of a suitable definition of ``fitness'', is transformed into a fitness landscape \cite{Holland1975,David1989,Holland1992,Reeves2002,haupt,Michalewicz2004}. In the case of models such as the pMSSM the optimisation in question is of course to find the minimum overall $\chi^2$, whose inverse can therefore serve directly as a measure of the fitness. There are several advantages of GAs that this study will highlight. The first is simply the extreme efficiency of such techniques versus traditional scanning techniques, or even more sophisticated Bayesian Inference techniques, such as that employed by MultiNest \cite{Feroz:2007kg,Feroz:2008xx,Feroz:2013hea}. Indeed, compared to the latter, GAs can find a best fit point orders of magnitude more quickly, because the number of models that need to be built is considerably smaller\footnote{We should remark that we checked the consistency of our procedure using MultiNest in the CMSSM.}. As a practical demonstration, we show that with this approach it is easily possible to exclude the pMSSM, and to identify the main culprit that apparently cannot be reconciled with experiment in any of the parameter space, namely $(g-2)_\mu$. It becomes clear that a GA can efficiently find regions of parameter space in which the $\chi^2$ of all other parameters are reasonable, with $(g-2)_\mu$ standing out as the dominant contribution. In the present case only $\sim 10^4$ models need to be evaluated in order to reach this conclusion\footnote{For comparison, it is worth point out that performing a flat scan even as rudimentary as taking one large and one small value for each parameter would require $10^7$ evaluations.}. 
While they are not exhaustive in the usual sense, GAs {\it do} probe the entire search-space, albeit in a highly non-linear way \cite{Holland1975}. Therefore, one can now be confident that the pMSSM does not have any remaining regions of parameter space that harbour better solutions for $(g-2)_\mu$. However, no other observable is particularly problematic to fit. 

Another advantage of GAs arises from the fact that they are a dynamical process.  It has been argued that whether a problem is ``GA-hard'' or ``GA-easy'' depends on the ``fitness-distance correlation'' in the parameter space \cite{Jones1995,Collard1998}. Problems that are GA-hard (or that are not tackled well) resemble ``needle-in-a-haystack'' problems, in which all incorrect solutions are equally bad and one has as much chance of landing on the correct solution as when performing a random scan. In this context, it is important that the GA is performed so that a ``fitness landscape'' is established as a  function of continuous parameters such as $\chi^2$. Any hard experimental exclusions are essentially step-functions in the fitness landscape that can locally weaken the fitness-distance correlation. 
This fitness-distance correlation is made manifest by the flow of the population as it evolves in the GA. By observing this flow over successive generations, one sees the pull of various observables. Naturally those that are well constrained experimentally within the parameter space, for example soft-terms such as $A_t$ that govern the Higgs mass, exert a strong pull (through the contribution to $\chi^2$), and the population evolves rapidly towards suitable values. 
Conversely, the limited precision in the measured Higgs couplings leads to less focused values for e.g. $\tan\beta$.
In the SUSY context, this can be thought of as a measure of the fine-tuning in the theory. Such flows {\it can} incidentally be understood by taking slices of the space of observables, where it becomes clear if a particular observable is becoming significantly focussed. The efficiency of GAs in this context compared to other techniques suggests that the problem of optimising $\chi^2$ for a multi-modal model such as the pMSSM is very ``GA-easy'': the fitness-distance correlation (by virtue of $\chi^2$) is very good.   

A final advantage of GAs lies in their end product, which (by construction) is a large population of models focussed around those regions of parameter space that are the most interesting given the current constraints. This provides a natural  tool with which new observables can be tested. The example we will consider here is the Fermi-LAT Galactic Centre excess. Given such a new observable, one could of course just fold it into the original study and start from the beginning. But one can also, either test the final population to see if it predicts the observed value, or even better add the new observable into the fitness of the final population and continue to evolve it to a new equilibrium. 
If the new best fit is considerably worse than the old one, then we can conclude that the new observable is in conflict with the model. 
This is a natural approach to take when new experimental results need to be taken into consideration. In this sense GAs are able to provide a (literally) evolving population of ``Snowmass points''.  

The paper is organized as follows. In Section~\ref{sec:GA} we briefly review the GA technique, and in Section~\ref{sec:pmssm} we explain how we apply it to the specific case of the pMSSM, with 19 parameters defined at the GUT scale and 4 nuisance parameters. We also discuss there the different experimental constraints that are included in our analysis. The results are presented in Section~\ref{sec:results}, analysing first the case of the muon anomalous magnetic moment and then the Galactic Centre gamma-ray excess. The conclusions are presented in Section~\ref{sec:conclusions}. Finally, Appendix~\ref{sec:app} contains some complementary plots that illustrate the evolution of the GA in the pMSSM parameter space.

We should mention where this work stands in relation to previous studies. In fact in our view the number of studies in the High Energy Physics arena employing GAs is still remarkably small considering the robustness and utility of the technique. It has been used in the model building context in Refs.~\cite{Blaback:2013ht,Abel:2014xta}. In those cases the construction of fitness landscape is more directly related to desirable properties such as small positive cosmological constant, number of generations and so-forth. As such one is looking for a small number of ``perfect solutions'', and the technique becomes more of a ``black-art''.
In the model-exclusion/profile-likelihood context it was discussed in Refs.~\cite{Yamaguchi:1999hq,Allanach:2004my,Akrami:2009hp}.  The main body of the study conducted here is most closely related to Ref.~\cite{Akrami:2009hp}\footnote{Our GA implementation is based on the publicly available \pikaia \ code, first introduced in Refs.~\citep{Charbonneau:1995,Charbonneau:1995b}.}, but in a much higher dimensionality.

\section{The Genetic Algorithm technique} 
\label{sec:GA}

We begin by briefly reviewing the GA technique with specific reference to the task at hand (for more pedagogical introductions see Refs.~\cite{Reeves2002,Allanach:2004my,Abel:2014xta}), namely surveying regions of model parameter space, excluding disfavoured regions and selecting favoured regions of some framework. The physical ``observable'' we wish to optimise in the parameter-space is the overall $\chi^2$. 
We shall focus in particular on the particular properties of the \pikaiav \ package which is used here to perform the GA \cite{Charbonneau:1995,Charbonneau:1995b,Charbonneau:2002,Charbonneau:2002b}.

Any GA is an optimisation based on evolving a population of $\npop$ trial individuals, typically 50-100. Each individual consists of a string of data (the so-called {\it chromosome}), that encodes the parameters defining a particular individual. This encoding can take various forms, and is referred to generically as the individual's {\it genotype}. In this case, it is simply all the input parameters collected together in one long string of data. The entries in the chromosome are called {\it alleles}. Often a binary encoding is preferred as it can work with smaller populations, however \pikaiav \  uses a decimal encoding. It is convenient to also introduce the notion of  uniformly sized small groups of alleles, called \emph{genes}, that each encode a single physical parameter, for example a soft-mass squared. The population is initially chosen with random genomes for the  $\npop$-individuals, and then the algorithm consists of repeated application of the following three basic elements: \\

\noindent {\bf Selection:} Individuals are first selected from the population to make ``breeding pairs''. If the population size is preserved (the usual scheme) then there will be $\npop$ breeding pairs, and the average individual will be selected for breeding twice. The first step in this process  is to assign to each individual a {\it fitness} based on its physical properties (the {\it phenotype}). In the present case, the phenotype is the collection of all the experimental observables of interest, for example Higgs masses, decay widths, and so forth. The fitness is a single function of all these variables whose theoretical maximum value corresponds to the perfect individual. In this study, the fitness functions is taken to be $1/\chi^2$ (typically the convergence to solutions is quite independent of this function). This step is usually the most case-dependent and time-intensive part of the whole procedure, because it is where the physics is bolted on. 

Once fitnesses have been assigned to the entire population, breeding pairs are formed by selecting individuals based on their fitness (with obviously fitter individuals being selected more often). Typically the fittest individual may breed a few times more than the average, but it is important that less fit individuals are allowed to mate.  The selection process may take many different forms, such as roulette-wheel, rank-weighting, tournament selection, and so on{\footnote{Note that one does not always have to take the fitness to be a continuous function of the phenotype. For example, for problems where this would give a very wide and shallow fitness plateau, it is preferable to base the fitness of individuals on their ranking. This reproduces some of the advantages of tournament selection but is much simpler to incorporate.}.   \\

\noindent {\bf Breeding/Cross-over:} A new population of individuals is formed by splicing together the chromosomes of the two individuals in each breeding pair. Again there are many different ways to do this, but a typical choice (uniform cross-over) might be to cut the chromosomes at two random points along their length and swop the middle sections. PIKAIA 1.2 uses both one- and two-point cross-over in roughly equal proportions to reduce end-point biasing. \\

\noindent {\bf Mutation:} With only the two previous elements, one would already observe convergence of the population around good solutions over generations. However, the real power of GAs comes from the third element which is {\it mutation}. This is the feature which is chiefly responsible for the orders-of-magnitude gain in efficiency over a simple Monte-Carlo. Once a new generation is formed, a small fraction (usually around a percent) of the  alleles have their values flipped at random. This prevents stagnation in the population, where the entire population clusters around a local maximum in the fitness, when there are better solutions globally. It is important to understand that mutation is not just an improvement to the convergence, but is absolutely integral to the entire process. Depending on the problem and the structure of the fitness landscape, the nett effect is a dramatic increase in the overall rate of convergence. (As can be seen practically by optimising the mutation rate.) One of the innovations of \pikaiav \  in this aspect is its use of {\emph {creep mutation}} in order to overcome the so-called Hamming walls, which occur when the population is close to an optimum solution in terms of phenotype, but far away in terms of Hamming distance: for example the number 0.999 versus 1.000 requires a change in all 4 digits, but this very large change in genotype produces a very small change in phenotype. In short, creep-mutation ``carries the 1'' if a ``9'' is mutated by adding $+1$. As this kind of mutation results in small moves in physical parameter space, \pikaiav \ invokes  creep-mutation and one-point mutation with equal probability. This modification is also expected to mitigate somewhat the drawbacks of using decimal instead of binary encoding.    \\

And then the process repeats. 
We should add that, so that the maximum fitness is monotonically increasing, it is common at this point to copy the fittest individual from the last generation into the new one and to kill the least fit new individual, known as {\it elitist selection}. The particular parameters used for this study are shown in Table~\ref{GAtable}.

In summary, a GA incorporates and balances competing forces. Selection and breeding tends to produce convergence around local maxima in the fitness landscape, drawing the population in over generations. On the other hand the effect of mutation is to push the population away from local maxima (on average), so that as a whole it can explore the entire parameter space. The power of GAs then is in their ability to keep performing, 
regardless of the dimensionality of the physical parameter-space, which can even as large as the chromosome itself (as was the case of Ref.~\cite{Abel:2014xta}), and in their ability to be sensitive to the entire landscape, but simultaneously respond to and converge on interesting regions. 
Note that there are many other practical elements, such as fitness ``crowding penalties'', and  ``niching'', that we do not discuss (or use). They are covered in the literature (see Refs.~\cite{Reeves2002,haupt}) along with the underlying reasons for the effectiveness of  GAs, such the Schema theorem. 

\begin{table}[h!]
 
 \centering
 {\renewcommand{\arraystretch}{1.3}%
 \begin{tabular}{|l|c|}
\hline
Parameter & Value \\ 
 \hline
  Population, $\npop$ & 100 \\
  Generations, $\ngen$ & 300 \\
  Length of gene & 5 \\
  Crossover probability & 0.85 \\
  Mutation mode  & 2 (adjustable rate based on fitness)  \\
  Initial mutation rate & 0.005 \\
  Min. mutation rate & 0.0005 \\
  Max. mutation rate & 0.25 \\
  Relative fitness differential & 1.0 (breeding probability equals to rank) \\
  Reproduction plan & 1 (full generational replacement) \\
  Elitism & 1 (keep best)\\
  \hline
 \end{tabular}
  }
\caption{\pikaiav \ control parameters used in this study.}  
\label{GAtable}  
\end{table}

\section{Application to the pMSSM}
\label{sec:pmssm}

We now turn to the object of study, which is the phenomenological MSSM (pMSSM), with its 19 fundamental parameters. Here we define them at the Grand Unification Theory (GUT) scale and take ${\rm sign}(\mu)=1$. (In its usual definition the pMSSM takes parameters at the weak scale, however as a GA is not frequentist there is essentially no difference except for the effect of running on flavour degeneracy and consequently flavour changing observables. These effects are expected to be negligible for this study given that experimental constraints ultimately favour very large soft-terms. Note that $\delta a_\mu$ {\it will} be important, but precise first/second generation degeneracy would have little bearing on it.)
As well as these parameters, we include four additional parameters to account for the SM parameters with the largest uncertainties that could have 
an impact on the final theoretical predictions. These nuisance parameters are: the electromagnetic coupling constant evaluated at the Z-boson
pole mass, $\alphEM$, the strong coupling constant at $\mZ$, $\alphs$, the pole mass of the top quark, $\mt$ and the pole mass of the 
bottom quark, $\mb$\footnote{For simplicity, we use the pole mass of the bottom quark as an input parameter instead of 
$\mb(\mb)^{\overline{MS}}$, since the bottom pole mass is an input of \Zfit, the package used to compute the SM contributions to 
the Z boson decay width (see text below). In fact, $\mb(\mb)^{\overline{MS}}$ and  $\mc(\mc)^{\overline{MS}}$ are calculated by \Zfit, therefore 
for consistency we used these running masses as input  parameters of the other packages mentioned throughout this work.}. Their central values and uncertainties are given in Table~\ref{tab:SMnparam}. 
Hence, there is a 23 dimensional parameter space, whose range of variation is listed in Table~\ref{tab:inputs}. 
We restrict the study to positive gaugino masses, due to convergence issues in the selected SUSY spectrum calculator which occurred when negative gaugino masses were present\footnote{Note that this analysis can be easily extended to regions of the parameter space with a negative $\mu$ parameter and negative values of the soft masses.}.

\begin{table}[h!]
  \centering
 {\renewcommand{\arraystretch}{1.3}%
\begin{tabular}{|l|c|}
\hline
 Observable &  Value \\
\hline
$\invalphEM$ & $127.950\pm 0.017$ \\
$\alphs$ & $0.1185\pm 0.0006$ \\ 
$\mb(\GeV)$ & $4.78\pm0.06$ \\
$\mt(\GeV)$ & $173.1\pm 0.6$ \\
\hline
\end{tabular}
}
\caption{Standard model nuisance parameters, central values and uncertainties \cite{Olive:2016xmw}.}
\label{tab:SMnparam}
\end{table}

\begin{table}[h!]
  \centering
 {\renewcommand{\arraystretch}{1.3}%  
\begin{tabular}{|l|c|}
\hline
Parameter & Range \\
\hline
\multicolumn{2}{|c|}{SM} \\
\hline
$\invalphEM$ & $[127.882,128.018]$ \\
$\alphs$ & $[0.1161,0.1209]$ \\ 
$\mb(\GeV)$ & $[4.54,5.02]$ \\
$\mt(\GeV)$ & $[170.1,175.5]$ \\
\hline
\multicolumn{2}{|c|}{pMSSM (GUT scale)} \\
\hline
$M_1,M_2,M_3(\GeV)$ & [50,10000] \\
$\mHu,\mHd(\GeV)$ & [50,10000] \\
$\msq{1,2}\msq{3}(\GeV)$ & [50,10000] \\
$\msu{1,2}\msu{3}(\GeV)$ & [50,10000] \\
$\msd{1,2}\msd{3}(\GeV)$ & [50,10000] \\
$\msl{1,2}\msl{3}(\GeV)$ & [50,10000] \\
$\mse{1,2}\mse{3}(\GeV)$ & [50,10000] \\
$\At,\Ab,\Atau(\TeV)$ & [-10,10] \\
$\tanb$ & [2,62] \\
\hline
\end{tabular} 
}
\caption{SM nuisance parameters and pMSSM input parameters defined at the GUT scale.}
\label{tab:inputs}
\end{table}

In order to evaluate the fitness as a function of the initial parameters, the pMSSM predictions were  
implemented in a joint likelihood comprising the following experimental constraints:

\begin{itemize}

\item {\bf Electroweak precision observables (EWPOs)}: i.e. $Z$ pole observables and $\mW$.  
The theoretical prediction for the W boson pole mass $\mW$ were calculated with \SSUSYv~\cite{Allanach:2001kg}, 
and the effective electroweak mixing angle for leptons $\swsqeff$ with \FHv~\cite{Heinemeyer:1998yj,Heinemeyer:1998np,Hahn:2013ria,Bahl:2016brp}. The SM contributions to
the total decay width of the Z boson $\GamZ$ and the Z invisible width $\GamZinv$ were computed with 
\Zfitv~\cite{Bardin:1999yd,Arbuzov:2005ma} and those of the MSSM with \microv~\cite{Belanger:2001fz}. 
$\like_{\rm EWPO}$, Eq.~(\ref{eq:likeEWPO}), contains a Gaussian probability distribution function  for each of these quantities, with central values and experimental and theoretical uncertainties added in quadrature (see Table~\ref{tab:constraints}):
\begin{equation}
 \lnlike_{\rm EWPO} ~=~ \lnlike_{\mW} + \lnlike_{\swsqeff} + \lnlike_{\GamZ}  + \lnlike_{\GamZinv}. \label{eq:likeEWPO}  
\end{equation}

\item {\bf Flavour observables from B physics}: These include $\brbsgamma$, $\brbsmumu$ and $\RBtaunu$ (Eq.~\ref{eq:likeB}). Theoretical predictions were calculated with \micro. As in the previous case, $\like_{B}$ includes Gaussian likelihoods for every B observable, with mean values and uncertainties given in Table~\ref{tab:constraints}:
\begin{equation}
 \lnlike_{B} ~=~ \lnlike_{\brbsgamma} + \lnlike_{\brbsmumu} + \lnlike_{\RBtaunu}. \label{eq:likeB}
\end{equation}

\item {\bf Constraints from the Higgs sector}: $\like_{\rm Higgs}$ accounts for the likelihood of the model predictions for the Higgs masses, branching ratios, production cross
sections and total decay widths of the Higgs sector computed with \FHv. These predictions were tested against exclusion bounds from Higgs searches at the LEP, Tevatron and
LHC experiments using \HBv~\cite{Bechtle:2008jh,Bechtle:2013wla} and \HSv~\cite{Bechtle:2013xfa}. 
$\like_{\rm Higgs}$ also includes a Gaussian likelihood around the central value of the Higgs mass, the experimental and theoretical uncertainties considered here can be found in Table~\ref{tab:constraints}:
\begin{equation}
 \lnlike_{\rm Higgs} ~=~ \lnlike_{\mh} + \lnlike_{\rm Higgs \ sector}. \label{eq:likeHiggs} 
\end{equation}

\item {\bf LEP bounds on chargino and slepton masses}:  $\mcha{1}$, $\msel{R}$, $\msmu{R}$, $\mstau{1}$ and sneutrino mass constraints are incorporated in $\like_{\rm LEP}$. Using the generic limits implemented in \micro~\cite{Barducci:2016pcb},  
smeared step-function likelihoods were constructed for each of them, 
at 95\% CL, as in Ref.~\cite{deAustri:2006jwj}. 
\begin{equation}
 \lnlike_{\rm LEP} = \lnlike_{\mcha{1}} + \lnlike_{\msel{R}} + \lnlike_{\msmu{R}} + \lnlike_{\mstau{1}} + \lnlike_{\msnu}. \label{eq:likeLEP}
\end{equation}

\item {\bf LHC results on SUSY searches}: These were incorporated using \smodelsv~\cite{Ambrogi:2017neo,Kraml:2013mwa}\footnote{Even though, the use of \smodels \ entails a set of underlying assumptions such as that only on-shell particles are considered in the cascade decay and virtual particles are replaced by an effective vertex, other tools available in the literature for the same purpose are less suited for extensive searches in multidimensional parameter spaces and for taking advantage of massive parallelism.}, which employs upper limits and efficiency maps 
provided by the experimental collaborations. To properly estimate this likelihood, we first calculated the SUSY spectrum and decay widths with \SSUSY~\cite{Allanach:2014nba,Allanach:2017hcf}, and the relevant SUSY cross sections at LO with \micro. These cross sections  were then improved with NLO+NLL contributions, using 
 \pythiav~\cite{Sjostrand:2006za,Sjostrand:2014zea} and \nllfastv~\cite{Beenakker:1996ch,Beenakker:1997ut,Kulesza:2008jb,Kulesza:2009kq,Beenakker:2009ha,Beenakker:2010nq,Beenakker:2011fu} for use by \smodels.
$\like_{\rm LHC}$ accounts for \smodels \ computed likelihoods calculated for efficiency map results and smeared step-function likelihoods  
implemented for upper bounds at  95\% CL as in the LEP case.  
We employed the most up-to-date \smodels \ and \fastlimv \ databases~\cite{Papucci:2014rja}, which include 8 and 13 TeV results.

\item {\bf Dark matter (DM) relic abundance}: The value of  $\relic$ was calculated with \micro, and we implemented a Gaussian likelihood as for the previous constraints. See Table~\ref{tab:constraints}, for the corresponding experimental values.

\end{itemize}

\begin{table}[h!] 
  \centering
{\renewcommand{\arraystretch}{1.3}%
    \begin{tabular}{|l|S|S|S|c|}
     \hline
     Observable &  {Mean value} & \multicolumn{2}{c|}{Standard deviation} & Ref. \\
     \cline{3-4}
                &   & {experimental} & {theoretical} &   \\
      \hline
      $\mW(\GeV)$   &  80.385 &  0.015 & 0.01 & \cite{Olive:2016xmw} \\
      $\swsqeff$ & 0.23153 & 0.00016 & 0.0001 & \cite{ALEPH:2005ab} \\
      $\GamZ(\GeV)$ &  2.4952 &  0.0023 & 0.001 & \cite{Schael:2013ita,Olive:2016xmw} \\
      $\GamZinv(\GeV)$ &  0.499 &  0.0015 & 0.001 &  \cite{Olive:2016xmw} \\
      $\mh(\GeV)$ & 125.09 & 0.24 & 2.0 & \cite{Aad:2015zhl} \\
      $\brbsgamma\times 10^4$ &  3.43 & 0.22 & 0.24 &  \cite{Amhis:2014hma} \\
      $\brbsmumu \times 10^{9}$ & 2.9 & 0.7 & 0.29 & \cite{CMS:2014xfa} \\
      $\RBtaunu$   &  1.04  & 0.34  & {-} & \cite{Amhis:2014hma,Akeroyd:2010qy} \\
      $\amu\times 10^{10}$ & 26.8 & 6.3 & 4.3 &   \cite{Olive:2016xmw} \\ 
      $\relic$ & 0.1186 & 0.0010 & 0.012 & \cite{Ade:2015xua} \\
      \hline
      & \multicolumn{3}{c|}{Limits (95\% CL)}  & Ref. \\
           \hline
      $\mcha{1}$  & \multicolumn{3}{c|}{LEP2} & \cite{Heister:2003zk}  \\
      $\msel{R}$, $\msmu{R}$, $\mstau{1}$  & \multicolumn{3}{c|}{LEP2} & \cite{LEPSUSYWG} \\
      $\msnu$  & \multicolumn{3}{c|}{LEP2} & \cite{Abdallah:2003xe} \\
     \hline
    \end{tabular}
    }
\caption{Experimental constraints used to implement the joint likelihood. Experimental uncertainties account for both systematic and 
statistical errors added in quadrature.
}
\label{tab:constraints}    
\end{table}

Now, let us describe the  pMSSM-GA implementation. As mentioned in Section~\ref{sec:GA}, the fitness function was chosen to be the inverse of the chi-squared (as of course the GA seeks to maximise the fitness). 
In detail, (for each model) first the input parameters were evolved from the GUT scale down to the electro-weak (EW) scale to compute the SUSY spectrum, 
branching ratios and decay widths using \SSUSY. Then, the Higgs sector was evaluated with \FH. Next, the DM relic abundance and the aforementioned observables were calculated as previously outlined. These data constitute the phenotype of each individual. Finally, the predictions were combined into a likelihood as in Eq.~(\ref{eq:jointlike}) to compute a total chi-squared and hence the fitness.

On a practical level, the value of the fitness function of each individual in a given population, which as mentioned in the Introduction is by far the most computationally 
intensive step of a GA, is of course independent for each individual, providing inherent parallelism
and an opportunity to improve the performance of the heuristic search. 
To take advantage of this, we used the public parallel version of \pikaiav~\cite{Metcalfe:2002pd}, which implements the 
Message Passing Interface (MPI) for a more efficient exploration of  parameter space. 
Every package for the calculation of physical observables was modified accordingly and properly interfaced to \pikaia \  to avoid data loss and disruption.

The number of individuals in a population, $\npop$, was fixed to be 100. We explored a wide range of possibilities for the number of generations  $\ngen$, and 
determined that for $\ngen > 300$, there was no significant improvement in the minimum $\chisq$. In other words, $\ngen = 300$ generations, 
and hence only  $\npop \times \ngen = 3\times 10^4$ evaluations of the fitness function, 
were sufficient to achieve a good convergence of the total $\chisq$. (The number of times a model has to be evaluated 
is one of the best indicators of the overall efficiency gain: as mentioned earlier a useful point of comparison is the most rudimentary approach, namely a flat scan with just 2 points in each of the 
23 dimensions, which would require  $10^7$ evaluations.)

The complete set of selected GA parameters is shown in Table~\ref{GAtable}.
Overall we performed 10 runs of this pMSSM-GA implementation, varying only the initial seed of the random number generator. The results did not change significantly between runs, or for longer runs.

\subsection{Muon Anomalous Magnetic Moment}
The measured muon anomalous magnetic moment \cite{Bennett:2006fi} shows a $3.5\sigma$ deviation from the SM value, which could potentially be explained by supersymmetric contributions.  
The value of $\amu$ for the MSSM was computed with \micro, and the latest experimental average used from Ref.~\cite{Olive:2016xmw}  (see Table~\ref{tab:constraints}) in a Gaussian probability distribution function, $\like_{\amu}$.
Thus, the joint likelihood function reads,
\begin{equation}
 \lnlike_{\rm Joint} ~=~ \lnlike_{\rm EWPO} + \lnlike_{B}  + \lnlike_{\rm Higgs} + \lnlike_{\rm LEP} + \lnlike_{\rm LHC} + \lnlike_{\relic}+ \lnlike_{\amu} ~. 
 \label{eq:jointlike}
\end{equation}

\subsection{The Galactic Center Excess}
For the later treatment of the Galactic Center Excess (GCE), we incorporated it into the joint likelihood as
\begin{equation}
 \lnlike_{\rm Joint} ~=~ \lnlike_{\rm EW} + \lnlike_{B}  + \lnlike_{\rm Higgs} + \lnlike_{\rm LEP} + \lnlike_{\rm LHC}+ \lnlike_{\relic}  + \lnlike_{\rm GCE}~.
\end{equation}
Note that here we do not now take into account the likelihood from $\amu$.

To evaluate $\chisqGCE$, the procedure outlined in Ref.~\cite{Cerdeno:2015ega} was followed. That is we convoluted the differential photon spectrum of a given point of the parameter space with the energy resolution of the LAT instrument. We  used the {\tt P8REP-SOURCE-V6} total (front and back) resolution of the reconstructed incoming photon energy as a function of the energy for normally incident photons.
Then $\chisqGCE$ was calculated as follows~\cite{Calore:2014xka}:
\begin{equation}
	\chisqGCE~=~ \sum_{ij}\left(\frac{d\bar{N}}{dE_i}(\boldsymbol{\theta})-\frac{dN}{dE_i}\right)
	\Sigma^{-1}_{ij}\left(\frac{d\bar{N}}{dE_j}(\boldsymbol{\theta})-\frac{dN}{dE_j}\right),
	\label{eq:chi2}
\end{equation}
where $\Sigma_{ij}$ is the covariance matrix containing the statistical errors and the diffuse model and residual systematics obtained in Ref.~\cite{Achterberg:2017emt} using the reprocessed Fermi-LAT Pass~8 data from 6.5 yr of observations. $dN/dE_i$ ($d\bar{N}/dE_i$) stands for the measured (predicted) flux in the $i$th energy bin. The measured flux corresponds to the GCE spectrum from Ref.~\cite{TheFermi-LAT:2017vmf}, derived using the Sample Model (see Section 2.2 of Ref.~\cite{TheFermi-LAT:2017vmf} for a complete description of this model). 
The vector $\boldsymbol{\theta}$ refers to the pMSSM parameters that determine the predicted photon flux.

\section{Results}
\label{sec:results}

\subsection{Muon Anomalous Magnetic Moment}
In Fig.\ref{fig:gen_amu}, we represent the evolution of the minimum $\chisq$ (associated with the maximum fitness) as a function of the generation number for each of the ten runs. As already mentioned, the maximum fitness is a monotonically increasing function (due to the elitism), which results in a monotonically decreasing $\chisq$. The evolution proceeds rapidly during the first iterations and stabilises after approximately 100 generations, with no apparent differences among the various runs.

\begin{figure}[h!]
\centering
\includegraphics[width=0.65\linewidth]{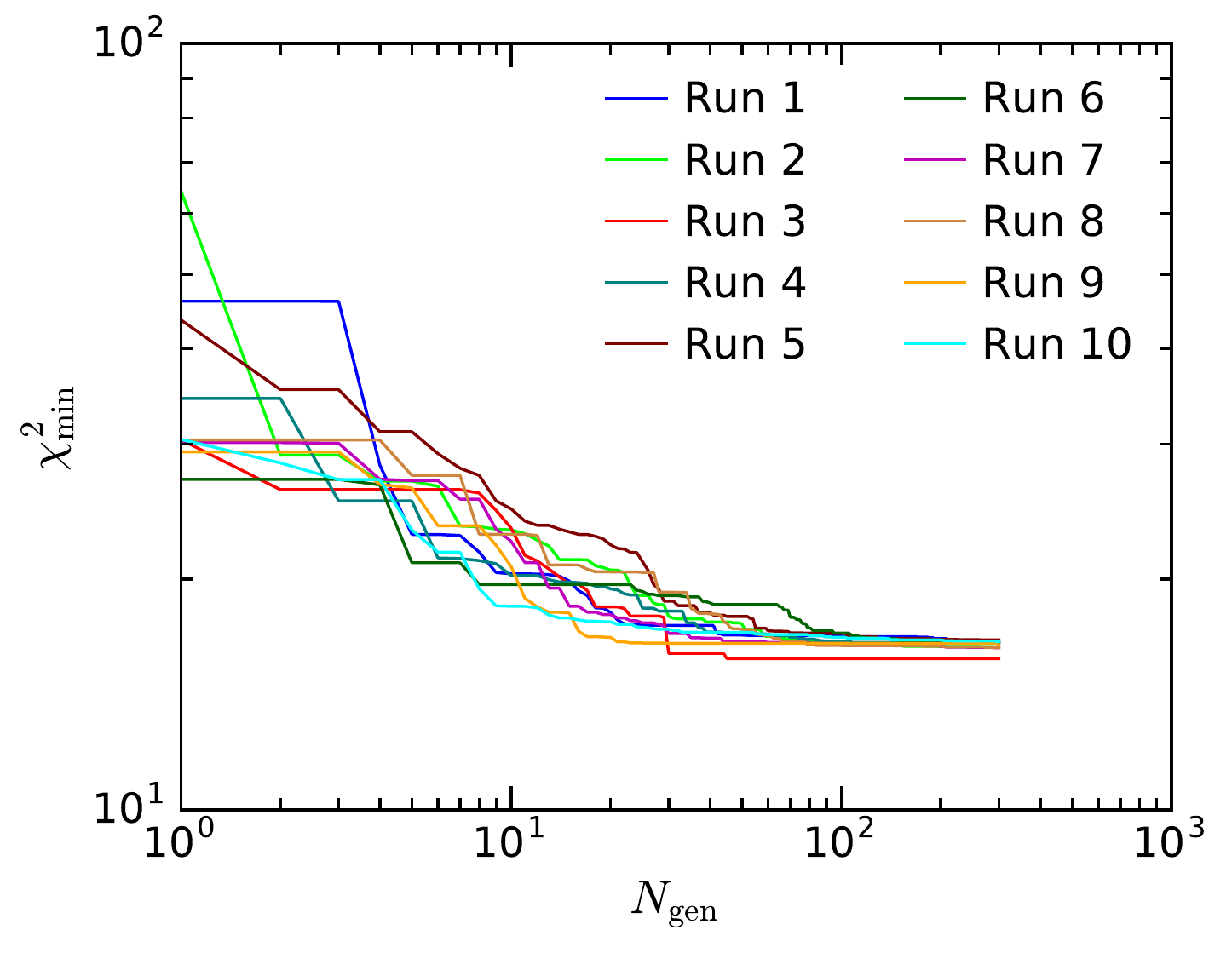}
\caption{$\chisqmin$ vs. number of generations for the ten runs.}
\label{fig:gen_amu}
\end{figure}

The goodness of the best-fit point for each run is shown in Table~\ref{tab:chi_amu}, where we also include the contribution from each observable. The total $\chisq$  is of order $\chisq\approx 16$ for the ten runs. The greatest contribution always comes from the muon anomalous magnetic moment ($\chisq_{\amu}\approx12$), while the predictions for the other observables are in good agreement with the experimental results.
 For example, the combination of Higgs observables leads to $\chisq_{\rm HiggsSignals}\approx 1.2$. The fit to the invisible $Z$-width, which leads to $\chisq_{\GamZ}$ is consistent with the SM prediction.
There is an evident tension between the muon anomalous magnetic moment and the rest of the observables. A good fit to the latter is only possible at the expense of a very small supersymmetric contribution to $a_\mu$.
Table~\ref{tab:bestfit_amu} shows the corresponding values of the observables for these best fit points, where we can observe that the resulting $\amu$ is always two orders of magnitude smaller than the observed $\amu=26.8^{+6.3}_{-4.3} \times 10^{-10}$.  
The tension between the observed value of the Higgs mass and the muon anomalous magnetic moment is well documented in the literature (see e.g. Ref.~\cite{Endo:2013bba}).

\begin{table}[h!]
  \centering
 {\small\setlength{\tabcolsep}{3pt}\renewcommand{\arraystretch}{1.2}%  
\begin{tabular}{|l|r|r|r|r|r|r|r|r|r|r|}
\hline
  & Run 1 & Run 2 & Run 3 & Run 4 & Run 5 & Run 6 & Run 7 & Run 8 & Run 9 & Run 10\\
\hline
$\chisq_{\relicneu}$ & 0.0067 & 0.0044 & 0.0174 & 0.0002 & 0.0045 & 0.0035 & 0.0096 & 0.0021 & 0.0000 & 0.0020\\
$\chisq_{\rm HiggsSignals}$ & 1.2950 & 1.2983 & 1.1452 & 1.2899 & 1.2902 & 1.2914 & 1.1579 & 1.2811 & 1.2804 & 1.2995\\
$\chisq_{\mh}$ & 0.1125 & 0.2174  & 0.0005 & 0.0921 & 0.0879 & 0.0782 & 0.3911 & 0.0656 & 0.1475 &  0.1331\\
$\chisq_{\mW}$  &  0.1190 & 0.0350 & 0.0008 & 0.1006 & 0.2500 & 0.0223 & 0.0004 & 0.1642 & 0.1205 & 0.2239\\
$\chisq_{\swsqeff}$  & 0.1538 & 0.1463 & 0.1569 & 0.1575 & 0.1552 & 0.1665 & 0.1639 & 0.1601 & 0.1567 & 0.1470\\
$\chisq_{\GamZ}$  & 0.0332 & 0.0121 &  0.0001 & 0.0602 & 0.0388 & 0.1175 & 0.0102 & 0.0451 & 0.0362 & 0.0561\\
$\chisq_{\GamZinv}$  & 2.3054 & 2.3027 & 2.2842 & 2.3056 & 2.3045 & 2.3089 & 2.2998 & 2.3028 & 2.3003 & 2.3024\\
$\chisq_{\brbsgamma}$  & 0.0664 & 0.0741 & 0.0596 & 0.0911 & 0.0689 & 0.1050 & 0.1664 & 0.0929 & 0.0717 & 0.0761\\
$\chisq_{\brbsmumu}$  & 0.1647 & 0.1818 & 0.1498 & 0.1707 & 0.1617 & 0.1623 & 0.1733 & 0.1888 & 0.1715 & 0.1593\\
$\chisq_{\RBtaunu}$  & 0.0140 & 0.0143 & 0.0142 & 0.0143 & 0.0141 & 0.0140 & 0.0142 & 0.0154 &  0.0143 & 0.0140\\
$\chisq_{\rm LEP}$  & 0.0000 & 0.0000 & 0.0000 & 0.0000 & 0.0000 & 0.0000 & 0.0000 & 0.0000 & 0.0000 & 0.0000\\
$\chisq_{\rm LHC}$  & 0.0000  & 0.0000 & 0.0000 & 0.0000 & 0.0000 & 0.0000 & 0.0000 & 0.0000 & 0.0000 & 0.0000\\
{\color{blue}${ \chisq_{\amu}}$}  & {\color{blue} 12.2691} & {\color{blue}12.0273} & {\color{blue}11.9275} & {\color{blue} 12.2113} & {\color{blue}12.2873} & {\color{blue}12.2926} & {\color{blue}11.8926} & {\color{blue}11.9721} & {\color{blue}12.1162} & {\color{blue}12.1683}\\
\hline
$\chisq_{\rm tot}$ & 16.5398 & 16.3138 & 15.7562 & 16.4935 & 16.6631 & 16.5621 & 16.2793 & 16.2904 & 16.4152 & 16.5816\\
\hline
\end{tabular} 
}
\caption{Contributions to the $\chisq$ of the best fit points. In blue, we show the leading contribution, which comes from the fit to the muon anomalous magnetic moment, $\amu$.}
\label{tab:chi_amu}
\end{table}

\begin{table}[h!]
  \centering
 {\footnotesize\setlength{\tabcolsep}{2.5pt}\renewcommand{\arraystretch}{1.3}%  
 \begin{tabular}{|l|r|r|r|r|r|r|r|r|r|r|}
 \hline 
 Observable & Run 1  & Run 2 & Run 3 & Run 4 & Run 5 & Run 6 & Run 7 & Run 8 & Run 9 & Run 10\\
 \hline 
 $\mh(\GeV)$  & 124.42 & 124.15 & 125.13 & 124.48 & 124.49 & 124.53 & 123.83 & 124.57 & 124.32 & 124.36\\ 
  $\mW(\GeV)$ & 80.379 & 80.382 & 80.386 & 80.379 & 80.376 & 80.382 & 80.385 & 80.378 & 80.379 &  80.377\\ 
 $\swsqeff$ & 0.23146 & 0.23146 & 0.23146 & 0.23146 & 0.23146 & 0.23145 & 0.23145 & 0.23146 & 0.23146 & 0.23146\\ 
 $\GamZ(\GeV)$ & 2.4947	& 2.4949 & 2.4952 & 2.4946 & 2.4947 & 2.4943 & 2.4950 & 2.4947 & 2.4947 & 2.4946\\
 $\GamZinv(\GeV)$ & 0.5017 & 0.5017 & 0.5017 & 0.5017 & 0.5017 & 0.5017 &  0.5017 & 0.5017 & 0.5017 & 0.5017\\ 
 $\brbsgamma\times10^{4}$ & 3.35 & 3.34 & 3.35 & 3.33 & 3.34 & 3.32 & 3.30 & 3.33 & 3.34 & 3.34\\
 $\brbsmumu\times10^{9}$  & 3.21 & 3.22 & 3.19 & 3.21 & 3.20 & 3.21 & 3.22 & 3.23 & 3.21 & 3.20\\
 $\RBtaunu$ & 1.00  & 1.00 & 1.00 & 1.00 & 1.00  & 1.00 & 1.00 & 1.00 & 1.00 & 1.00\\
 $\relicneu$ & 0.1178 & 0.1180 & 0.1204 & 0.1190 & 0.1180 & 0.1195 & 0.1200 & 0.1194 & 0.1188 & 0.1193\\
 {\color{blue}${\amu\times10^{10}}$} & {\color{blue}0.0827} & {\color{blue}0.3472} & {\color{blue}0.4572} & {\color{blue}0.1457} & {\color{blue}0.0063} & {\color{blue}0.0057} & {\color{blue}0.4958} & {\color{blue}0.4081} & {\color{blue}0.2497} & {\color{blue}0.1927}\\ 
 \hline
 \end{tabular}
}
\caption{Observable values for the best fit points. In blue, we display the results for $\amu$, which show a large discrepancy with the observed value.}
\label{tab:bestfit_amu}
\end{table}

The top plot of Fig.~\ref{fig:massspectrum_all_amu} shows the resulting SUSY spectrum for the particular case of run~3. The colour code is a visual aid to illustrate the evolution of the GA towards a final result. Blue corresponds to early generations, green to late ones, and the final generation, 300, is shown in yellow. The same colour map will be used throughout all the plots in this paper. 
Note that it is entirely expected that there will still be unfit individuals in the population exhibiting a large $\chisq$. For this reason, a useful approach is to collate the best fit points from all the different runs. 
The bottom plot of  Fig.~\ref{fig:massspectrum_all_amu} includes the information from all the ten runs, together with the corresponding best fit points. For convenience, these are also listed in Table~\ref{tab:bestfit_spect_amu}. As the population evolves, one can observe clustering around certain solutions. Whereas the best fit points seem to favour specific ranges of masses in the lightest neutralino and chargino, they appear more spread in the squark and slepton sector. A pattern emerges where $\mneu{1}\approx\mcha{1}\approx 2$~TeV, the squark masses are generally above 6~TeV (except for the lightest stop, for which $\mstop{1}\approx2-3$~TeV), and slepton masses show a wide range of variation $2-10$~TeV. For completeness, the pMSSM input parameters (19 soft supersymmetry-breaking terms and four nuisance parameters) for the best fit points of each run are listed in Table~\ref{tab:bestfit_ini_amu}.

\begin{table}[h!]
  \centering
 {\footnotesize\setlength{\tabcolsep}{2pt}\renewcommand{\arraystretch}{1.3}%  
\begin{tabular}{|l|r|r|r|r|r|r|r|r|r|r|}
\hline
Parameter & Run 1  & Run 2  & Run 3 & Run 4 & Run 5 & Run 6 & Run 7 & Run 8 & Run 9 & Run 10 \\
\hline
\multicolumn{11}{|c|}{SM} \\
\hline
$\invalphEM$ & 128.0152 & 128.0177 & 128.0127 & 128.0171 & 128.0161 & 128.0167 & 128.0119 & 128.0138 & 128.0119 & 128.0161\\
$\alphs$ & 0.1171 & 0.1174 & 0.1183 & 0.1169 & 0.1171 &  0.1164 & 0.1175 & 0.1170 & 0.1171 & 0.1168\\ 
$\mb(\GeV)$ & 4.5417 & 4.5560 & 4.7420 & 4.5973 & 4.5680 & 4.5552 & 4.5468 & 4.5943 & 4.5403 & 4.5407\\
$\mt(\GeV)$ & 175.4984 & 175.4853 & 175.3661 & 175.4839 & 175.4868 & 175.4799 & 175.4380 & 175.4675 & 175.4268 & 175.4351\\
\hline
\multicolumn{11}{|c|}{pMSSM (GUT scale)} \\
\hline
$M_1(\TeV)$ & 9.5703 & 9.7113 & 3.9085 & 7.5518 & 5.7486 & 8.6224 & 7.4272 & 7.8098 & 9.2195 & 6.2369\\
$M_2(\TeV)$ & 2.5890 & 2.4750 & 2.1502 & 2.5057 & 2.6390 & 2.6583 & 2.2613 & 2.2840 & 2.4634 & 2.6338\\
$M_3(\TeV)$ & 3.2780 & 2.2800 & 1.1040 & 3.2561 & 4.0505 & 3.1197 & 1.3771 & 2.1184 & 2.4783 & 4.0718\\
$\mHu(\TeV)$ & 1.6667 & 1.3993 & 4.1690 & 5.8189 & 5.0971 & 2.2366 & 0.3387 & 6.8716 & 0.9580 & 6.3635\\
$\mHd(\TeV)$ & 7.1708 & 5.2775 & 8.2714 & 1.1057 & 6.3784 & 8.0392 & 6.9334 & 1.6798 & 3.1264 & 7.2671\\
$\msq{3}(\TeV)$ & 4.7038 & 4.9000 & 9.0737 & 7.0818 & 5.8858 & 5.9333 & 2.2149 & 6.0595 & 5.8636 & 6.6930\\
$\msq{1,2}(\TeV)$ & 5.3936 & 7.6163 & 3.6165 & 8.0554 & 4.6423 & 9.7422 & 4.8321 & 1.0849 & 7.1282 & 9.5621\\
$\msu{3}(\TeV)$ & 0.1784 & 3.9889 & 6.6280 & 2.1355 & 1.9408 & 3.0753 & 4.0231 & 4.4181 & 3.1066 & 0.6947\\
$\msu{1,2}(\TeV)$ & 1.4621 & 2.2405 & 8.0701 & 4.2783 & 6.5618 & 7.0730 & 1.6677 & 1.5040 & 6.5059 & 6.8387\\
$\msd{3}(\TeV)$ & 0.7708 & 0.4281 & 5.4714 & 7.2873 & 1.4363 & 0.3643 & 0.1945 & 0.0882 & 1.8254 & 0.7754\\
$\msd{1,2}(\TeV)$ & 0.1395 & 4.1474 & 7.2591 & 0.4960 & 3.9536 & 3.1839 & 0.7612 & 0.9395 & 1.3482 & 5.1211\\
$\msl{3}(\TeV)$ & 6.7797 & 8.9868 & 4.8607 & 8.7979 & 7.6243 & 9.8535 & 7.6822 & 6.3356 & 7.6277 & 8.3796\\
$\msl{1,2}(\TeV)$ & 7.5105 & 0.8366 & 2.6651 & 4.1286 & 9.0621 & 8.6661 & 1.0870 & 1.7923 & 1.6668 & 3.3159\\
$\mse{3}(\TeV)$ & 1.3255 & 4.6784 & 3.1272 & 3.3281 & 0.7113 & 4.5917 & 8.9473 & 6.9846 & 7.7331 & 3.0309\\
$\mse{1,2}(\TeV)$ & 9.4672 & 1.2755 & 1.8426 & 8.3371 & 9.0375 & 9.8130 & 1.1521 & 5.6990 & 6.7869 & 5.9095\\
$\At(\TeV)$ & -9.4740 & -9.8184 & -9.6588 & -9.9870 & -9.9878 & -9.7382 & -7.3392 & -9.3542 & -9.8716 & -9.6492\\
$\Ab(\TeV)$ & -7.7118 & -6.1908 &  8.5946 & 0.3908 & 9.4784 & -6.1000 & -1.2494 & -2.1864 & 2.3702 & -1.1100\\
$\Atau(\TeV)$ & 8.6880 & 6.2710 & -6.5096 & -0.0063 & -9.9068 & -0.9948 & 8.6904 & 8.9710 & -9.2448 & -1.9952\\
$\tanb$ & 22.6238 & 29.3282 & 28.4906 & 21.5324 & 19.4162 & 18.4574 & 26.7140 & 22.5170 & 24.8102 & 20.9054\\
\hline
\end{tabular} 
}
\caption{Input parameters for the best fit points.}
\label{tab:bestfit_ini_amu}
\end{table}

\begin{figure}[h!]
\centering
\includegraphics[width=1\linewidth]{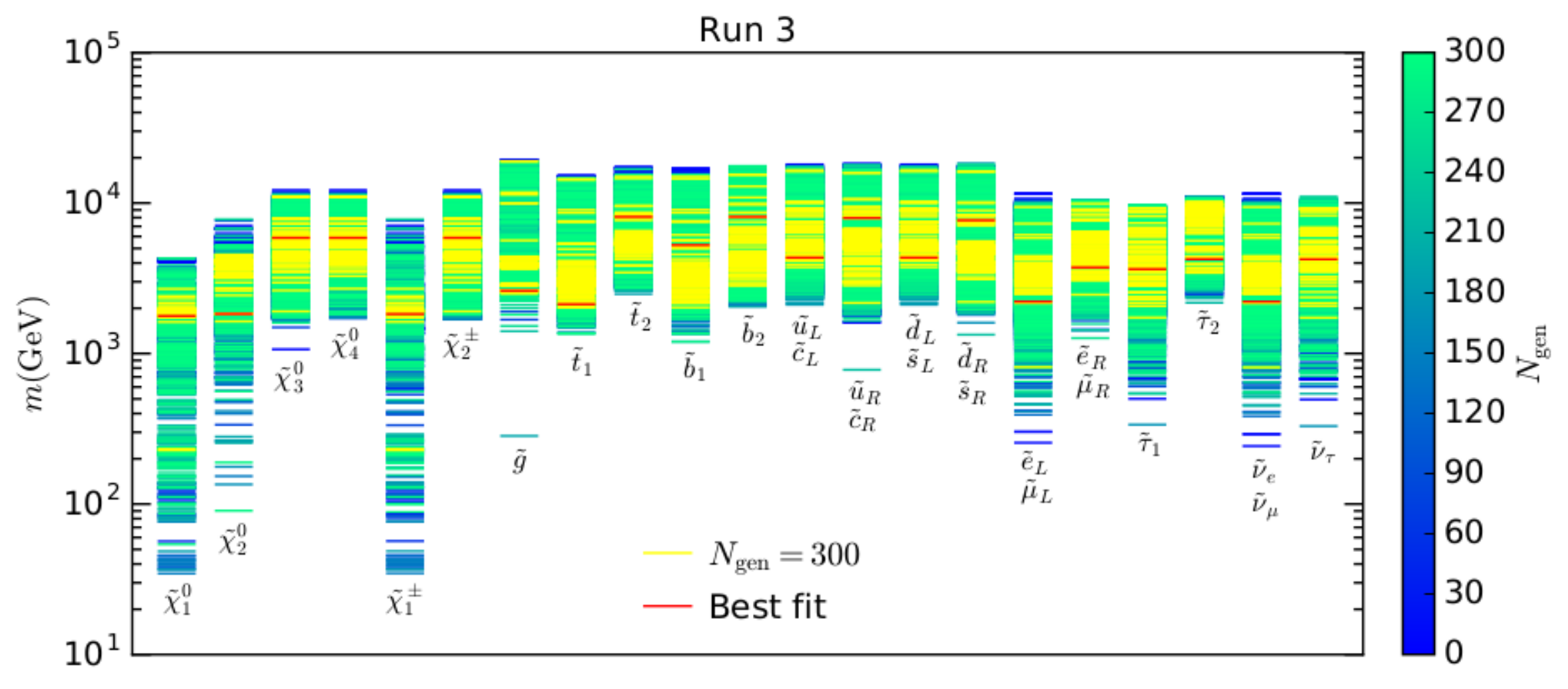}
\includegraphics[width=1\linewidth]{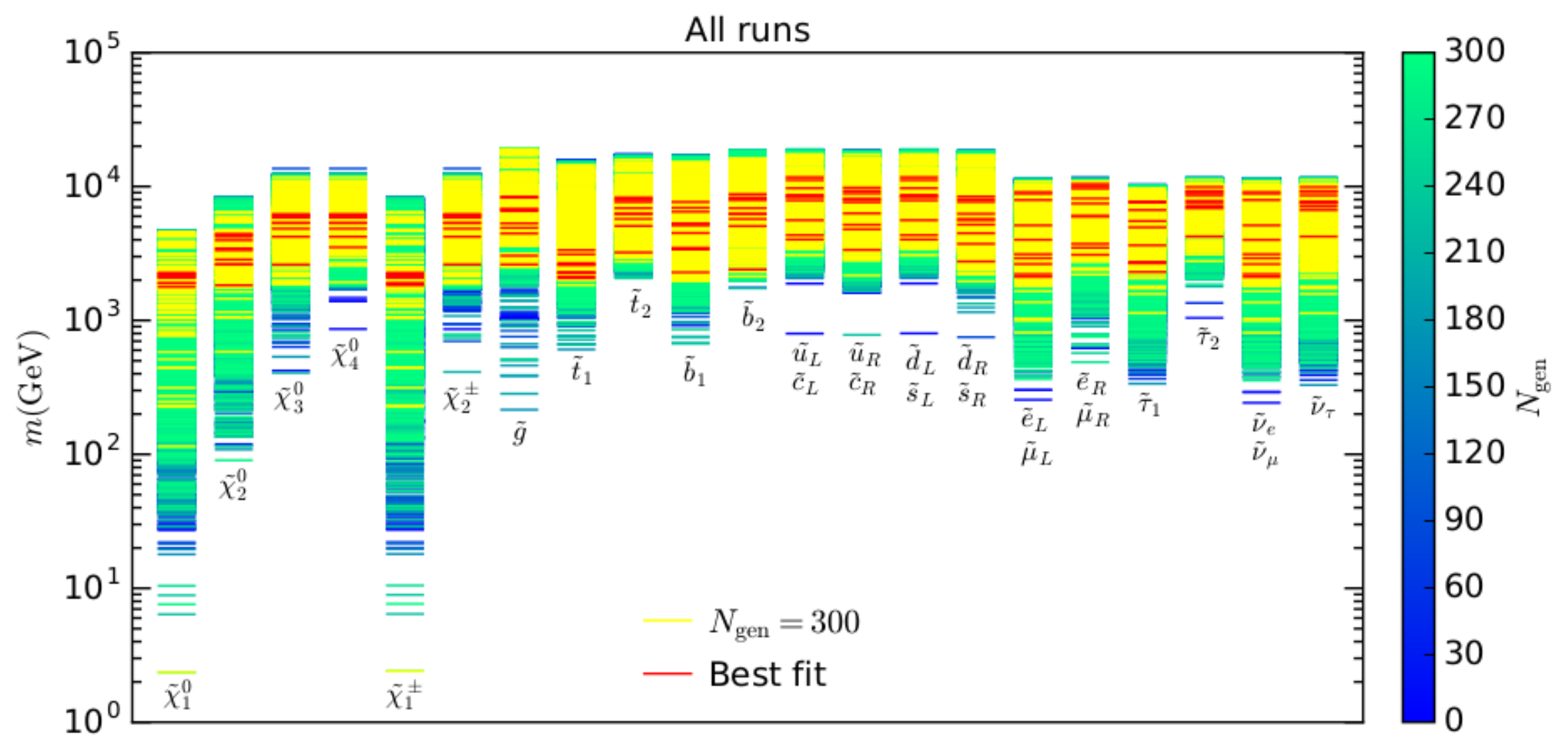}
\caption{(Top) SUSY spectrum for all the generations in run 3. Yellow represents the results for the last generation and the red line corresponds to the best fit point. (Bottom) The same, but including the results for the ten runs.}
\label{fig:massspectrum_all_amu}
\end{figure}

\begin{table}[h!]
  \centering
 {\renewcommand{\arraystretch}{1.3}%  
\begin{tabular}{|c|r|r|r|r|r|r|r|r|r|r|}
\hline
m(TeV) & Run 1  & Run 2 & Run 3 & Run 4 & Run 5 & Run 6 & Run 7 & Run 8 & Run 9 & Run 10\\
 \hline
$\mneu{1}$ & 2.1899 & 2.1016 & 1.7779 & 2.1262 & 2.2249 & 2.2701 & 1.8967 & 1.9043 & 2.0891 & 2.2252\\
$\mneu{2}$ & 4.3778 & 4.4140 & 1.8323 & 3.4508 & 2.6267 & 3.9700 & 3.3381 & 2.6099 & 4.2132 & 2.8475\\
$\mneu{3}$ & 6.0329 & 5.8963 & 5.8774 & 5.3433 & 5.3503 & 6.1583 & 4.2183 & 2.6122 & 6.2057 & 4.8463\\
$\mneu{4}$ & 6.0340 & 5.8975 & 5.8779 & 5.3444 & 5.3512 & 6.1593 & 4.2205 & 3.5199 & 6.2066 & 4.8475 \\
$\mcha{1}$ & 2.1901 & 2.1017 & 1.8308 & 2.1264 & 2.2250 & 2.2702 & 1.8968 & 1.9045 & 2.0893 & 2.2254\\
$\mcha{2}$ & 6.0341 & 5.8974 & 5.8784 & 5.3446 & 5.3516 & 6.1595 & 4.2200 & 2.6147 & 6.2067 & 4.8478\\

\hline
$\mglu$ & 6.7304 & 4.9071 & 2.6147 & 6.8046 & 8.2574 & 6.5645 & 3.0447 & 4.4649 & 5.3003 & 8.3863\\
$\mstop{1}$ & 3.1168 & 3.3535 & 2.1280 & 2.2745 & 2.5939 & 2.6113 & 2.2757 & 2.0702 & 2.2747 & 2.6774\\ 
$\mstop{2}$ & 6.2486 & 5.0756 &  8.1006 & 7.6622 & 7.9175 & 6.8952 & 3.2181 & 5.7306 & 6.2510 & 8.2790\\ 
$\msbot{1}$ & 5.2788 & 3.4830 & 5.2605 & 7.6589 & 6.9380 & 5.1184 & 2.2860 & 3.3888 & 4.4942 & 6.4800\\ 
$\msbot{2}$ & 6.2432 & 5.0657 & 8.0991 & 8.7702 & 7.9142 & 6.8912 & 2.4021 & 5.7259 & 6.2467 & 8.2761\\ 
$\msup{L}$ & 7.8708  & 8.6567 & 4.3350 & 9.7225 & 8.4026 & 11.1512 & 5.6043 & 4.0216 & 8.3820 & 11.7328\\
$\msup{R}$ & 6.3070 & 5.1789 & 7.9673 & 7.6343 & 9.2697 & 8.8657 & 3.3666 & 4.7984 & 8.1069 & 9.7987\\
$\msdw{L}$ & 7.8711 & 8.6569 & 4.3356 & 9.7226 & 8.4029 & 11.1513 & 5.6047 & 4.0222 & 8.3822 & 11.7329\\
$\msdw{R}$ & 5.6090 & 5.7292 & 7.6764 & 5.1876 & 8.0160 & 6.2496 & 2.7580 & 3.7299 & 4.4788 & 8.4131\\
$\msc{L}$ & 7.8708  & 8.6566 & 4.3349 & 9.7224 & 8.4026 & 11.1511 & 5.6042 & 4.0216 & 8.3819 & 11.7327\\
$\msc{R}$ & 6.3070  & 5.1789 & 7.9673 & 7.6343 & 9.2697 & 8.8657 & 3.3666 & 4.7984 & 8.1069 & 9.7986\\
$\mss{L}$ & 7.8710  & 8.6568 & 4.3355 & 9.7226 & 8.4028 & 11.1513 & 5.6046 & 4.0222 & 8.3822 & 11.7329\\
$\mss{R}$ & 5.6088  & 5.7290 & 7.6763 & 5.1875 & 8.0159 & 6.2495 & 2.7577 & 3.7299 & 4.4787 & 8.4130\\

\hline
$\mstau{1}$ & 2.6846 & 4.9582 & 3.6410 & 2.3003 & 2.6804 & 5.5212 & 7.5905 & 6.6731 & 7.7304 & 2.7379\\ 
$\mstau{2}$ & 7.0855 & 9.1210 & 4.2356 & 9.2520 & 7.4738 & 9.9211 & 8.8642 & 6.8868 & 7.8166 & 8.5809\\ 
$\msel{L}$ & 7.9354 & 2.6358 & 2.2153 & 5.1338 & 9.0696 & 8.8432 & 2.1177 & 3.1214 & 2.9146 & 4.0030\\
$\msel{R}$ & 9.9813 & 3.4901 & 3.7335 & 8.1401 & 9.6424 & 10.4569 & 3.0946 & 5.9648 & 7.4653 & 6.0597\\
$\msmu{L}$ & 7.9348 & 2.6351 & 2.2140 & 5.1334 & 9.0692 & 8.8428 & 2.1168 & 3.1210 & 2.9139 & 4.0025\\
$\msmu{R}$ & 9.9803 & 3.4891 & 3.7320 & 8.1396 & 9.6416 & 10.4562 & 3.0934 & 5.9644 & 7.4648 & 6.0590\\
$\msneu{e}$ & 7.9347 & 2.6343 & 2.2136 & 5.1329 & 9.0689 & 8.8425 & 2.1160 & 3.1201 & 2.9132 & 4.0019\\
$\msneu{\mu}$ & 7.9341 & 2.6336 & 2.2123 & 5.1325 & 9.0685 & 8.8421 & 2.1151 & 3.1197 & 2.9125 & 4.0014\\
$\msneu{\tau}$ & 7.0855 & 9.1202 & 4.2321 & 9.2513 & 7.4730 & 9.9204 & 7.5898 & 6.6725 & 7.7339 & 8.5802\\
\hline
$\mH$ & 9.2213 & 7.6930 & 9.1800 & 5.8055 & 7.9619 & 9.9571 & 7.8982 & 3.4162 & 6.9123 & 8.6629\\
$\mA$ & 9.2205 & 7.6926 & 9.1809 & 5.8055 & 7.9620 & 9.9566 & 7.8985 & 3.4162 & 6.9121 & 8.6630\\
$\mCH$ & 9.2210 & 7.6931 & 9.1791 & 5.8061 & 7.9624 & 9.9571 & 7.8992 & 3.4172 & 6.9126 & 8.6634\\
\hline
\end{tabular} 
}
\caption{SUSY spectrum for the best fit points (represented by red lines in Fig.~\ref{fig:massspectrum_all_amu}).}
\label{tab:bestfit_spect_amu}
\end{table}

\clearpage

Let us discuss the results more in detail. We will use run 3 as an example, but the results for other runs are qualitatively similar.

First, it is clear that the dark matter relic density is one of the main drivers of the evolution of the fitness function, as we can see from the right panel of Fig.~\ref{fig:chirelic_amu}, which shows  the correlation between the total $\chisq$ and $\chisq_{\relicneu}$. This is due to the high precision of the observed value of the dark matter relic abundance, but also to the fact that the relic density of the neutralino is in general very large. In order to reproduce the observed value, resonant annihilation (generally through the pseudoscalar Higgs, when $2\mneu{1}\approx m_{A^0}$) or coannihilation with the next-to-lightest supersymmetric particle (NLSP) is required \cite{Griest:1990kh}. 
The flexible structure of the pMSSM allows for various forms of coannihilation, where the NLSP can be either the lightest stau \cite{Ellis:1998kh,Ellis:1999mm}, the lightest stop \cite{Boehm:1999bj,Arnowitt:2001yh,Ellis:2001nx}, electroweakinos  (such as the second lightest neutralino or the lightest chargino) \cite{Mizuta:1992qp,Edsjo:1997bg,Baer:2002fv,BirkedalHansen:2002sx,Baer:2005jq}. 
The latter can occur in the so-called focus point region, where both the neutralino and chargino are 1~TeV Higgsino-like particles \cite{Olive:1989jg,Olive:1990qm} or $2-3$~TeV wino-like particles \cite{Hisano:2006nn}. The choice of non-universal soft parameters at the GUT scale \cite{Cerdeno:2004zj} facilitates obtaining these various solutions, contrary to more constrained scenarios such as the CMSSM.

The values of the wino soft mass parameter at the GUT scale, $M_2\approx2.5$~TeV (see Table~\ref{tab:bestfit_ini_amu}) and the hierarchy of the gaugino masses $M_2<M_3<M_1$ ensure that the lightest neutralino and the lightest chargino are both wino-like and with very similar masses (degenerate to order $1\%$). This facilitates coannihilation effects, without introducing a large fine-tuning in the dark matter sector \cite{Cabrera:2016wwr}, and is the clearest characteristic of all the runs. The composition of the lightest neutralino is shown in Fig.~\ref{fig:neut_amu}, clearly showing that the last generation corresponds to wino-like neutralino, with a subleading Higgsino component. 
The GUT values of the gaugino masss parameters are represented in Fig.~\ref{fig:gauginos_amu}. This feature occurs for all the runs. It is well known, from previous studies in non-universal SUSY models \cite{Olive:1989jg,Olive:1990qm}, that a wino-like neutralino can have the correct relic abundance for a range of masses around $2-3$~TeV.

The final generations of all the runs cluster around the observed value of the dark matter relic abundance, as the left panel of Fig.~\ref{fig:chirelic_amu} shows.
Satisfying the dark matter relic density while fulfilling all the other experimental constraints requires in general a careful choice of the initial parameters,
only possible in narrow bands of the parameter space. Finding these solutions in scans of the parameter space is therefore very costly, and it  is here that the GA excels, by the population quickly condensing on the relevant subspaces. It is indeed remarkable how easily these are obtained by a GA, requiring a relatively small number of generations. 
As we mentioned, each of the runs required approximately $10^4$ model evaluations.  
Refs.~\cite{Akrami:2009hp,Workgroup:2017htr} concluded that evolutionary algorithms can outperform  Bayesian inference tools even in relatively low dimensional models such as the CMSSM, but we find here that in broader models such as the 
pMSSM they become orders of magnitude more efficient. 

It is indeed interesting to compare these results in more detail with the previous GA scans performed in the context of the Constrained version of MSSM (CMSSM) \cite{Akrami:2009hp}, which only contains five free parameters and in which gaugino masses are assumed to be universal at the GUT scale. In that case, after applying the corresponding RGEs, one obtains $M_2>M_1$ at low-energy. Thus, the lightest neutralino cannot be wino-like, and instead, the best fit point is obtained for Higgsino-like neutralinos (with an approximate mass of 1~TeV).

\clearpage

\begin{figure}[h!]
\centering
\includegraphics[width=\linewidth]{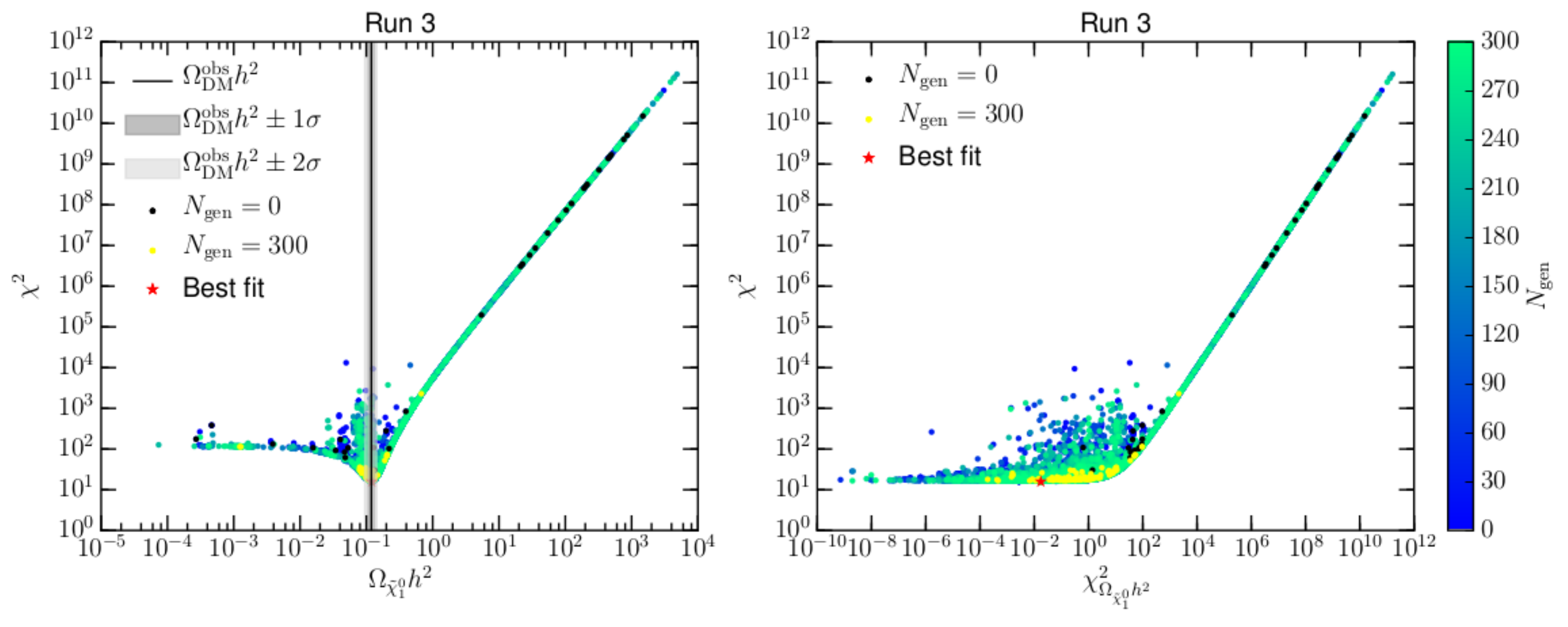}
\caption{Left: $\chisq$ vs. $\relicneu$. The solid black line corresponds to the $\relic$ mean value, see Table~\ref{tab:constraints}. As a reference, we show the $1\sigma$ and $2\sigma$ regions around the mean value in grey and light grey, respectively.
Right: $\chisq$ vs. $\chisq_{\relicneu}$.
The colour map denotes the evolution from number from generation 0 up to 300, the initial guesses ($\ngen=0$) are  depicted in black and the final generation ($\ngen=300$) in yellow. The red star corresponds to the best fit. }
\label{fig:chirelic_amu}
\end{figure}

\begin{figure}[h!]
\centering
\includegraphics[width=\linewidth]{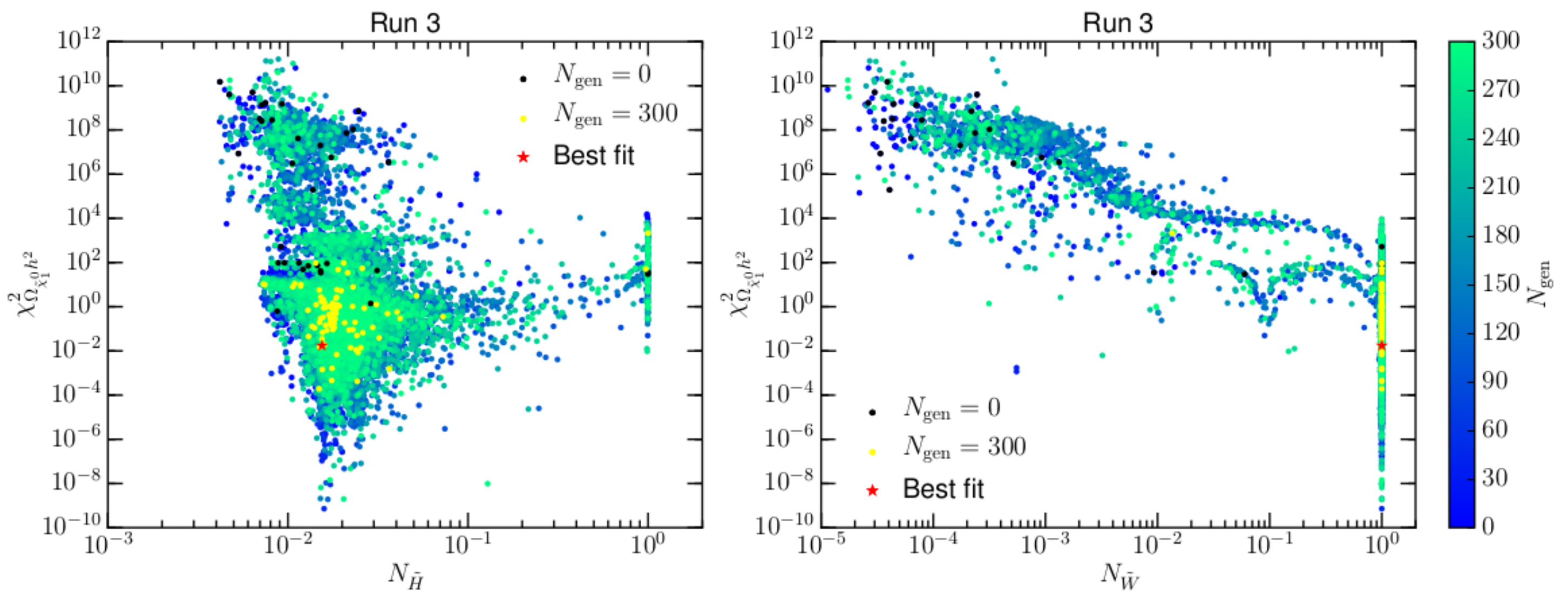}
  \caption{$\chisq_{\relicneu}$ vs. the Higgsino (left) and wino (right) component of the lightest neutralino.}
  \label{fig:neut_amu}
\end{figure}

A wino-like neutralino is not particularly easy to find through direct detection techniques (as the elastic scattering cross section with nuclei is generally dominated by Higgs exchange diagrams which are enhanced by the Higgsino component). In Fig.~\ref{fig:dm_amu}, we show the predicted contribution to the spin-independent (SI) and spin-dependent (SD) scattering cross section for all the different runs and in Table~\ref{tab:neut_detection} we include the values obtained for the best fit points. 
Note that these plots only include points with $\relicneu\leq\relic+1\sigma$: solutions with $\relicneu<\relic$ have been weighted by $\xi={\rm min}[1,\relicneu/\relic]$ as indicated in each panel. It is interesting to observe that all the best fit points are nicely grouped around the same solution, with $\sip\approx10^{-11}$~pb and $\mneu{1}\approx 2$~TeV. This is just below the projected sensitivity of LZ and potentially within the reach of the planned Darwin experiment. Notice, however, that it is extremely close to the region where the background due to coherent neutrino scattering becomes important. The spin-dependent contribution is negligible for these points. 
Regarding indirect detection, the predicted thermal averaged annihilation cross section at zero velocity is also shown in this table. It is of the order of $\sigmav\approx10^{-26}$~cm$^3$s$^{-1}$, just within the reach of the future CTA \cite{Acharya:2017ttl}, as we can see in the lower panel of Fig.~\ref{fig:dm_amu}.

\begin{figure}[h!]
\centering
\includegraphics[width=\linewidth]{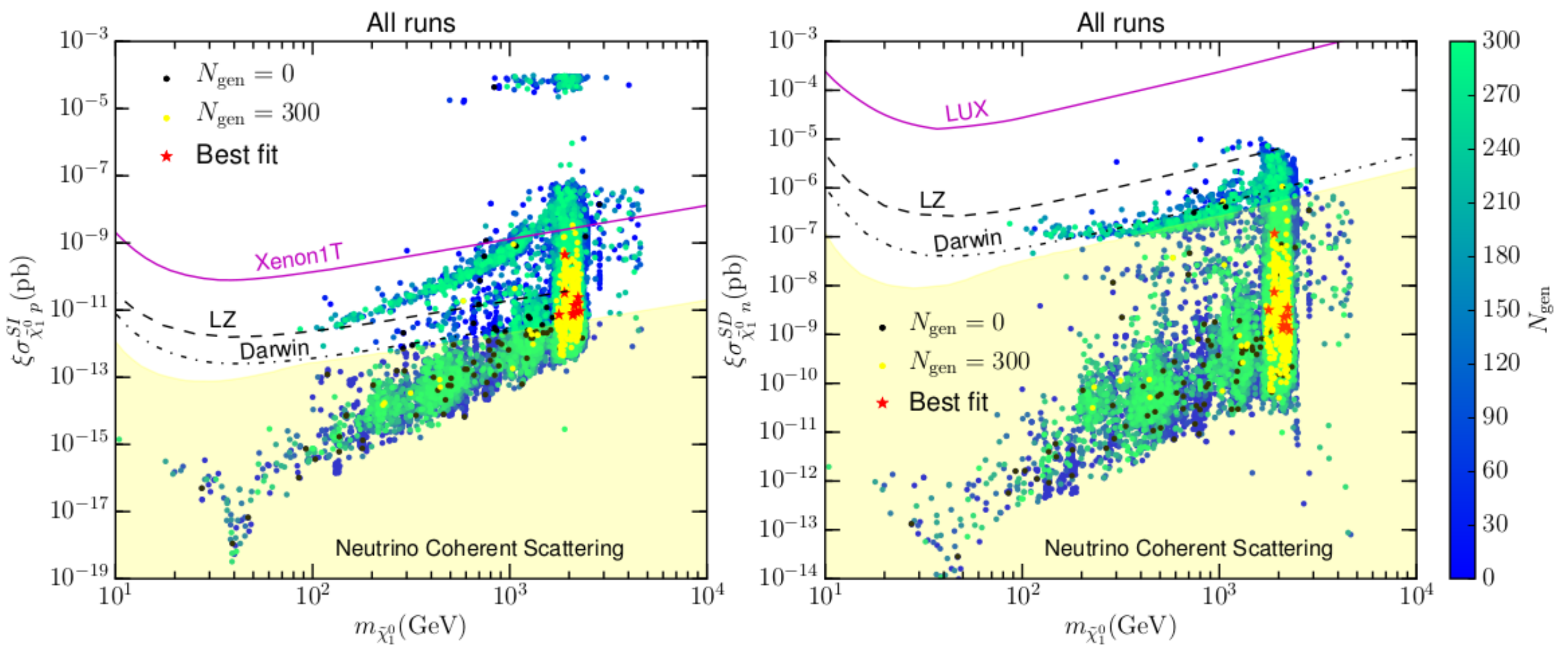}
\includegraphics[width=0.55\linewidth]{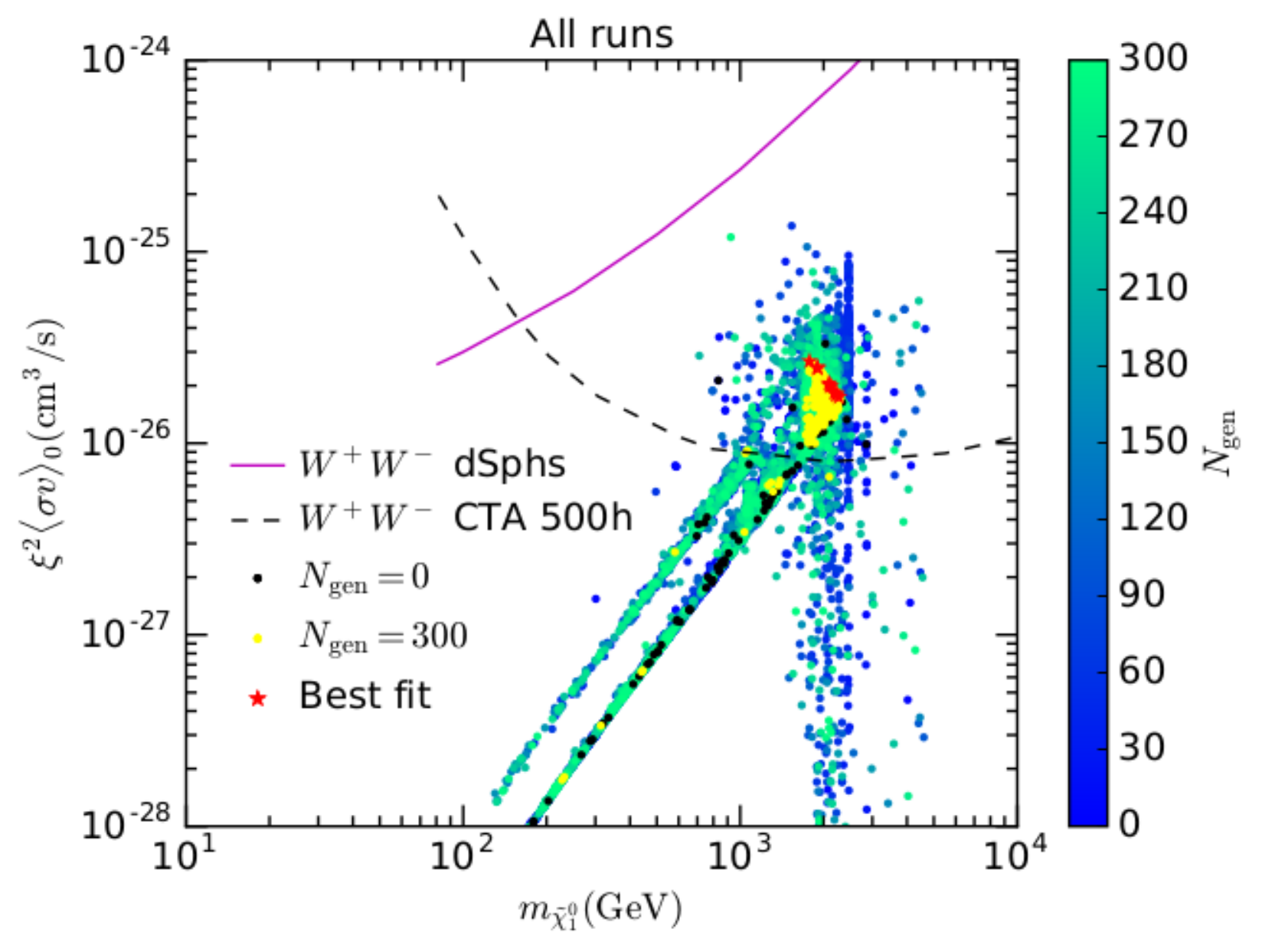}
\caption{Top left: Theoretical predictions for $\xi\sip$ as a function of $\mneu{1}$ for all the runs. 
Top right: Predictions for $\xi\sdn$ vs. $\mneu{1}$. 
The solid violet lines represent the leading constraint on SI and SD interactions from XENON1T~\cite{Aprile:2017iyp} and LUX~\cite{Akerib:2017kat}, respectively. The dashed and dot-dashed lines correspond to the sensitivity projections for LUX-ZEPLIN (LZ)~\cite{Akerib:2018lyp} and DARWIN~\cite{Aalbers:2016jon}. As a reference, we also show the irreducible neutrino background for a xenon target in yellow for SI (proton) and SD (neutron) cross sections \cite{Ruppin:2014bra}.
Bottom: Thermally averaged neutralino annihilation cross section in the Galactic halo, $\xi^2\sigmav$, as a function of the lightest neutralino mass. The upper bound on $\sigmav$ for the $W^+W^-$ annihilation channel derived from an analysis of 15 dwarf spheroidal (dSph) galaxies  using the Fermi-LAT Pass 8 reprocessed data set \cite{Ackermann:2015zua} is depicted in violet. The dashed line corresponds to the expected sensitivity of CTA for the same annihilation channel~\cite{Acharya:2017ttl}.
}
\label{fig:dm_amu}
\end{figure}

\begin{table}[h!]
  \centering
 {\renewcommand{\arraystretch}{1.3}
\begin{tabular}{|l|c|c|c|c|c|}
\hline
 & $\sip$ (pb) & $\sineu$ (pb) & $\sdp$ (pb) & $\sdn$ (pb) & $\sigmav$ (cm$^3$s$^{-1}$)\\
\hline
Run 1 & $8.77 \times 10^{-12}$ & $8.95 \times 10^{-12}$ & $1.05 \times 10^{-9}$ & $1.53 \times 10^{-9}$ & $1.87\times 10^{-26}$\\
Run 2 & $8.07 \times 10^{-12}$ & $8.23 \times 10^{-12}$ & $1.23 \times 10^{-9}$ & $1.55 \times 10^{-9}$ & $2.04\times 10^{-26}$\\
Run 3 & $7.24 \times 10^{-12}$ & $7.40 \times 10^{-12}$ & $3.70 \times 10^{-10}$ & $3.16 \times 10^{-9}$ & $2.68\times 10^{-26}$\\
Run 4 & $1.39 \times 10^{-11}$ & $1.42 \times 10^{-11}$ & $2.18 \times 10^{-9}$ & $2.27 \times 10^{-9}$ & $1.99\times 10^{-26}$\\
Run 5 & $1.60 \times 10^{-11}$ & $1.63 \times 10^{-11}$ & $2.11 \times 10^{-9}$ & $2.45 \times 10^{-9}$ & $1.81\times 10^{-26}$\\
Run 6 & $9.30 \times 10^{-12}$ & $9.48 \times 10^{-12}$ & $1.15 \times 10^{-9}$ & $1.22 \times 10^{-9}$ & $1.75\times 10^{-26}$\\
Run 7 & $3.24 \times 10^{-11}$ & $3.30 \times 10^{-11}$ & $5.32 \times 10^{-9}$ & $7.38 \times 10^{-9}$ & $2.50\times 10^{-26}$\\
Run 8 & $4.45 \times 10^{-10}$ & $4.54 \times 10^{-10}$ & $1.12 \times 10^{-7}$ & $1.17 \times 10^{-7}$ & $2.46\times 10^{-26}$\\
Run 9 & $7.01 \times 10^{-12}$ & $7.15 \times 10^{-12}$ & $9.36 \times 10^{-10}$ & $1.30 \times 10^{-9}$ & $2.06\times 10^{-26}$\\
Run 10 & $2.38 \times 10^{-11}$ & $2.43 \times 10^{-11}$ & $3.88 \times 10^{-9}$ & $3.50 \times 10^{-9}$ & $1.81\times 10^{-26}$\\
\hline
\end{tabular} 
}  
\caption{Predictions for DM direct and indirect detection observables for the  best fit points of the ten runs}
\label{tab:neut_detection}
\end{table}

The Higgs sector is of course another important source of constraints. The Higgs boson mass is properly recovered and, as Fig.~\ref{fig:higgsmass} shows, it is an important influence in the evolution of the likelihood. The final population of models is grouped around the observed value. 
In contrast, the resulting $\chisq_{\rm HiggsSignals}$ is always smaller than 2, which shows that the values of the Higgs couplings are never too far from the observed experimental values (compatible with the SM Higgs) and are thus not relevant in minimising the total $\chisq$.
In general, the predicted Higgs mass in the pMSSM is below the observed value, and in order to maximise the one-loop contributions, the stop trilinear coupling has to lead to maximal LR mixing in the stop mass matrix \cite{Baer:2011ab,Hall:2011aa,Arbey:2011ab,Akula:2011aa}. The GUT values of these quantities for the best fit points (Table~\ref{tab:bestfit_ini_amu}) are such that this relation is fulfilled at low energy. This pushes $A_t$ to large values, whereas $\tan\beta\approx20$ is favoured, as we can see in Fig.~\ref{fig:atop_tanb}. The best fit points feature typical values of the $\mu$ parameter in the range of $5 - 6$~TeV, which leads to an EW fine-tuning of the order of thousands \cite{Casas:2014eca}.

\begin{figure}[h!]
\centering
\includegraphics[width=0.55\linewidth]{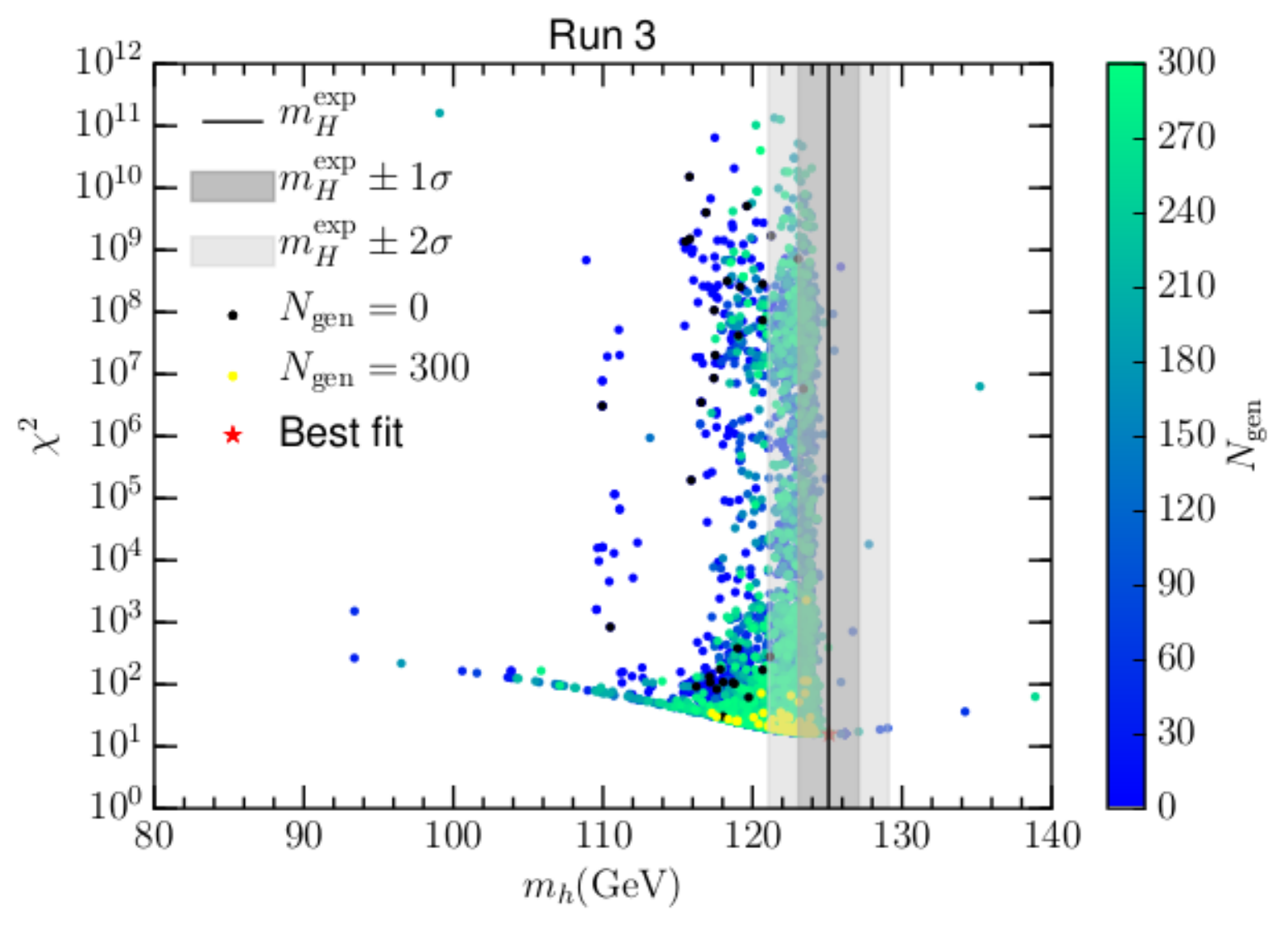}
\caption{$\chisq$ vs. Higgs mass. As a reference, we show the $\mh$ mean value (solid black line),  
the $1\sigma$ (grey) and $2\sigma$ (light grey) regions, see Table~\ref{tab:constraints} for the exact values.}
\label{fig:higgsmass}
\end{figure}

\begin{figure}[h!]
\centering
\includegraphics[width=\linewidth]{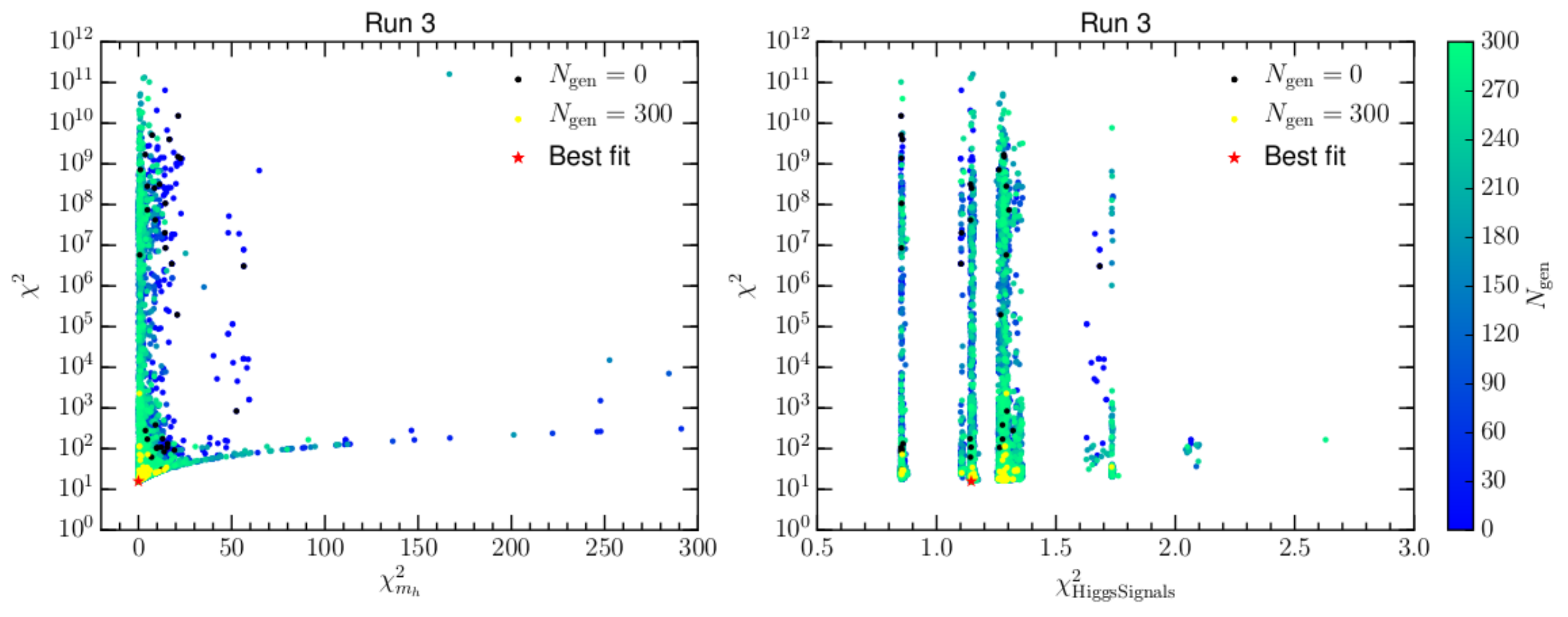}
\caption{Left: $\chisqmin$ vs. $\chisq_{\mh}$. Right: $\chisq$ vs. $\chisq_{\rm HiggsSignals}$.}
\label{fig:higgsfit}
\end{figure}

\begin{figure}[h!]
\centering
\includegraphics[width=\linewidth]{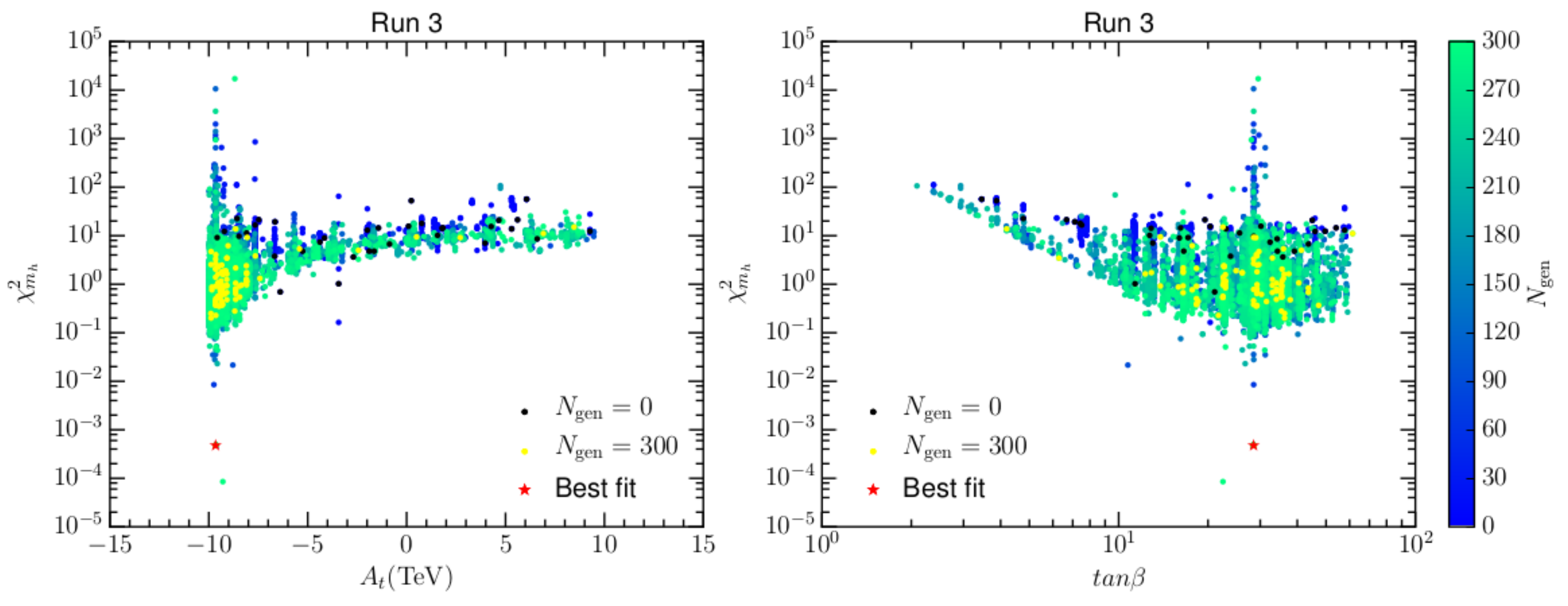}
\caption{Left: $\chisq_{\mh}$ vs. $\At$. Right: $\chisq_{\mh}$ vs. $\tanb$.}
\label{fig:atop_tanb}
\end{figure}

Figure \ref{fig:ew} contains the fit to EW observables. We  can see that both $M_W$ and $\swsqeff$ have an important influence on the fitness function. All these observables are properly reproduced in the final generation of points. 
The $Z$ boson invisible width is only due to decay into neutrinos, as the neutralino mass is in general very large, and therefore compatible with that of the SM.

\begin{figure}[h!]
\centering
\includegraphics[width=\linewidth]{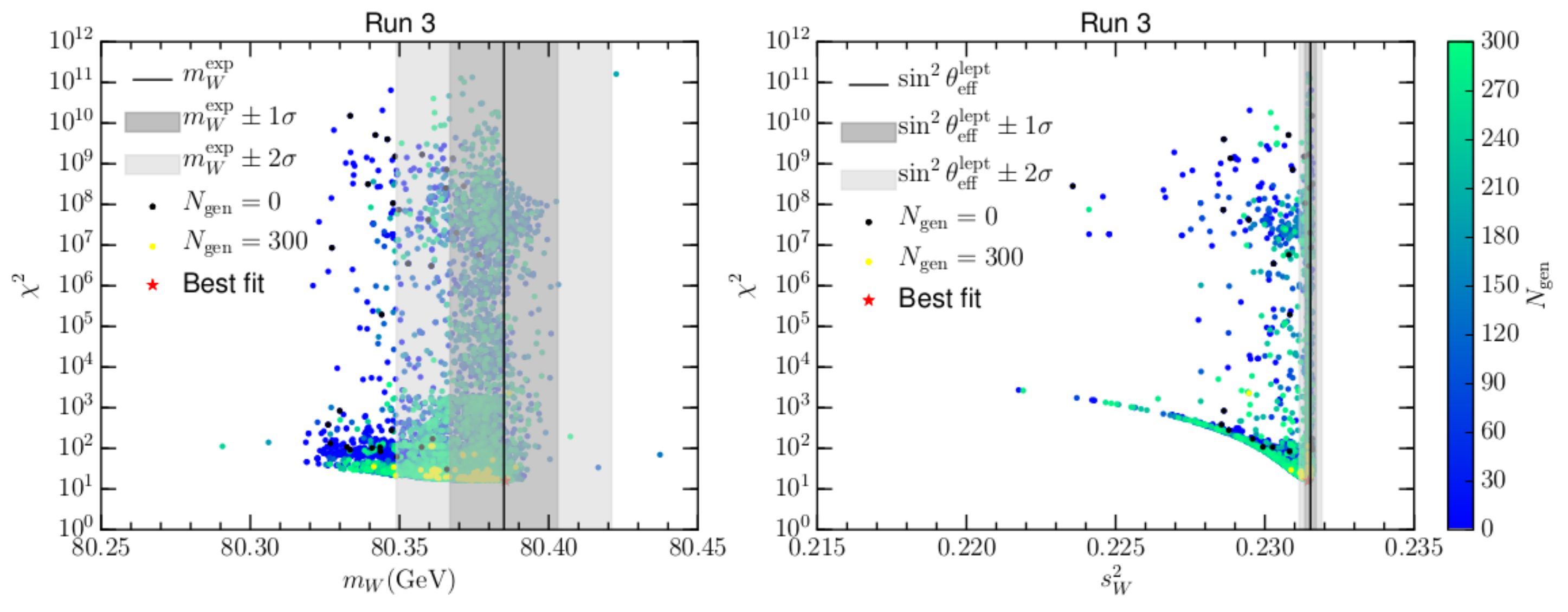}  
\includegraphics[width=\linewidth]{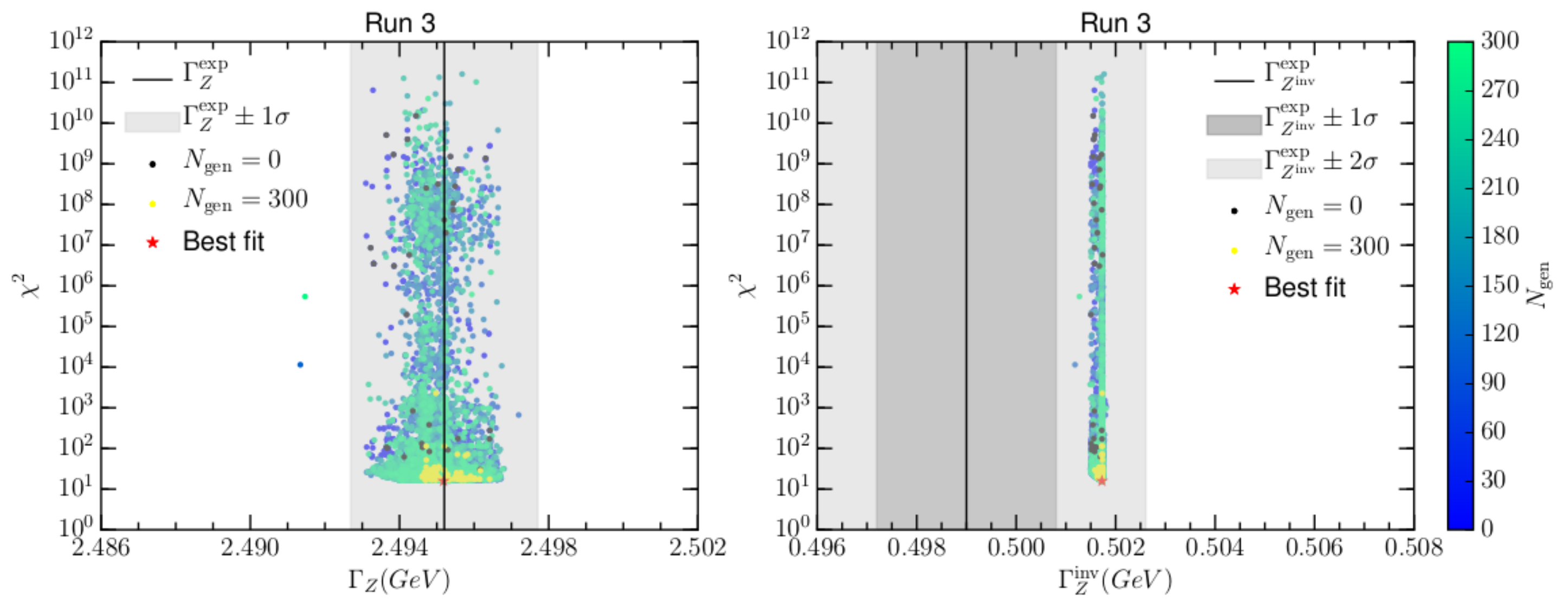}  
  \caption{Electroweak observables: Left: $\chisq$ vs. $\mW$. Right:  $\chisq$ vs. $\swsqeff$. Left: $\chisq$ vs. $\GamZ$. Right:  $\chisq$ vs. $\GamZinv$. The solid black line corresponds to the mean value of each observable, and the shaded areas to the $1\sigma$ (grey) and $2\sigma$ (light grey) regions around that value according to Table~\ref{tab:constraints}.}
  \label{fig:ew}
\end{figure}

Finally, the goodness of the fit to the muon anomalous magnetic moment is shown in Fig.~\ref{fig:amu}. It is evident from this plot that the observed value of $\amu$ is not properly reproduced and that $\amu\lesssim10^{-10}$ throughout the whole evolution (which is almost equivalent to having just the SM contribution). The supersymmetric contribution to this observable is very small, thus resulting in a $3\sigma$ discrepancy with respect to the observed value, and $\chisq_{\amu}\approx12$ for all the points.
As we can see in the right-hand side plot in Fig.~\ref{fig:amu}, $\amu$ has no impact in the GA evolution and $\chisq_{\amu}$ does not vary through the different generations\footnote{Notice that, as the $\mu$-term is taken to be positive, $\amu$ is positive and always adds to the SM contribution. Thus, the fit is always marginally better than the SM discrepancy with the observed experimental value.}. This may seem counterintuitive but is in fact a general expectation: attempting to fit this observable would degrade the fitness of  the population much more than ignoring it altogether. Consequently the fitness of the entire population is degraded equally by including it in the likelihood, but the {\it relative} fitness (which is what determines the evolution) is relatively unaffected. We conclude that this observable simply cannot be fit within the model without severely degrading the $\chi^2$. 

These results evidence the well-known tension between the muon anomalous magnetic moment and the rest of the observables. Whereas the former requires a light spectrum (in particular, light sleptons and neutralinos or charginos), LHC bounds and the value of the Higgs mass favour much heavier supersymmetric particles. There are very many LHC constraints, and in fact calculating them is the most costly part of determining the likelihood and hence fitness of a particular model. 

\begin{figure}[h!]
\centering
\includegraphics[width=\linewidth]{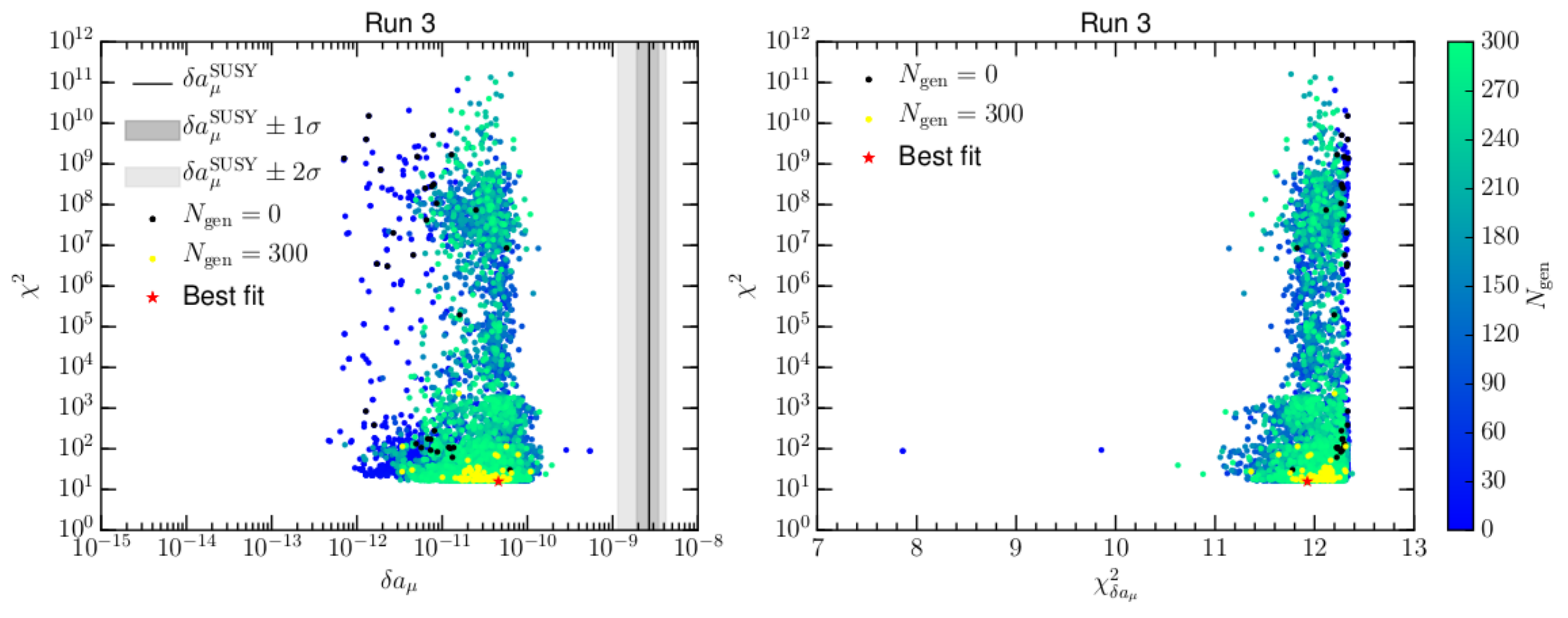}
\caption{Left: $\chisq$ vs. $\amu$. The solid black line corresponds to the $\amu$ mean value, see Table~\ref{tab:constraints}. The $1\sigma$ and $2\sigma$ regions (light grey) around the mean value are shaded in grey and light grey, respectively. Right: $\chisq$ vs. $\chisq_{\amu}$.}
\label{fig:amu}
\end{figure}

In contrast, other observables, such as $\brbsmumu$  are properly recovered, and in fact, contribute to minimising the total $\chisq$, as shown in the left-hand side of Fig.~\ref{fig:brbsmumu} ($\brbsgamma$ and $\RBtaunu$, shown in the middle and right panels of Fig.~\ref{fig:brbsmumu}, are in general in very good agreement with the experimental result).
\begin{figure}[h!]
\centering
\includegraphics[width=\linewidth]{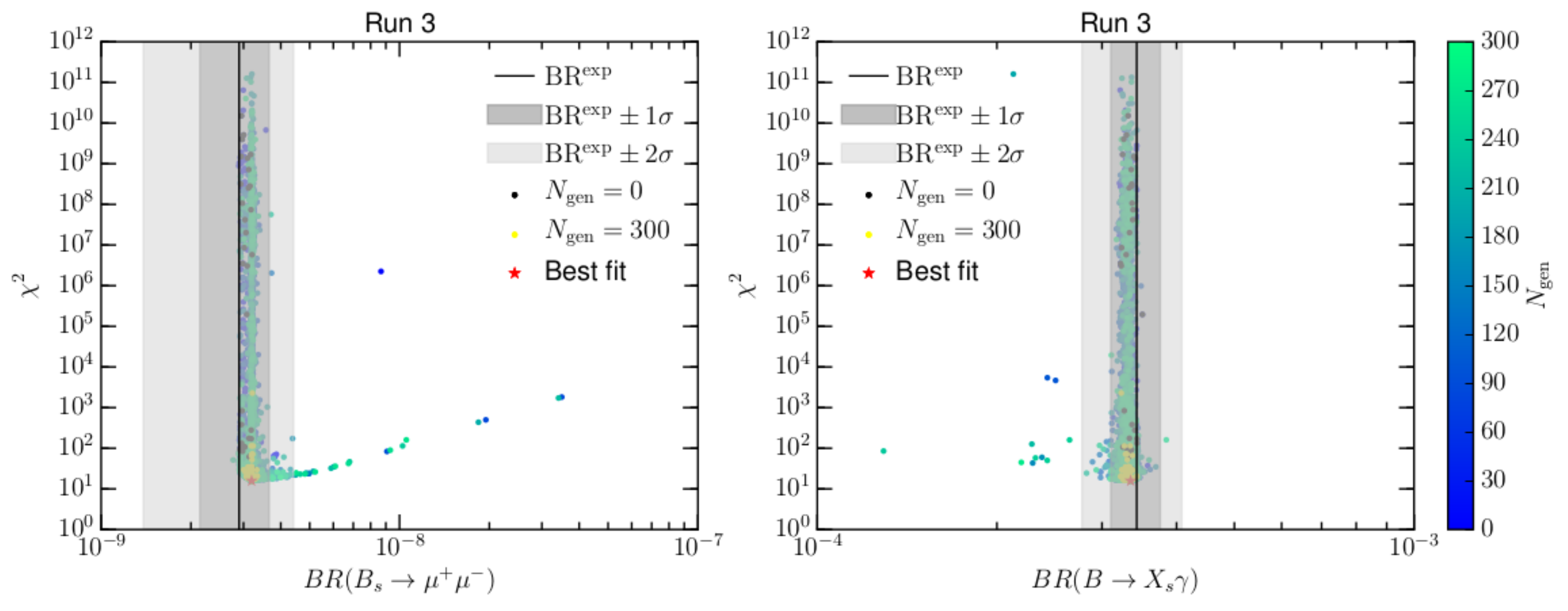}
\includegraphics[width=0.55\linewidth]{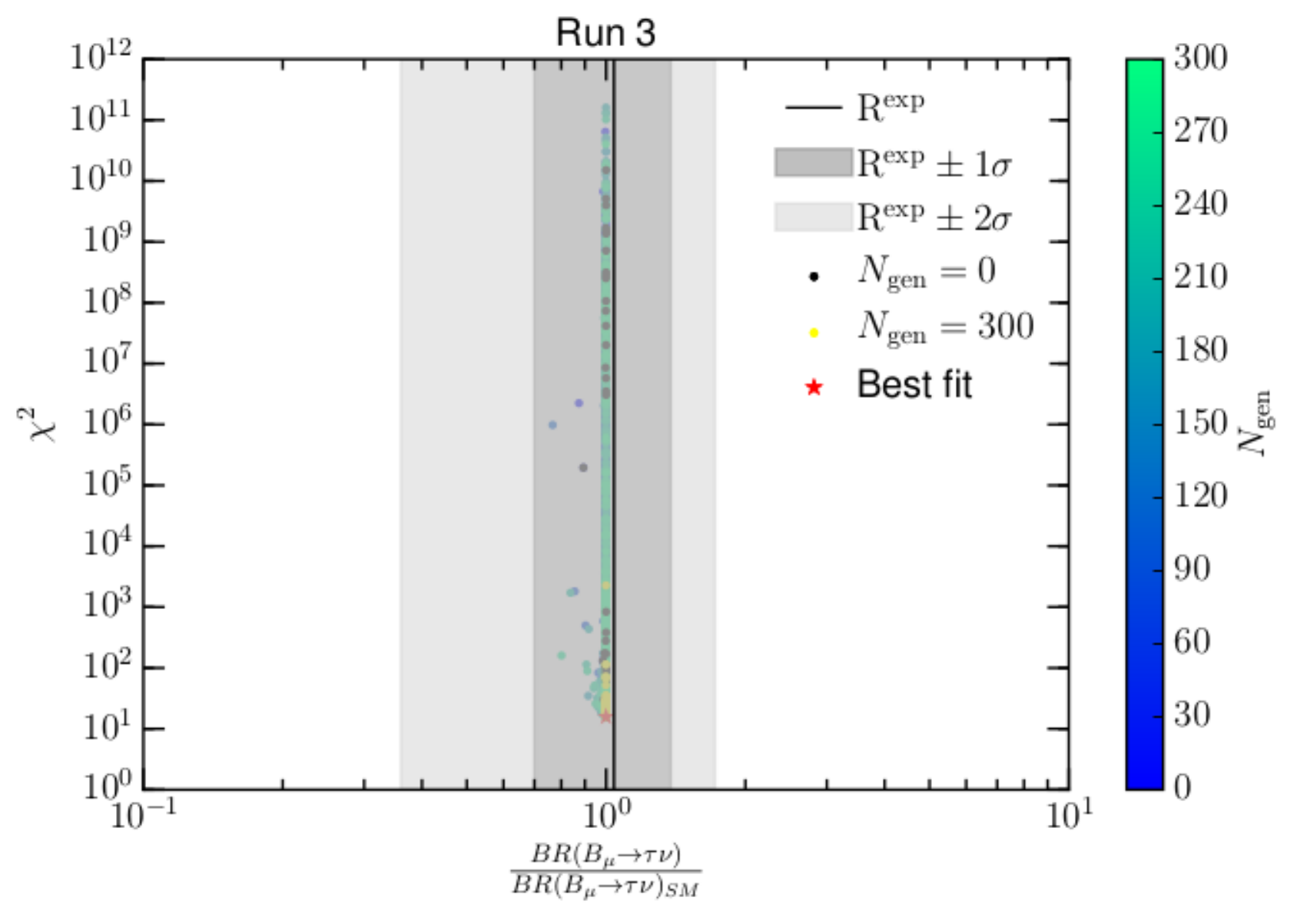}
  \caption{Top left: $\chisq$ vs. $\brbsmumu$. Top right:  $\chisq$ vs. $\brbsgamma$. Bottom: $\chisq$ vs. $\RBtaunu$. As a reference, we depict the mean value (solid black line) of each observable, and the $1\sigma$ (grey) and $2\sigma$ (light grey) regions around that value according to Table~\ref{tab:constraints}.}
  \label{fig:brbsmumu}
\end{figure}
Although we have used run 3 as an example, it should be pointed out that the other runs produce similar results.

\begin{figure}[h!]
\centering
  \includegraphics[width=\linewidth]{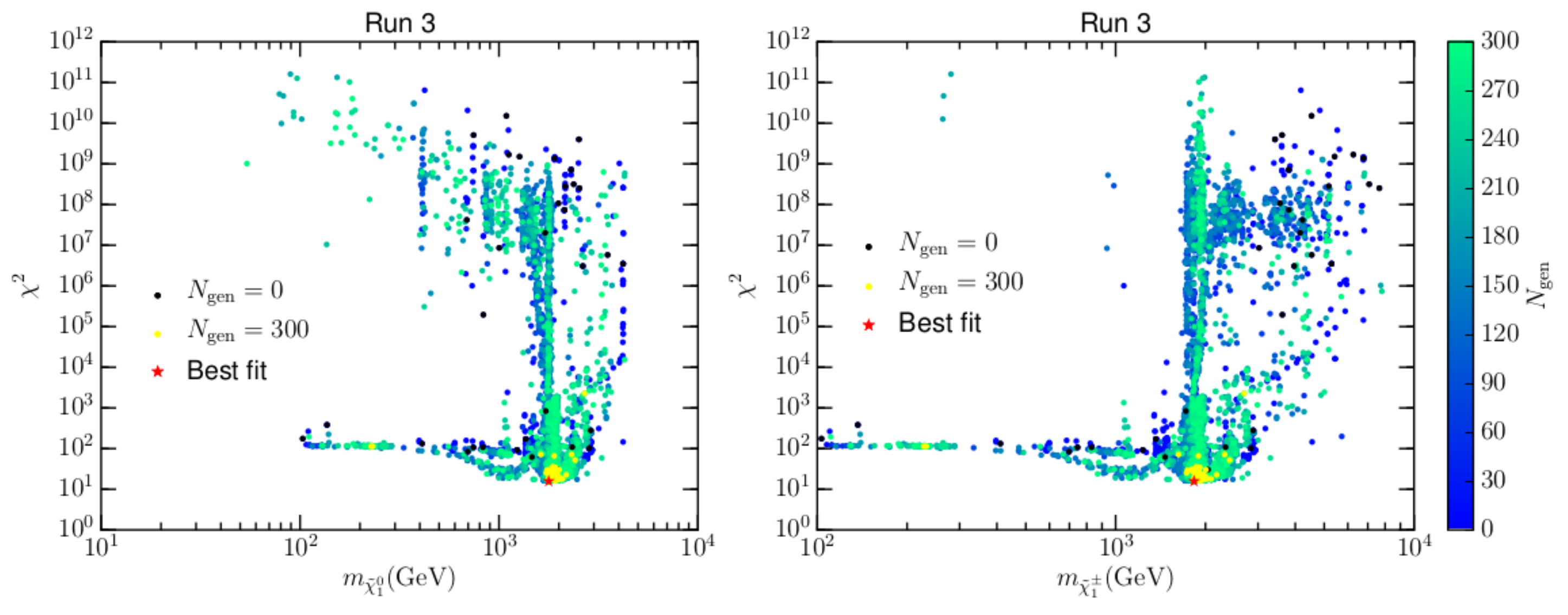}
  \includegraphics[width=\linewidth]{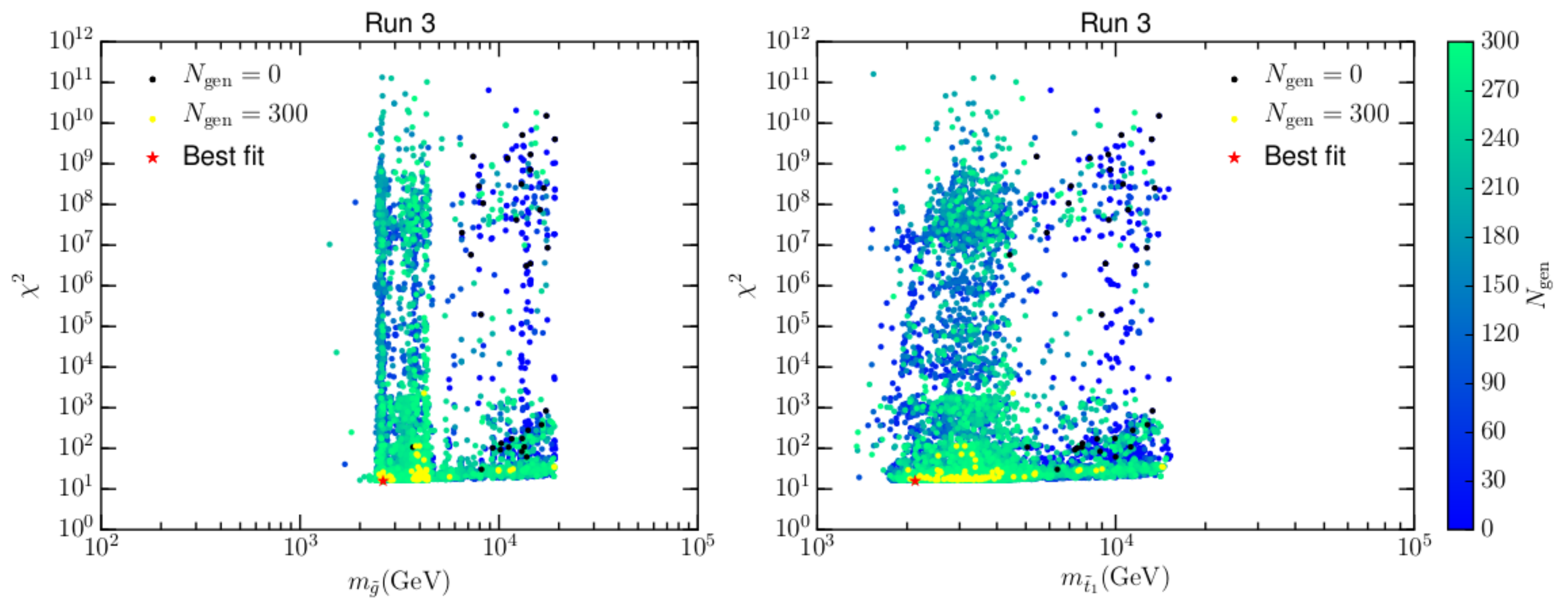}
  \caption{Top left: $\chisq$ vs. $\mneu{1}$. Top right:  $\chisq$ vs. $\mcha{1}$. 
  Bottom left: $\chisq$ vs. $\mglu$. Bottom right:  $\chisq$ vs. $\mstop{1}$}
\end{figure}

\begin{figure}[h!]
\centering
\includegraphics[width=0.55\linewidth]{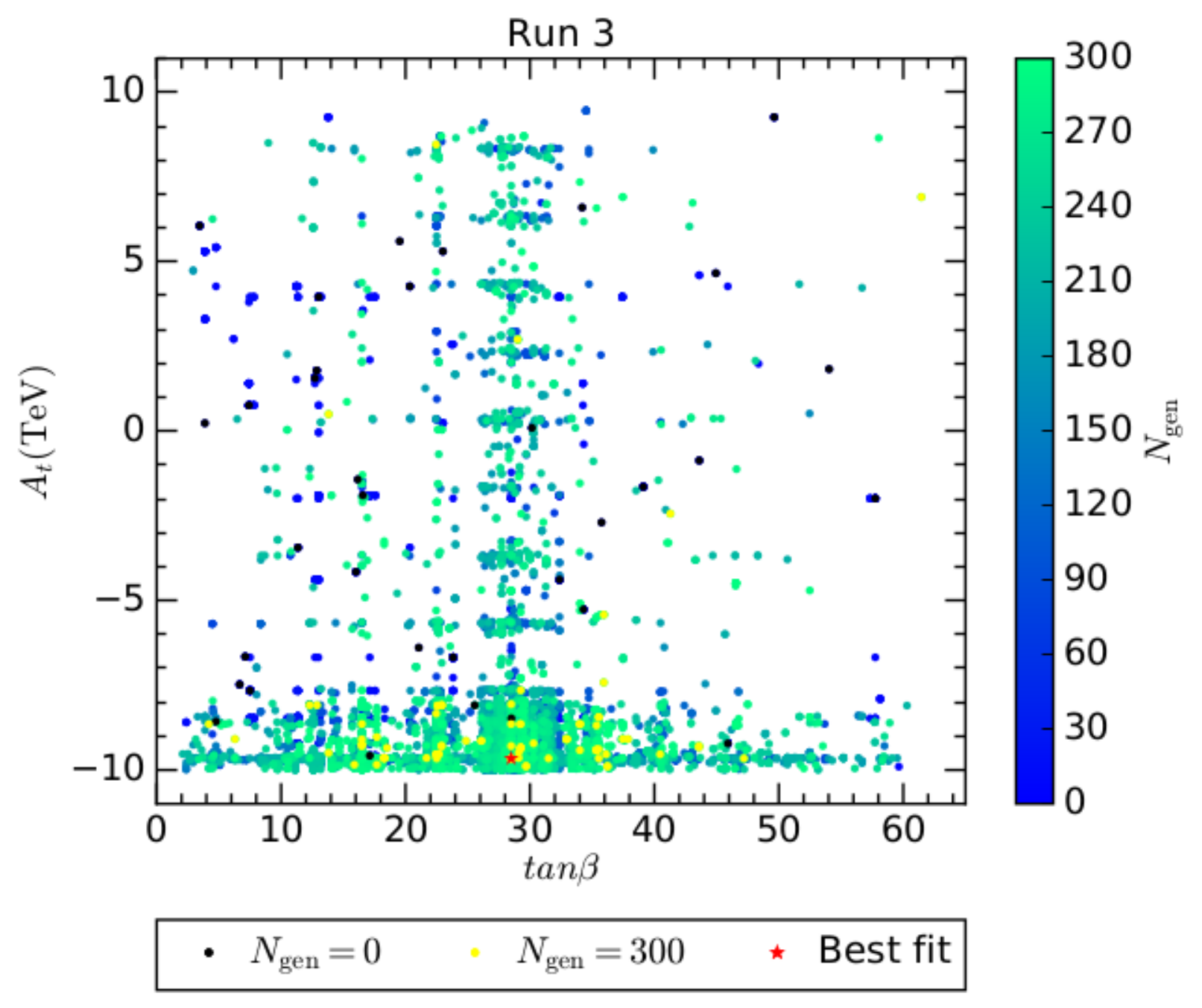}
\caption{$\At$ vs. $\tanb$}
\end{figure}

\begin{figure}[h!]
\centering
\includegraphics[width=\linewidth]{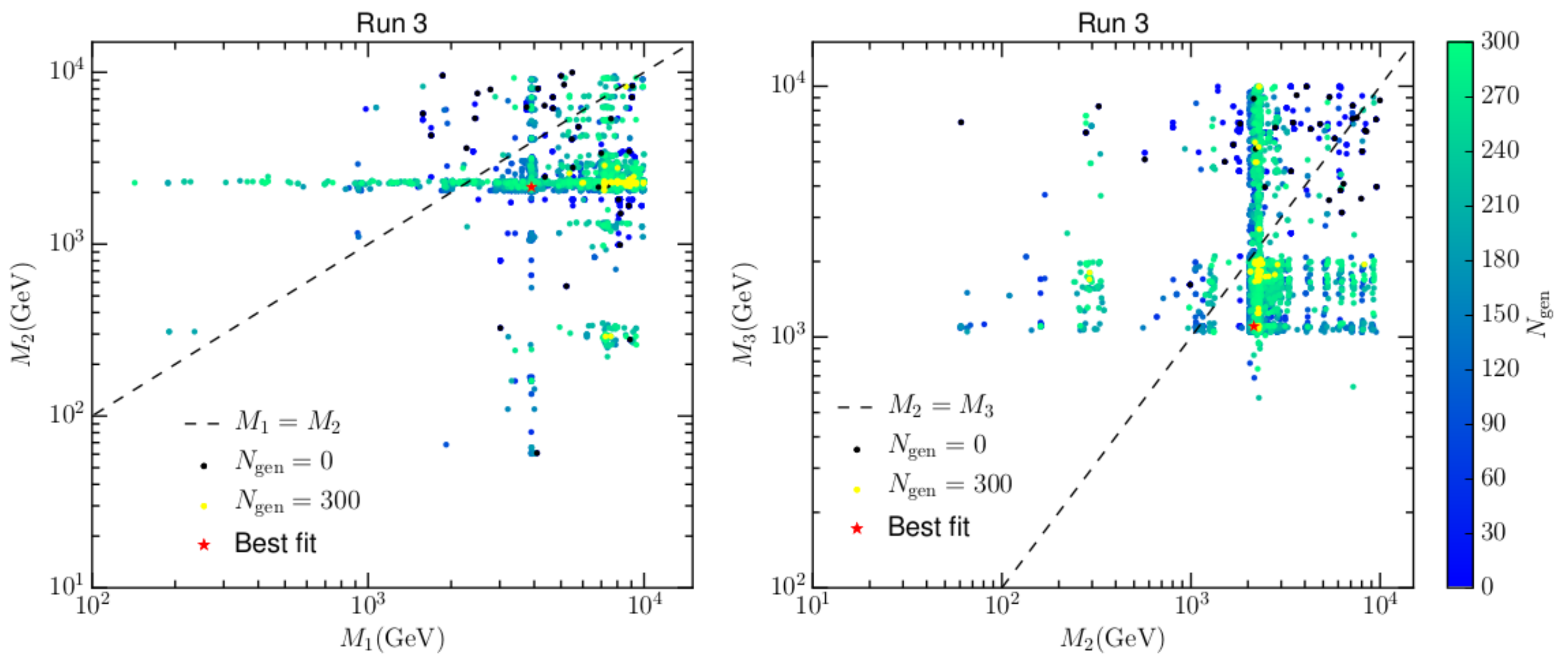}
\caption{Left: $M_2$ vs. $M_1$. Right: $M_3$ vs. $M_2$}
\label{fig:gauginos_amu}
\end{figure}

\clearpage

\subsection{The Galactic Centre excess}

As a second example, let us consider fitting the observed GCE in the context of the pMSSM. In order to reproduce the measured gamma-ray spectrum, a small range of values for dark matter pair-annihilation cross section are required, around $\sigmav\approx 10^{-26}$ cm$^{3}$s$^{-1}$, which is roughly consistent with the expected value to obtain the correct relic abundance. Interestingly, this leads to an upper bound on $\relic$, which in the previous example was not constrained from below. 
We can observe in Fig.~\ref{fig:chi_relic_gce} an increase of the global $\chisq$ for points where the neutralino relic density is too small. The requirement of fitting the GCE is consistent with recovering the correct relic abundance as well, which contributes to the clustering of solutions.

In Fig.~\ref{fig:chi_gce} we can see how the GCE contributes to the total $\chisq$. We can identify two types of behaviour. Points on the vertical branch correspond to those in which the annihilation cross section is too small and the neutralino relic abundance is too large, whereas points along the horizontal branch are those where the annihilation cross section is too large (thus the relic density is too small).

\begin{figure}[h!]
\centering
\includegraphics[width=\linewidth]{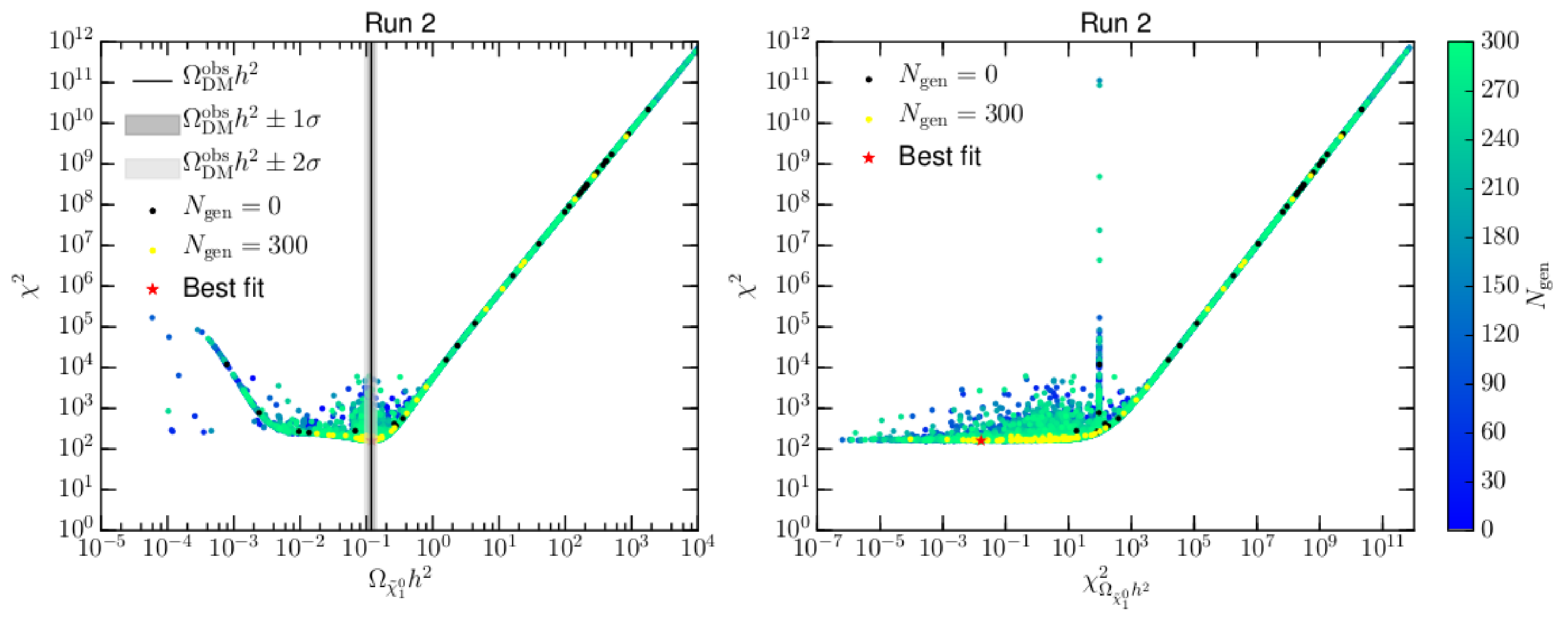}
\caption{Left: $\chisq$ vs. $\relicneu$. The solid black line corresponds to the $\relic$ mean value, see Table~\ref{tab:constraints}. As a reference, we show the $1\sigma$ and $2\sigma$ regions around that value in grey and light grey, respectively.
Right: $\chisq$ vs. $\chisq_{\relicneu}$. (Compare with Fig.~\ref{fig:chirelic_amu})}
\label{fig:chi_relic_gce}
\end{figure}

\begin{figure}[h!]
\centering
\includegraphics[width=0.55\linewidth]{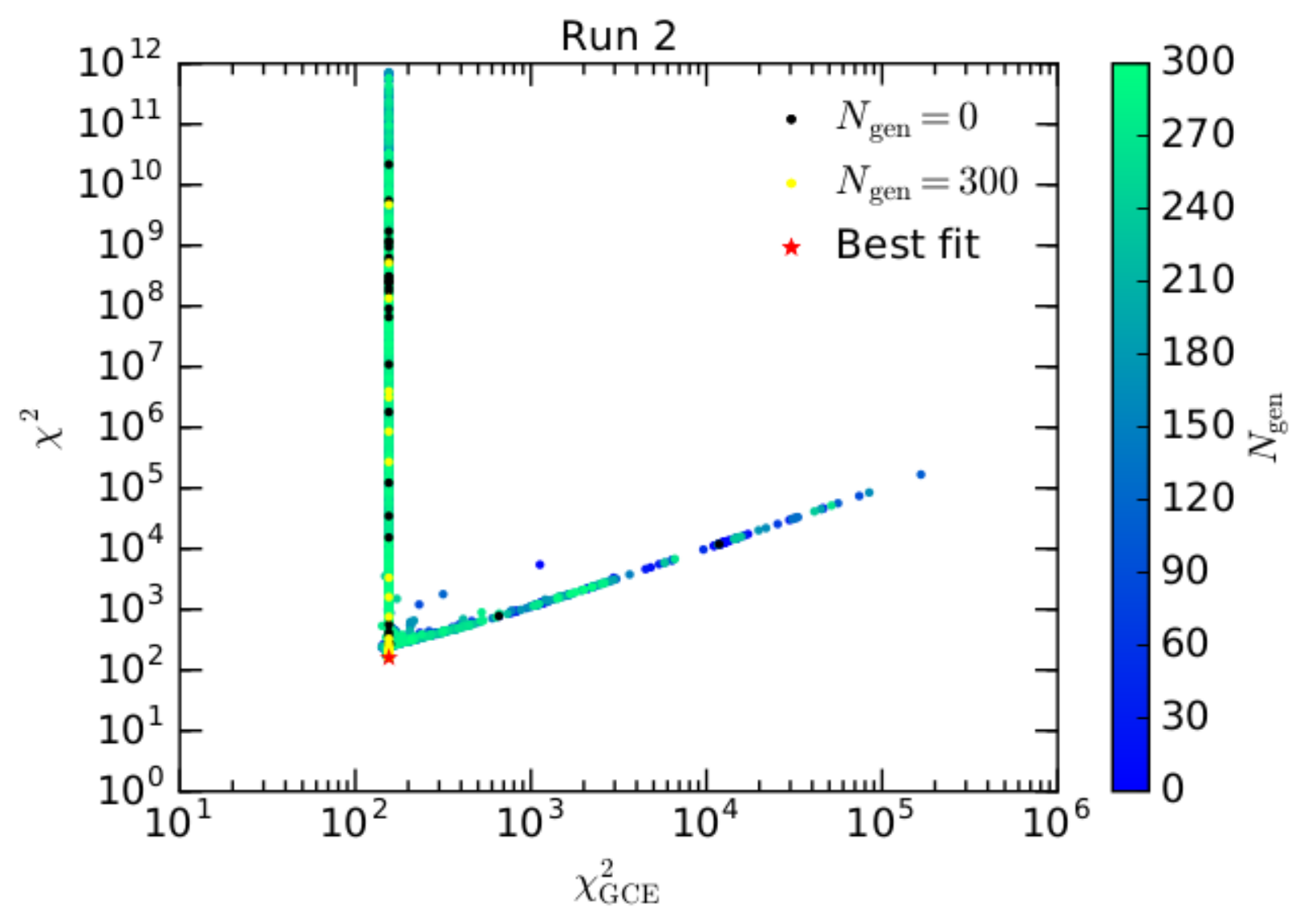}
\caption{$\chisq$ vs. $\chisq_{\rm GCE}$.}
\label{fig:chi_gce}
\end{figure}

\begin{table}[h!]
  \centering
 {\footnotesize\setlength{\tabcolsep}{3pt}\renewcommand{\arraystretch}{1.2}%  
\begin{tabular}{|l|r|r|r|r|r|r|r|r|r|r|}
\hline
  & Run 1 & Run 2 & Run 3 & Run 4 & Run 5 \\ 
\hline
$\chisq_{\relicneu}$ & 0.0040 & 0.0165 & 0.0015 & 0.0003 & 0.0472\\
$\chisq_{\rm HiggsSignals}$ & 1.2913 & 1.2812 & 1.2896 & 1.2938 & 1.2879 \\
$\chisq_{\mh}$ & 0.1411 & 0.1630 & 0.2316 & 0.5483 & 0.2214\\
$\chisq_{\mW}$  & 0.2283 & 0.1652 & 0.3996 & 0.0999 & 0.2341\\
$\chisq_{\swsqeff}$  & 0.1509 & 0.1459 & 0.1266 & 0.1527 & 0.1416\\
$\chisq_{\GamZ}$  & 0.1208 & 0.0668 & 0.1584 & 0.0076 & 0.0425\\
$\chisq_{\GamZinv}$  &  2.3020 & 2.2954 & 2.2874 & 2.2985  & 2.2955\\
$\chisq_{\brbsgamma}$  & 0.1087 & 0.1283 & 0.1563 & 0.1016 & 0.0951\\
$\chisq_{\brbsmumu}$  & 0.1575 & 0.1534 & 0.1508 & 0.1577 & 0.1547\\
$\chisq_{\RBtaunu}$  & 0.0140 & 0.0140 &  0.0142 & 0.0141 & 0.0140\\
$\chisq_{\rm LEP}$  & 0.0000 & 0.0000 & 0.0000 & 0.0000 & 0.0000\\
$\chisq_{\rm LHC}$  & 0.0000 & 0.0000 & 0.0000 & 0.0000 & 0.0000\\
{\color{blue}${\chisq_{\rm GCE}}$}  & {\color{blue}155.3494} & {\color{blue}155.3459} & {\color{blue}155.3588} & {\color{blue}155.3656} & {\color{blue} 155.3490}\\
\hline
$\chisq_{\rm tot}$ & 159.8680 & 159.7755 & 160.1747 & 160.0401 & 159.8832\\
\hline
\end{tabular} 
}
\caption{GCE: Contributions to the best fit $\chisq$. In blue, we show the leading contribution, which comes from the fit to the Galactic Centre excess.}
\label{tab:chi_gce}
\end{table}

\begin{table}[h!]
  \centering
 {\footnotesize\setlength{\tabcolsep}{2.5pt}\renewcommand{\arraystretch}{1.3}%  
 \begin{tabular}{|l|r|r|r|r|r|r|r|r|r|r|}
 \hline 
 Observable & Run 1 & Run 2  & Run 3 & Run 4 & Run 5 \\ 
 \hline 
 $\mh(\GeV)$  & 124.33 & 124.28 & 124.12 & 123.60 & 124.14\\ 
  $\mW(\GeV)$ & 80.376 & 80.378 & 80.374 & 80.379 & 80.376\\ 
 $\swsqeff$ & 0.23146 & 0.23146 & 0.23146 & 0.23146 & 0.23146\\ 
 $\GamZ(\GeV)$ & 2.4943 & 2.4946 & 2.4942 & 2.4950 & 2.4947\\
 $\GamZinv(\GeV)$ & 0.5017 & 0.5017 & 0.5017 & 0.5017 & 0.5017\\ 
 $\brbsgamma\times10^{4}$ & 3.32 & 3.31 & 3.30 & 3.33 & 3.33\\
 $\brbsmumu\times10^{9}$  & 3.20 & 3.20 & 3.19 & 3.20 & 3.20\\
 $\RBtaunu$ & 1.00 & 1.00 & 1.00 & 1.00 & 1.00\\
 $\relicneu$ & 0.1196 &  0.1203 & 0.1193 & 0.1186 & 0.1214\\
 {\color{blue}${\mneu{1}(\TeV)}$}&{\color{blue}2.2366} & {\color{blue}2.2643} & {\color{blue}2.1660} & {\color{blue}2.1124} & {\color{blue}2.2386}\\
 {\color{blue}${\sigmav \times 10^{26} (\mathrm{cm}^3\mathrm{s}^{-1})}$} & {\color{blue}$2.11$} & {\color{blue}$2.06$} & {\color{blue}$2.24$} & {\color{blue}$2.35$} & {\color{blue}$2.10$}  \\
 \hline
 \end{tabular}
}
\caption{GCE: Best fit observables. In blue, we show the mass of the DM and $\sigmav$.}
\label{tab:bestfit_gce}
\end{table}

The $\chisq$ for the best fit points is shown in Table~\ref{tab:chi_gce}, together with the contribution for each individual observable. 
The total $\chisq\approx160$ is quite large in this example, but the main contribution is solely due to the fit to the GCE, whereas the rest of the observables are properly fit.
It is illustrative to compare this table with Table~\ref{tab:chi_amu}, which shows that the goodness of the fit to all observables is similar and that there is only one outlier (the fit to either $\amu$ or the GCE). 
Likewise, the best fit to the different observables is shown in Table~\ref{tab:bestfit_gce}. We have included in this table the mass of the DM (the neutralino) and its annihilation cross section in the halo. We can observe that, although the annihilation cross section is of the right order of magnitude, the neutralino mass is approximately 2.2~TeV, too heavy compared to the best fit to the GCE, which requires DM masses of the order of 100~GeV or below, depending on the leading annihilation channel (see for example Ref.~\cite{Calore:2014xka}). This is the main reason for the high value of $\chisq_{\rm GCE}$.

As a consequence, the input parameters for the best fit points, shown in Tab.~\ref{tab:bestfit_ini_gce}, are indistinguishable from those obtained in the previous section, and the same holds for the low-energy supersymmetric spectrum of Tab.~\ref{tab:bestfit_spect_gce}. 
The spectrum for all the generations is shown in Fig.~\ref{fig:massspectrum_all_gce}, in which the best fit points (red lines) seem to show more clustering than in the previous section (Fig.~\ref{fig:massspectrum_all_amu}).
Once more, the rest of the observables drive the evolution of the GA and we are left with a heavy SUSY spectrum, featuring wino-like $2.2~\TeV$ neutralinos/charginos (see Fig.~\ref{fig:neut_gce} for the neutralino composition), with a heavy colour sector and where  slepton masses vary in the range of $3-10~\TeV$. As we already observed in the previous section, the GA has singled out one observable (the GCE) which cannot be fit.
As in the previous section, the Higgs mass is contributing to the GA evolution (see Fig.~\ref{fig:higgsmass_gce}).

\begin{table}[h!] 
  \centering
 {\scriptsize\setlength{\tabcolsep}{3pt}\renewcommand{\arraystretch}{1.3}%  
\begin{tabular}{|l|r|r|r|r|r|r|r|r|r|r|}
\hline
Parameter & Run 1 & Run 2  & Run 3 & Run 4 & Run 5 \\ 
\hline
\multicolumn{6}{|c|}{SM} \\
\hline
$\invalphEM$ & 128.0066 & 127.9962 & 127.9986 & 128.0143 & 127.9983\\
$\alphs$ & 0.1164 &  0.1170 & 0.1164 & 0.1177 & 0.1170\\ 
$\mb(\GeV)$ & 4.5603 &  4.6775 & 4.6551 & 4.6499 & 4.5627\\
$\mt(\GeV)$ & 175.3844 & 175.3538 & 175.18631 & 175.4313 & 175.3560\\
\hline
\multicolumn{6}{|c|}{pMSSM (GUT scale)} \\
\hline
$M_1(\TeV)$ & 5.5273 & 9.1160 & 8.3262 & 9.0043 & 9.6308\\
$M_2(\TeV)$ & 2.6422 & 2.6594 & 2.5554 & 2.4849 & 2.6406\\
$M_3(\TeV)$ & 3.6930 & 3.2241 & 3.7310 & 1.8545 & 3.1646\\
$\mHu(\TeV)$ & 2.7860 & 1.8275 & 4.5553 & 3.3882 & 0.7279\\
$\mHd(\TeV)$ & 7.7773 & 6.7757 & 4.5548 & 7.7847 & 9.2033\\
$\msq{3}(\TeV)$ & 5.4058 & 5.6435 & 7.6979 & 4.1138 & 4.3414\\
$\msq{1,2}(\TeV)$ & 7.1555 & 8.8009 & 9.8729 & 6.1029 & 2.6357\\
$\msu{3}(\TeV)$ & 0.2778 & 0.1488 & 1.2877 & 7.8291 & 2.4574\\
$\msu{1,2}(\TeV)$ & 8.4336 & 2.7263 & 8.5276 & 8.9409 & 7.3785\\
$\msd{3}(\TeV)$ & 3.0870 & 0.0584 & 3.1804 & 5.6241 & 3.7087\\
$\msd{1,2}(\TeV)$ & 8.1973 & 5.1068 & 4.2454 & 3.3341 & 0.5251\\
$\msl{3}(\TeV)$ & 8.6012 & 9.6593 & 9.7190 & 7.1871 & 9.7510\\
$\msl{1,2}(\TeV)$ & 6.6443 & 7.0022 & 2.9389 & 7.9204 & 9.3869\\
$\mse{3}(\TeV)$ & 2.6329 & 2.8818 & 3.6777 & 7.6730  & 2.6357\\
$\mse{1,2}(\TeV)$ & 8.7945 & 7.9453 & 8.7390 & 4.1183 & 8.5795\\
$\At(\TeV)$ &  -8.9734 & -9.8566 & -9.4894 & -9.5476 & -7.6226\\
$\Ab(\TeV)$ & -1.1680 & 4.7920 & 6.0644 & 1.1036 & -0.3166\\
$\Atau(\TeV)$ & -7.1090 & 0.5658 & 0.4036 & -1.1898 & 7.1546\\
$\tanb$ & 20.2322 & 19.3898 & 23.4440 & 23.2843 & 22.4006\\
\hline
\end{tabular} 
}
\caption{GCE: Best fit input parameters.}
\label{tab:bestfit_ini_gce}
\end{table}

\begin{table}[h!] 
  \centering
 {\footnotesize\renewcommand{\arraystretch}{1.3}%  
\begin{tabular}{|c|r|r|r|r|r|r|r|r|r|r|}
\hline
m(TeV) & Run 1 & Run 2 & Run 3 & Run 4 & Run 5 \\
 \hline
$\mneu{1}$ & 2.2366 & 2.2643 & 2.1660 & 2.1124 & 2.2386\\
$\mneu{2}$ & 2.5307 & 4.1854 & 3.8210 & 4.1222 & 4.4207\\
$\mneu{3}$ & 5.8278 & 6.4265 & 6.1551 & 4.7343 & 4.8861\\
$\mneu{4}$ & 5.8285 & 6.4274 & 6.1560 & 4.7367 & 4.8893\\
$\mcha{1}$ & 2.2367 & 2,2645 & 2.1661 & 2.1126 & 2.2388\\
$\mcha{2}$ & 5.8289 & 6.4276 & 6.1562 & 4.7357 & 4.8876\\

\hline
$\mglu$ & 7.6747 & 6.7363 & 7.7679 & 4.1441 & 6.6191\\
$\mstop{1}$ & 2.6531 & 2.6939 & 2.3329 & 3.9136 & 2.5218\\ 
$\mstop{2}$ & 7.2814 & 6.8084 & 8.7598 & 5.1231 & 6.1932\\ 
$\msbot{1}$ & 6.7616 & 5.1535 & 6.6804 & 3.9245 & 6.1878\\ 
$\msbot{2}$ & 7.2782 & 6.8041 & 8.7576 & 6.9004 & 6.6923\\ 
$\msup{L}$ & 9.5393 & 10.3778 & 11.6758 & 7.2910 & 9.2023\\
$\msup{R}$ & 10.3845 & 6.6924 & 10.7659 & 8.8563 & 8.8440\\
$\msdw{L}$ & 9.5395 & 10.3780 & 11.6759 & 7.2913 & 9.2025\\
$\msdw{R}$ & 10.2820 & 7.3770 & 7.4603 & 5.4592 & 5.8807\\
$\msc{L}$ & 9.5392 & 10.3777 & 11.6757 & 7.2909 & 9.2022\\
$\msc{R}$ & 10.3845 & 6.6923 & 10.7659 & 8.8563 & 8.8440\\
$\mss{L}$ & 9.5394 & 10.3779 & 11.6759 & 7.2912 & 9.2024\\
$\mss{R}$ & 10.2819 & 7.3768 & 7.4601 & 5.4590 & 5.8805\\

\hline
$\mstau{1}$ & 3.0421 & 3.2649 & 3.9357 & 5.5936 & 5.2167\\ 
$\mstau{2}$ & 8.5760 & 9.9785 & 9.9128 & 6.5254 & 9.5018\\ 
$\msel{L}$ & 6.7806 & 7.5375 & 3.7867 & 7.4626 & 9.2988\\
$\msel{R}$ & 9.1847 & 8.2899 & 9.0837 & 7.1857 & 10.0071\\
$\msmu{L}$ & 6.7801 & 7.5371 & 3.7860 & 7.4621 & 9.2982\\
$\msmu{R}$ & 9.1839 & 8.2892 & 9.0831 & 7.1847 & 10.0060\\
$\msneu{e}$ & 6.7798 & 7.5367 & 3.7856 & 7.4619 & 9.2982\\
$\msneu{\mu}$ & 6.7793 & 7.5364 & 3.7848 & 7.4614 & 9.2975\\
$\msneu{\tau}$ & 8.5753 & 9.9779 & 9.9122 & 6.5243 & 9.5011\\
\hline
$\mH$ & 9.4674 & 9.5080 & 7.5931 & 8.3477 & 9.9123\\
$\mA$ & 9.4671 & 9.5074 & 7.5931 & 8.3477 & 9.9121\\
$\mCH$ & 9.4675 & 9.5079 & 7.5935 & 8.3482 & 9.9126\\
\hline
\end{tabular} 
}
\caption{GCE: Best fit SUSY spectrum.}
\label{tab:bestfit_spect_gce}
\end{table}

\begin{figure}[h!]
\centering
\includegraphics[width=\linewidth]{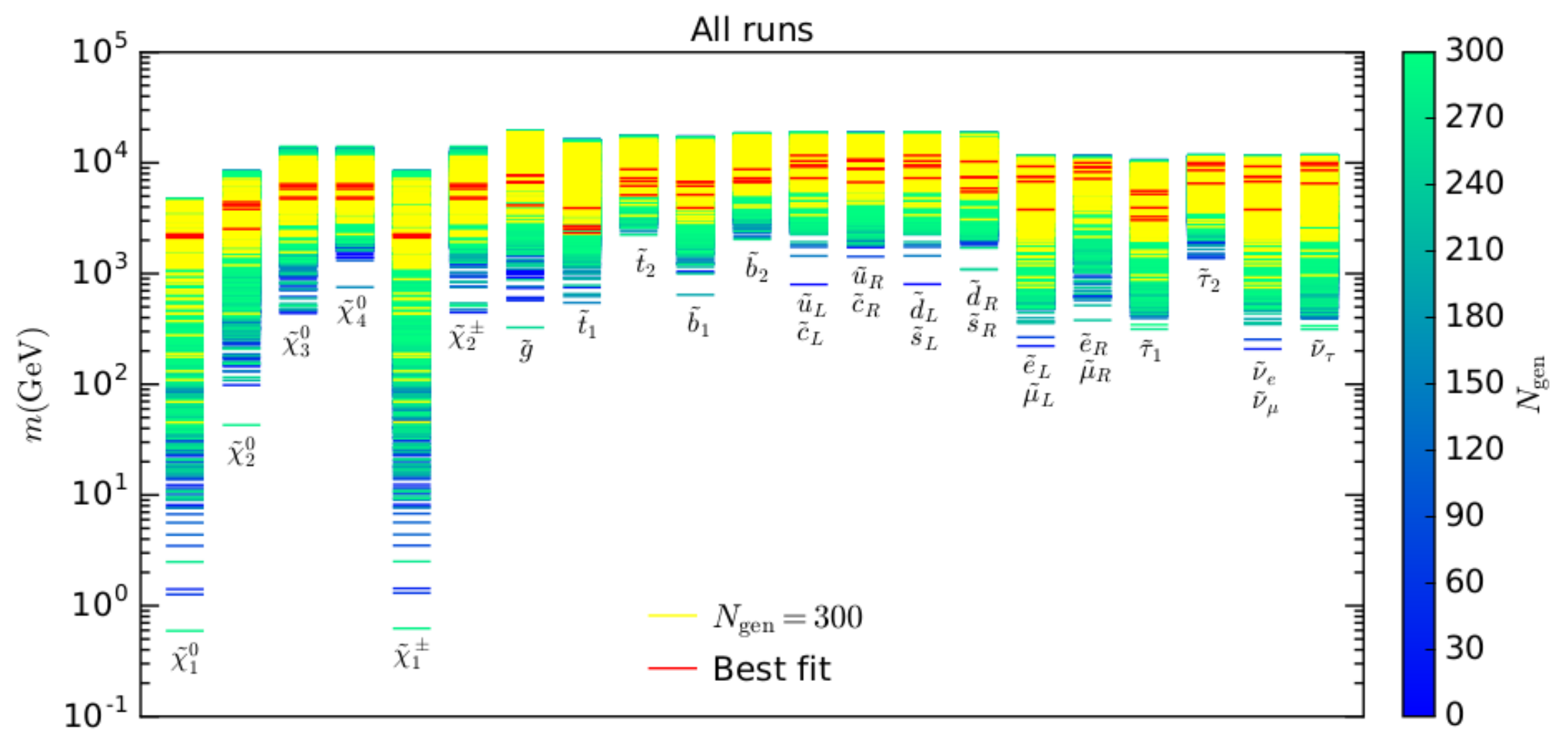}
\caption{SUSY spectrum for all the generations in the five runs. Yellow represents the results for the last generation and the red line corresponds to the best fit point. (Compare with Fig.~\ref{fig:massspectrum_all_amu})}
\label{fig:massspectrum_all_gce}
\end{figure}

\begin{figure}[h!]
\centering
\includegraphics[width=\linewidth]{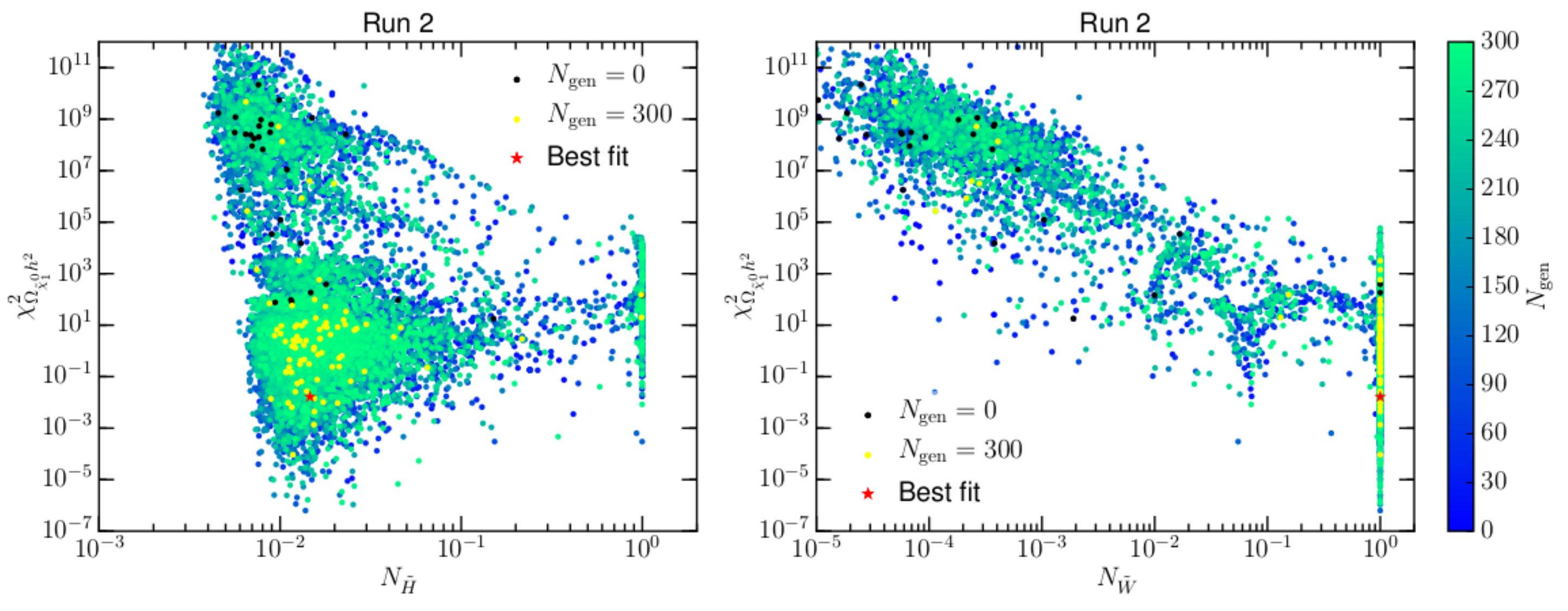}
  \caption{$\chisq_{\relicneu}$ vs. the Higgsino (left) and wino (right) component of the lightest neutralino. (Compare with Fig.~\ref{fig:neut_amu})}
  \label{fig:neut_gce}
\end{figure}

\begin{figure}[h!]
\centering
\includegraphics[width=0.55\linewidth]{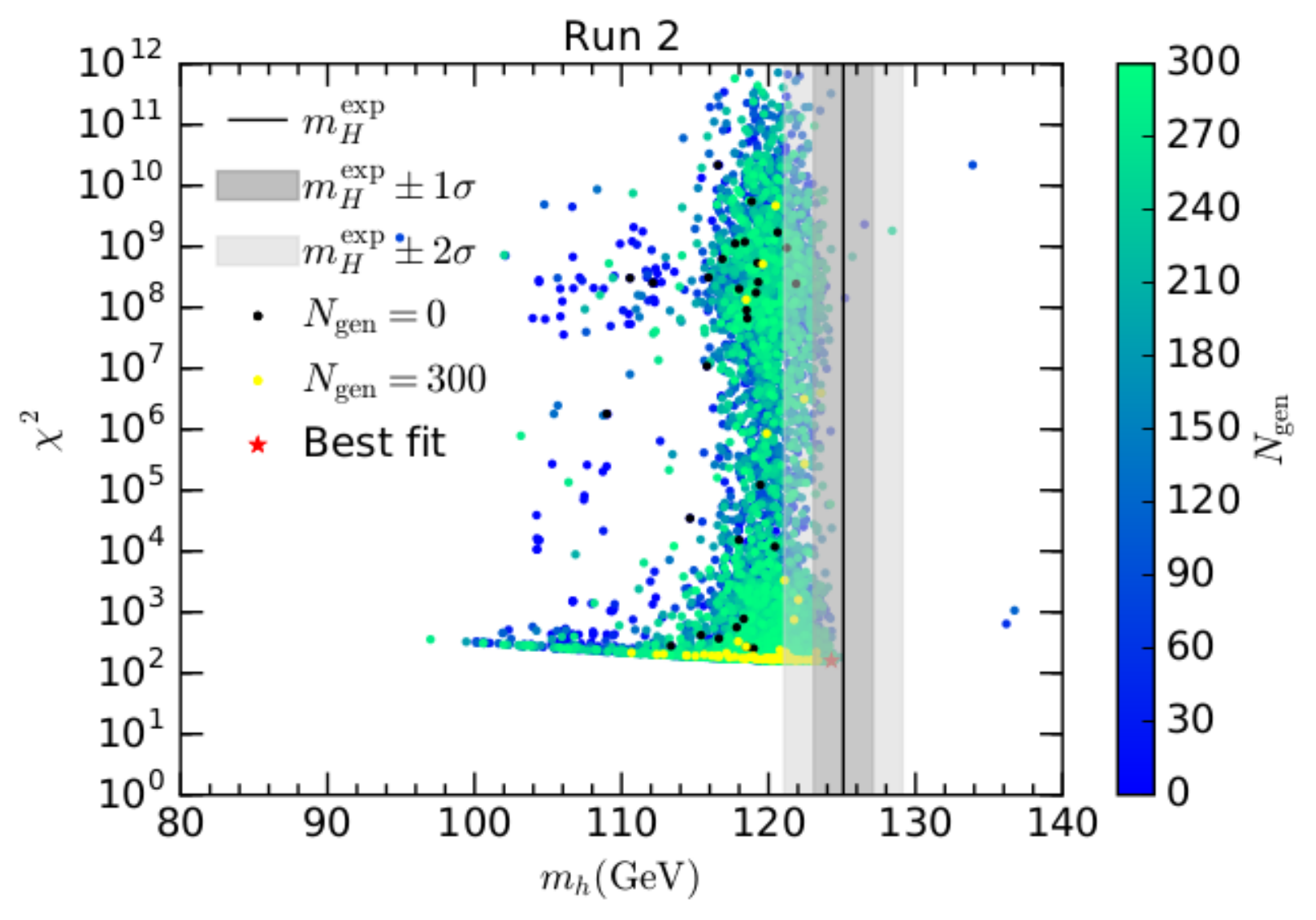}
\caption{$\chisq$ vs. Higgs mass. (Compare with Fig.~\ref{fig:higgsmass})}
\label{fig:higgsmass_gce}
\end{figure}

Finally, Fig.~\ref{fig:dm_gce} shows the predictions for direct and indirect dark matter detection. The results for direct detection are very similar to those of the previous section (Fig.~\ref{fig:dm_amu}), with neutralinos marginally within the sensitivity of future multi-ton xenon and argon experiments. The plot of the annihilation cross-section in Fig.~\ref{fig:dm_gce} still shows the best fit point with heavy neutralinos and $\sigmav\approx10^{-26}$~cm$^3$s$^{-1}$. As mentioned above, this is far from the preferred region that would explain the GCE in terms of DM with masses of order 100~GeV.

\begin{figure}[h!]
\centering
\includegraphics[width=\linewidth]{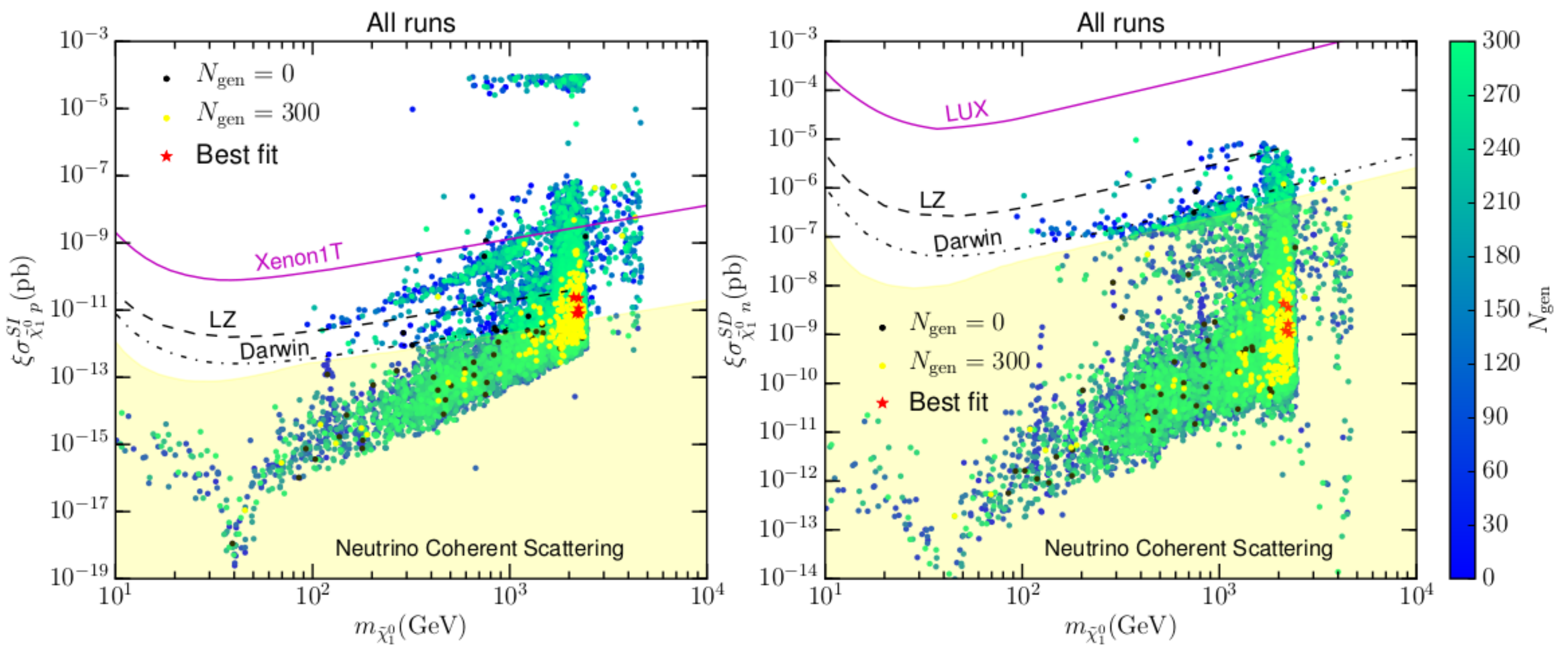}
\includegraphics[width=0.55\linewidth]{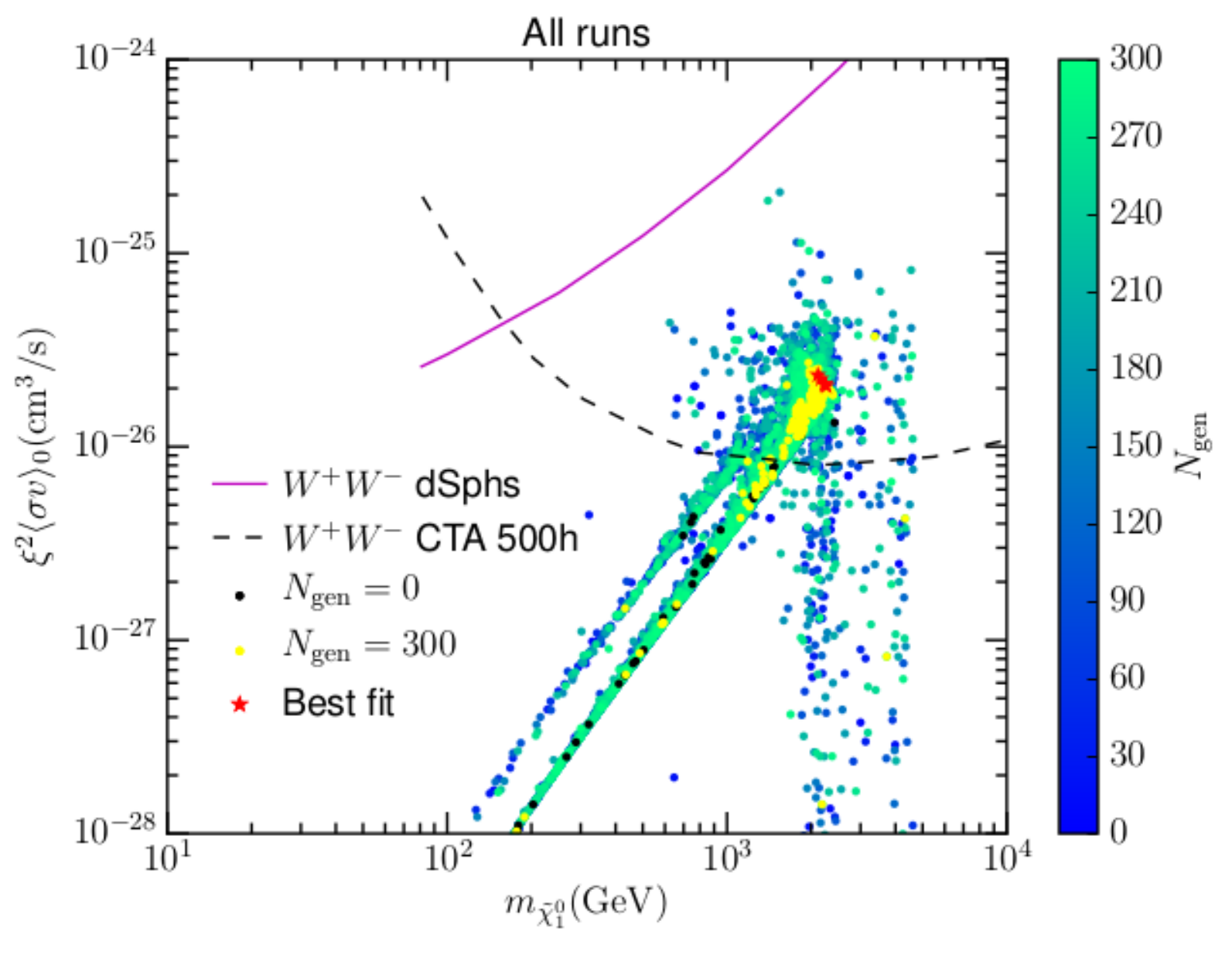}
\caption{Top left: Theoretical predictions for $\xi\sip$ as a function of $\mneu{1}$ for all the runs. Top right: Predictions for $\xi\sdn$ vs. $\mneu{1}$. Bottom: Thermally averaged neutralino annihilation cross section in the Galactic halo, $\xi^2\sigmav$, as a function of the lightest neutralino mass. The colour code and lines are as in  Fig.~\ref{fig:dm_amu}.}
\label{fig:dm_gce}
\end{figure}

\clearpage
\section{Conclusions}
\label{sec:conclusions}

In this article, we have investigated the use of Genetic Algorithms (GAs) to study the cross-compatibility of experimental constraints in high-dimensional models.
We have focused on the pMSSM, which features 19 input parameters (soft supersymmetry-breaking terms) defined at the GUT scale, and 4 nuisance parameters (the electromagnetic coupling constant evaluated at the Z-boson
pole mass,
the strong coupling constant at $\mZ$,
the pole mass of the top quark, 
and the pole mass of the 
bottom quark), 
for a total of 23 parameters.

GAs seem to be extremely effective in finding a best fit point that minimises the total $\chisq$. With only $10^4$ model evaluations, solutions could be found that were consistent with results that employ MCMC scans to probe the whole parameter space, and that require many more model evaluations. 
The GA leads to a final population of models with a roughly $ 2\, \TeV$ wino-like neutralino, which has the correct relic abundance due to coannihilations with a quasi-degenerate chargino. The resulting SUSY spectrum is shown in Fig.~\ref{fig:massspectrum_all_amu} (Table~\ref{tab:bestfit_spect_amu}) and Fig.~\ref{fig:massspectrum_all_gce} (Table~\ref{tab:bestfit_spect_gce}). The coloured sector is predicted to be heavy, $\mglu>5\,\TeV$, except for the lightest stop, for which $\mstop{1}\approx2.3\,\TeV$. We find that the pMSSM  does not give a clear prediction for the slepton sector, and the masses span a wide range, $\mstau{1}\sim 2.3-8\,\TeV$.
The neutralino relic abundance and the Higgs mass are the most important constraints driving the GA evolution.

We also demonstrated how one can deal with potential signals for new physics, by considering  the muon anomalous magnetic moment (which shows a large deviation with respect to the SM value) and the Fermi-LAT excess in the gamma ray spectrum from the Galactic Centre (which can be interpreted as a hint for DM pair-annihilation). A GA proves to be an excellent tool for assessing the compatibility of these observations with all the other experimental constraints, including LHC and LEP bounds on SUSY masses and on the Higgs sector, Planck measurement of the DM relic abundance, and constraints on low-energy observables. 
Moreover, it also yields a good diagnosis of which are the problematic observables. In both these examples, the main contribution to the final $\chisq$ was due to either the  muon anomalous magnetic moment, $\chisq_{\amu}\approx 12$, or the Galactic Centre excess, $\chisq_{\rm GCE}\approx155$, whereas the fit to all the other observables was good. 
This is an indication that the pMSSM, despite its large number of free parameters, cannot successfully include these potential hints for new physics. (A compromise could in principle have been possible, in which they were fit reasonably well by sacrificing $\chi^2$ elsewhere, but this turned out to be impossible.)

In our view, GAs offer a superior approach to probing BSM physics, especially in an era when the underlying principles are less clear, but when there are nevertheless definite hints of new physics. The technique we discussed here could for example be easily applied to the most general form of MSSM with its 124 parameters, 
as well as more general Higgs sectors, with no obvious impediment. Compared to other more conventional techniques, GAs are able (by sacrificing a little statistical rigour) to divine patterns of interesting models, and assess their consistency exceedingly quickly. 

\vspace*{1cm}
\noindent{\bf \large Acknowledgements}\\

\noindent We thank G. G\'omez-Vargas, D. V. Malyshev and M.A. S\'anchez Conde. SR was partially supported by MINECO, Spain, under contract FPA2016-78022-P,  Centro  de  excelencia  Severo  Ochoa  Program  under  grant  SEV-2014-0398, the Campus of Excellence UAM+CSIC and the Australian Research Council.

\clearpage

\appendix
\section{Additional plots}
\label{sec:app}

In this appendix, we include some extra figures that show the GA evolution of some of the 19 input parameters of the pMSSM for the example discussed in Section~\ref{sec:results}A, where we attempt to fit the muon anomalous magnetic moment. As in the other plots of this paper, these figures should not be understood as ``scatter plots" or general explorations of the parameter space, but rather as an indication of the genome evolution over different generations. 
In Figures \ref{fig:runs1237_1}-\ref{fig:runs1237_6}, we compare the results for runs 1, 2, 3, and 7 showing how different runs lead to compatible results.
We have not included plots corresponding to Section~\ref{sec:results}B, as they look very similar.

\begin{figure}[h!]
\centering
  \includegraphics[width=\linewidth]{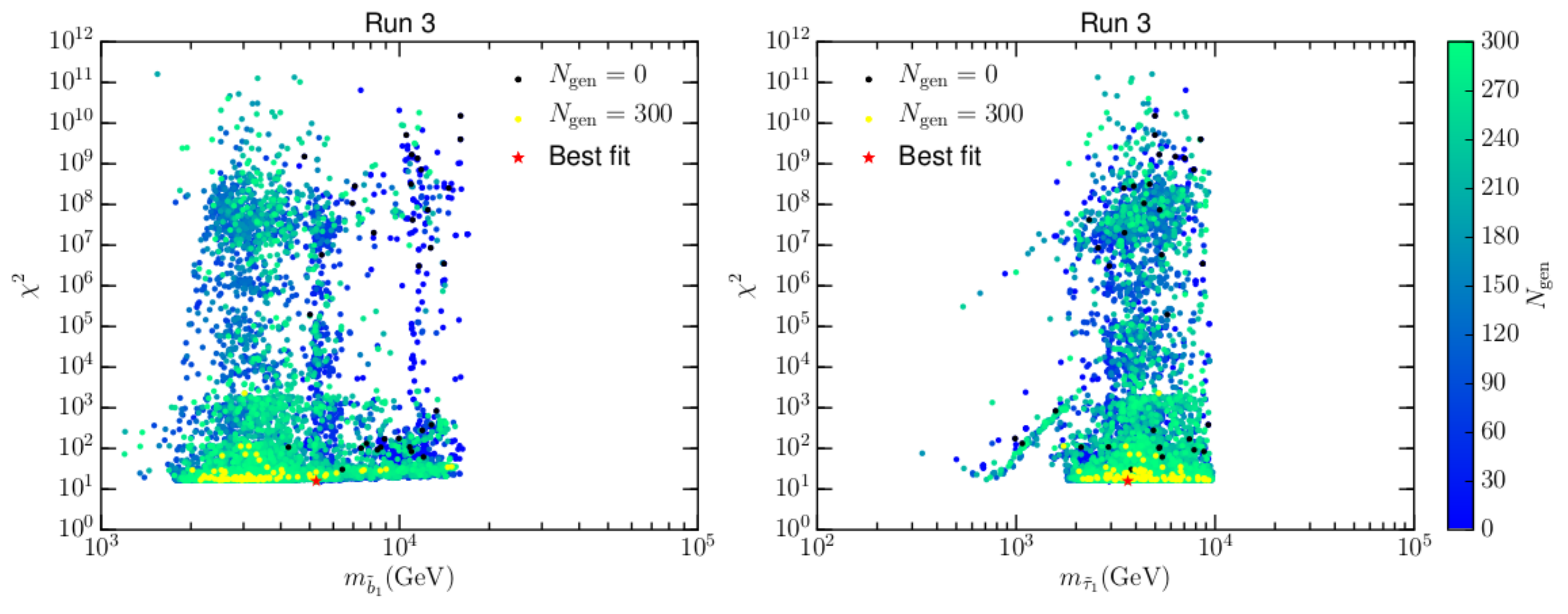}
  \caption{Left: $\chisq$ vs. $\msbot{1}$. Right:  $\chisq$ vs. $\mstau{1}$.
  The colour map represents the generation number from 0 to 300. The initial guesses (generation~0) are  depicted in black and the final generation in yellow. The red star corresponds to the best fit.}
\end{figure}

\begin{figure}[h!]
\centering
  \includegraphics[width=\linewidth]{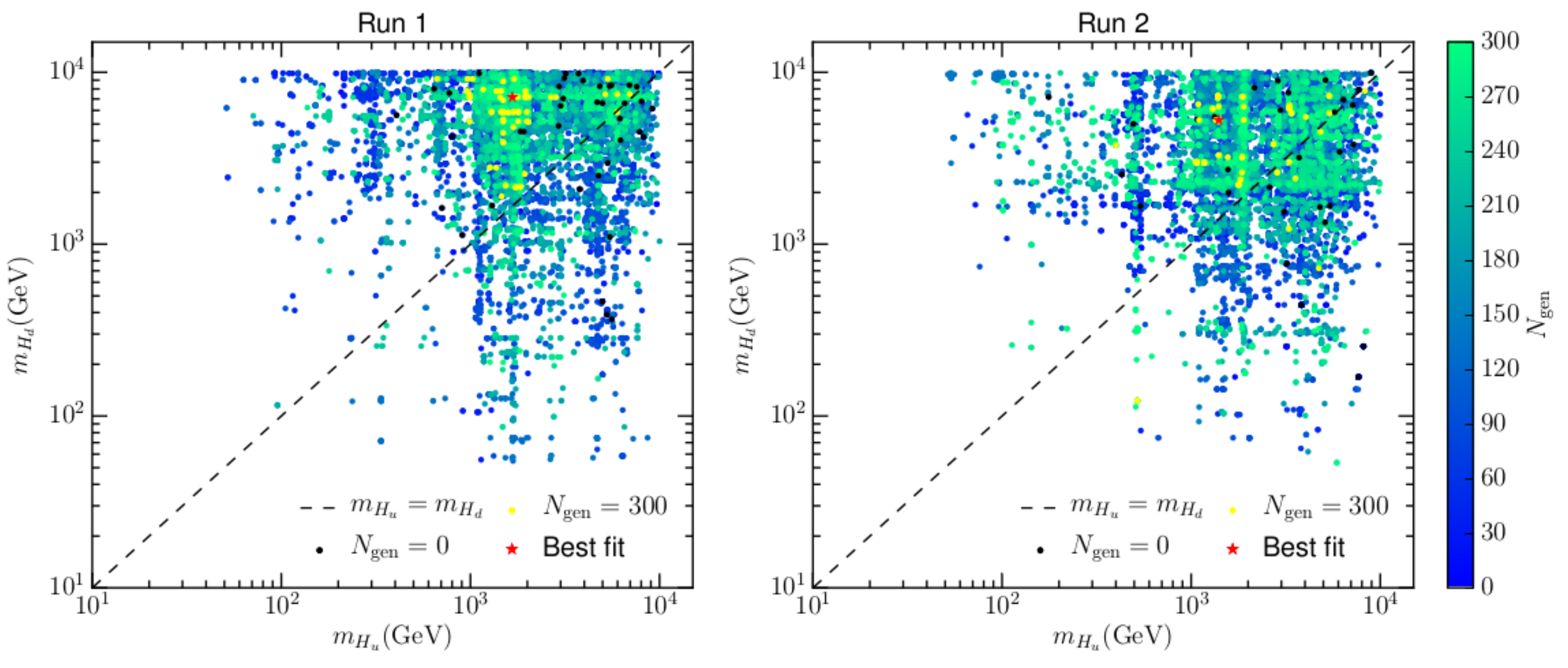}
  \includegraphics[width=\linewidth]{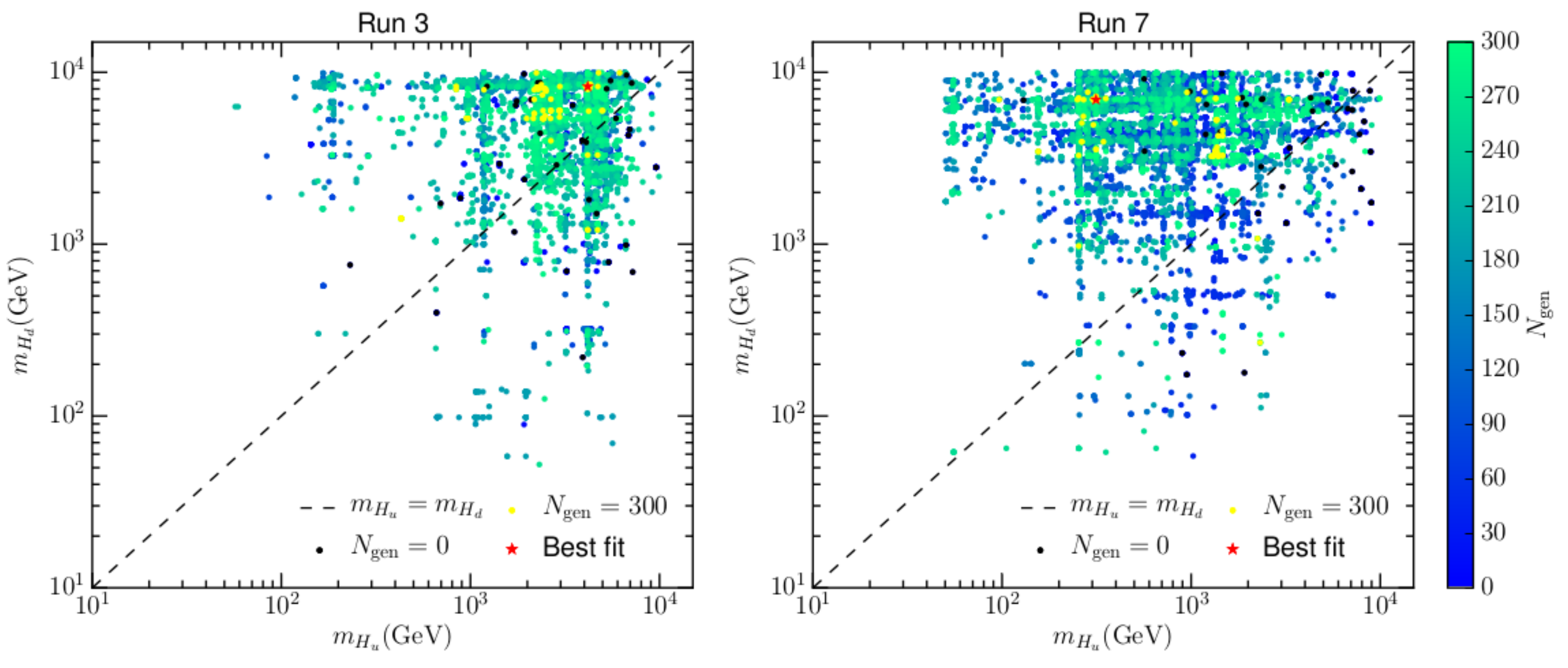}    
\caption{$\mHd$ vs. $\mHu$, top left (run 1), top right (run 2), bottom left (run 3), bottom right (run 7).}
\label{fig:runs1237_1}
\end{figure}

\begin{figure}[h!]
\centering
  \includegraphics[width=\linewidth]{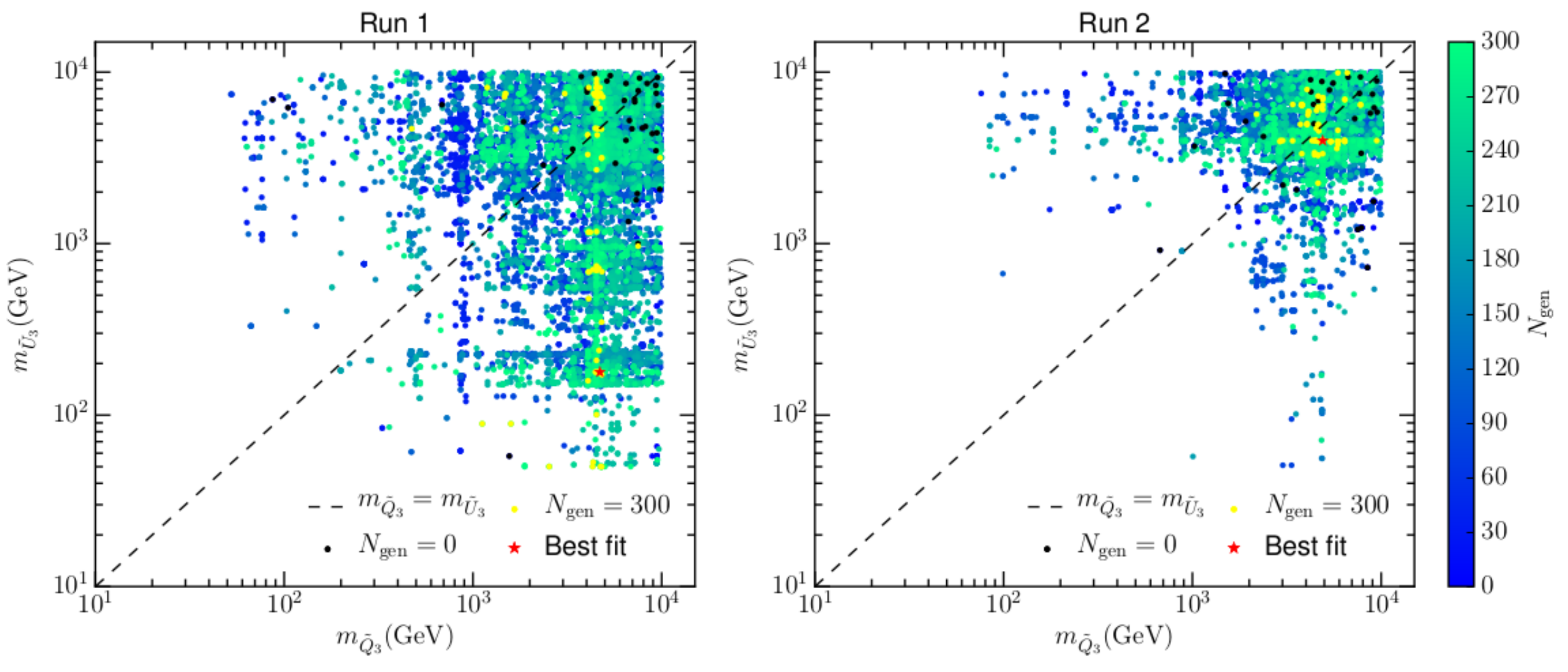}
  \includegraphics[width=\linewidth]{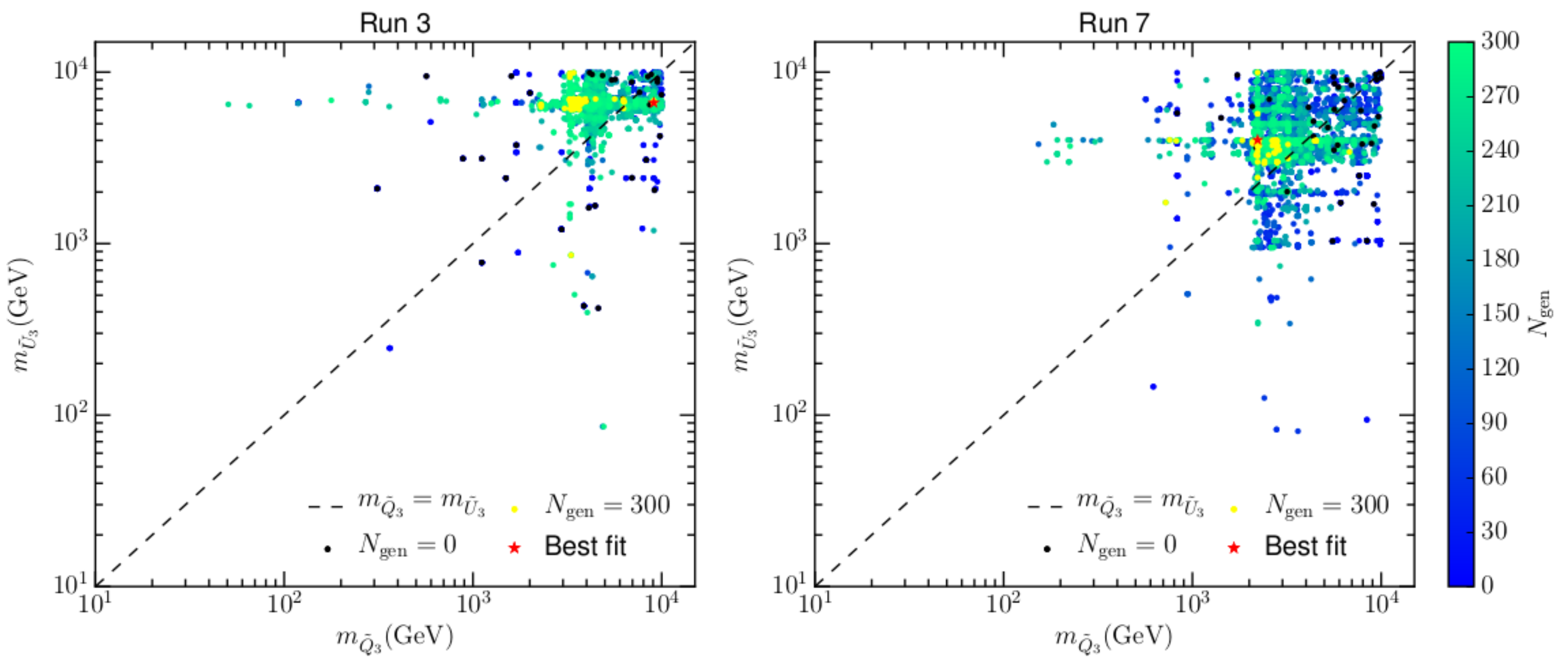}    
\caption{$\msu{3}$ vs. $\msq{3}$, top left (run 1), top right (run 2), bottom left (run 3), bottom right (run 7).}
\label{fig:runs1237_2}
\end{figure}

\begin{figure}[h!]
\centering
  \includegraphics[width=\linewidth]{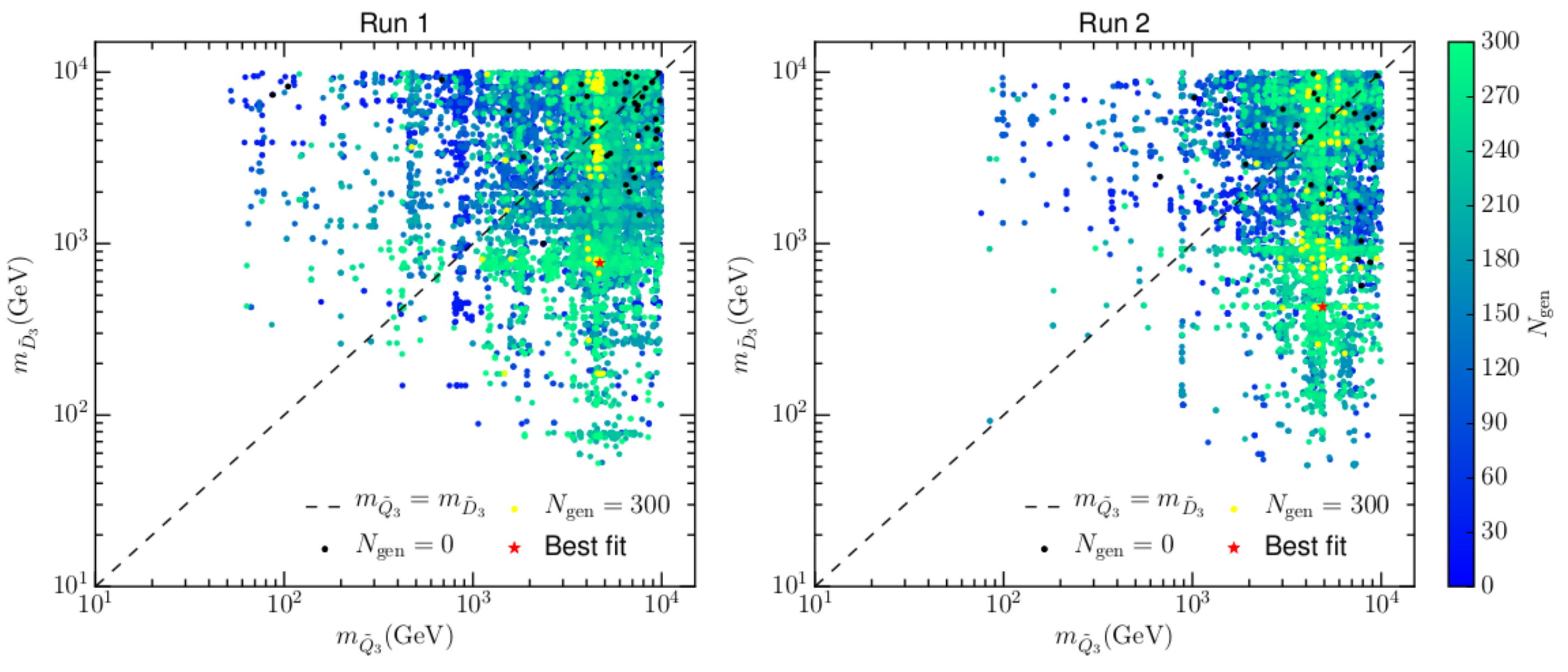}
  \includegraphics[width=\linewidth]{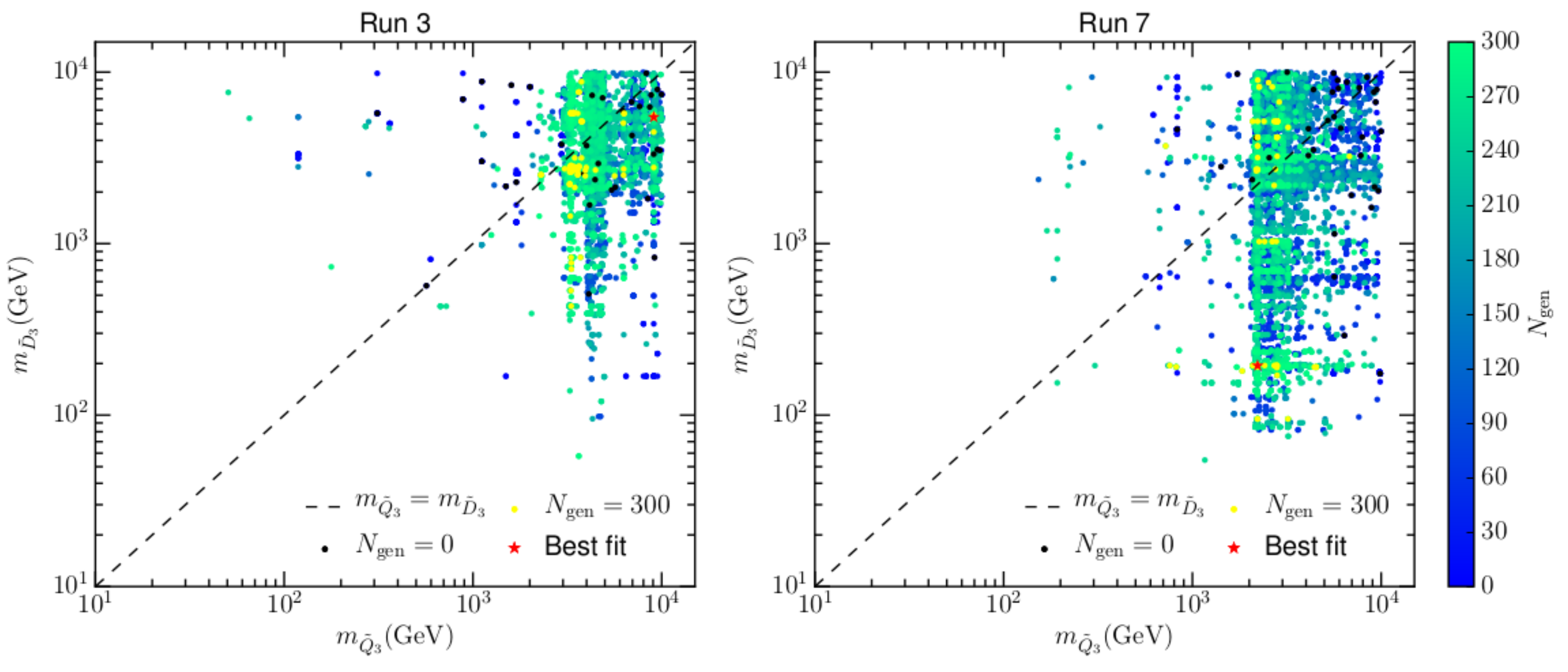}    
\caption{$\msd{3}$ vs. $\msq{3}$, top left (run 1), top right (run 2), bottom left (run 3), bottom right (run 7).}
\label{fig:runs1237_3}
\end{figure}

\begin{figure}[h!]
\centering
  \includegraphics[width=\linewidth]{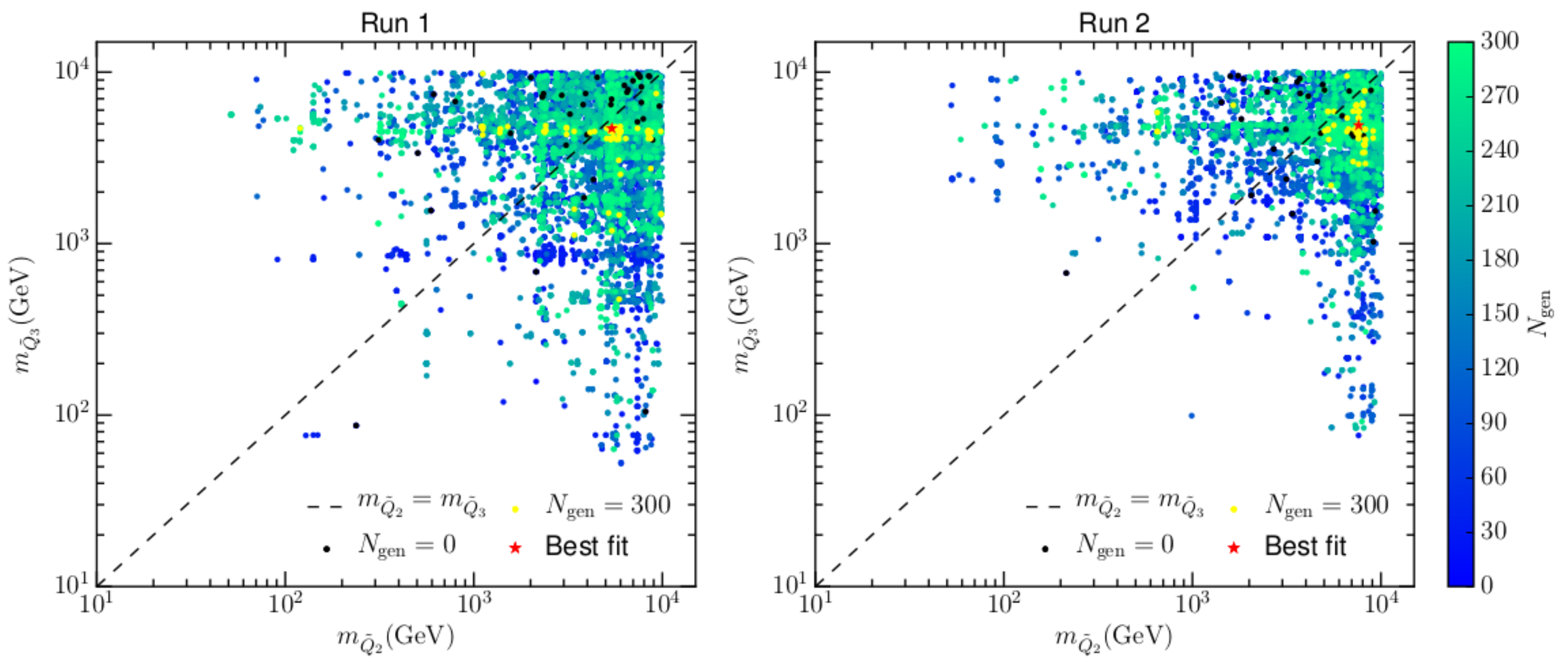}
  \includegraphics[width=\linewidth]{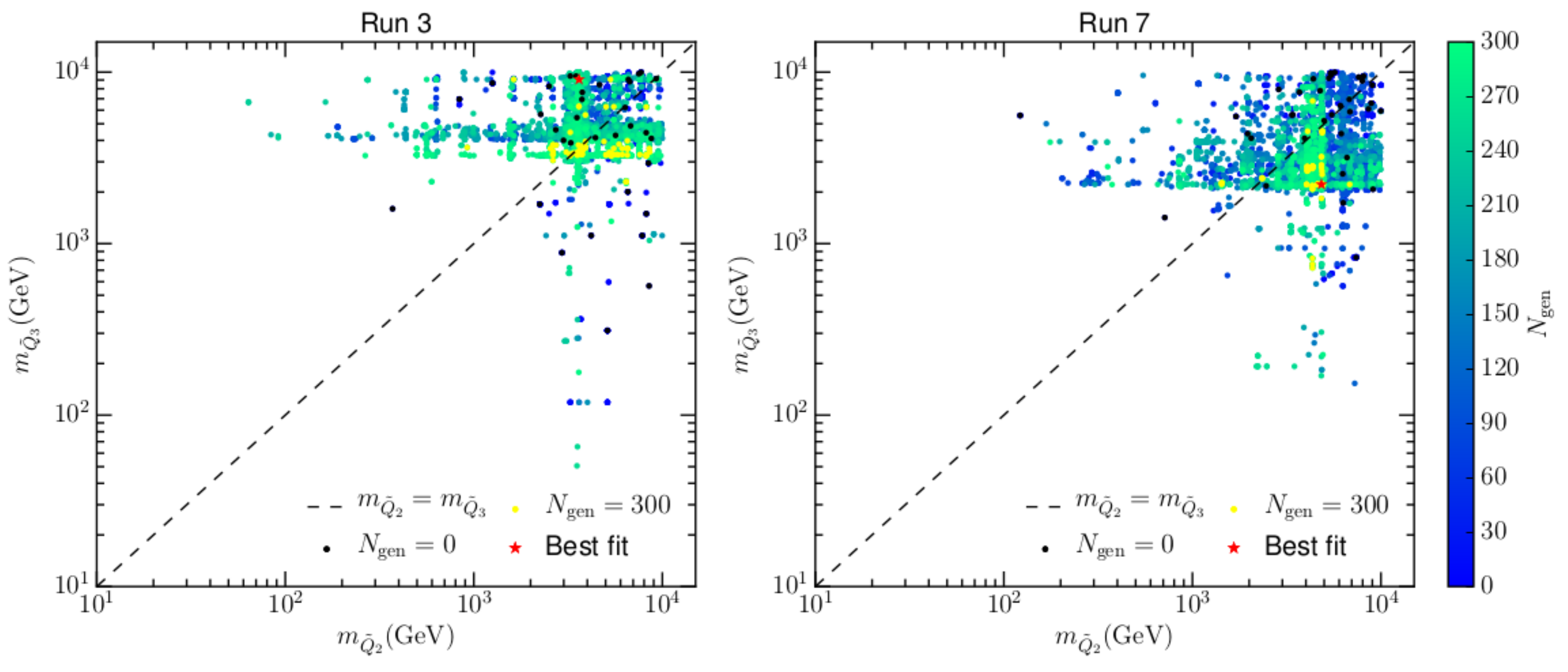}    
\caption{$\msq{3}$ vs. $\msq{2}$, top left (run 1), top right (run 2), bottom left (run 3), bottom right (run 7).}
\label{fig:runs1237_4}
\end{figure}

\begin{figure}[h!]
\centering
  \includegraphics[width=\linewidth]{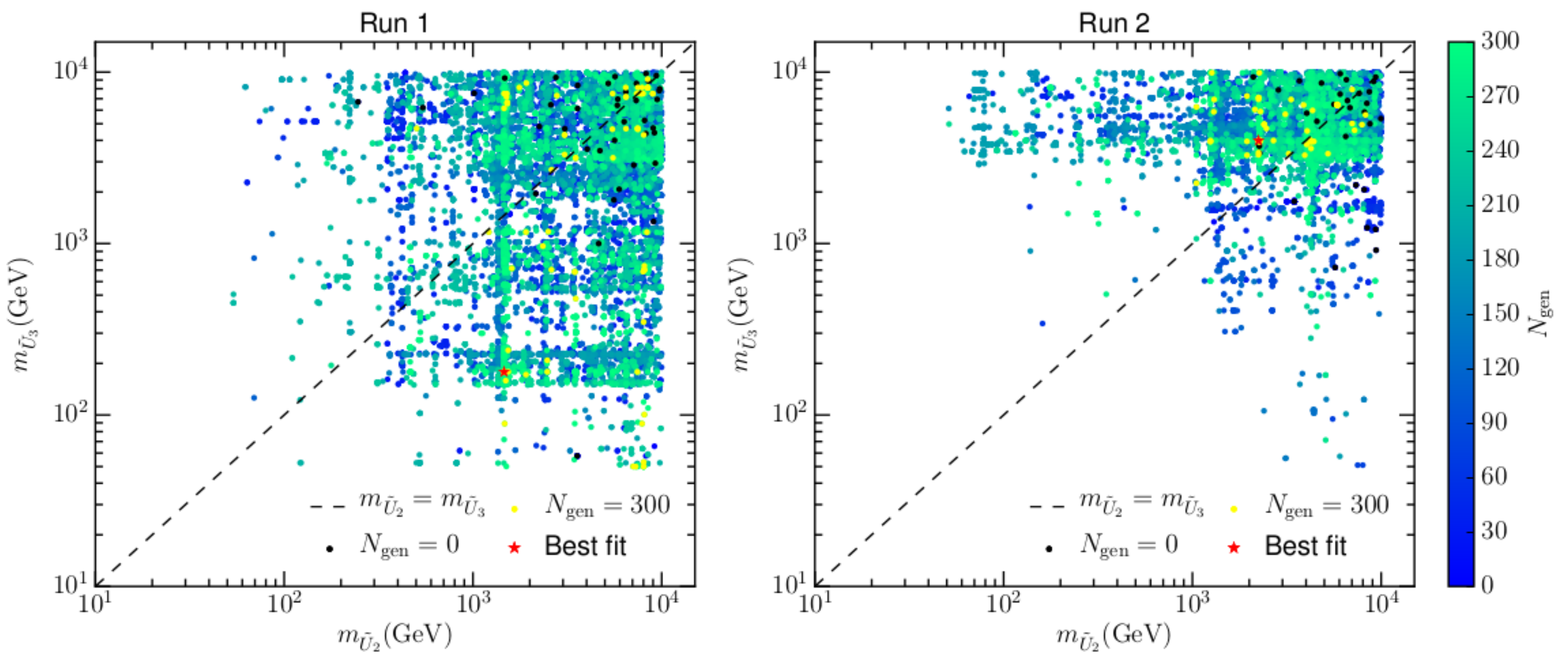}
  \includegraphics[width=\linewidth]{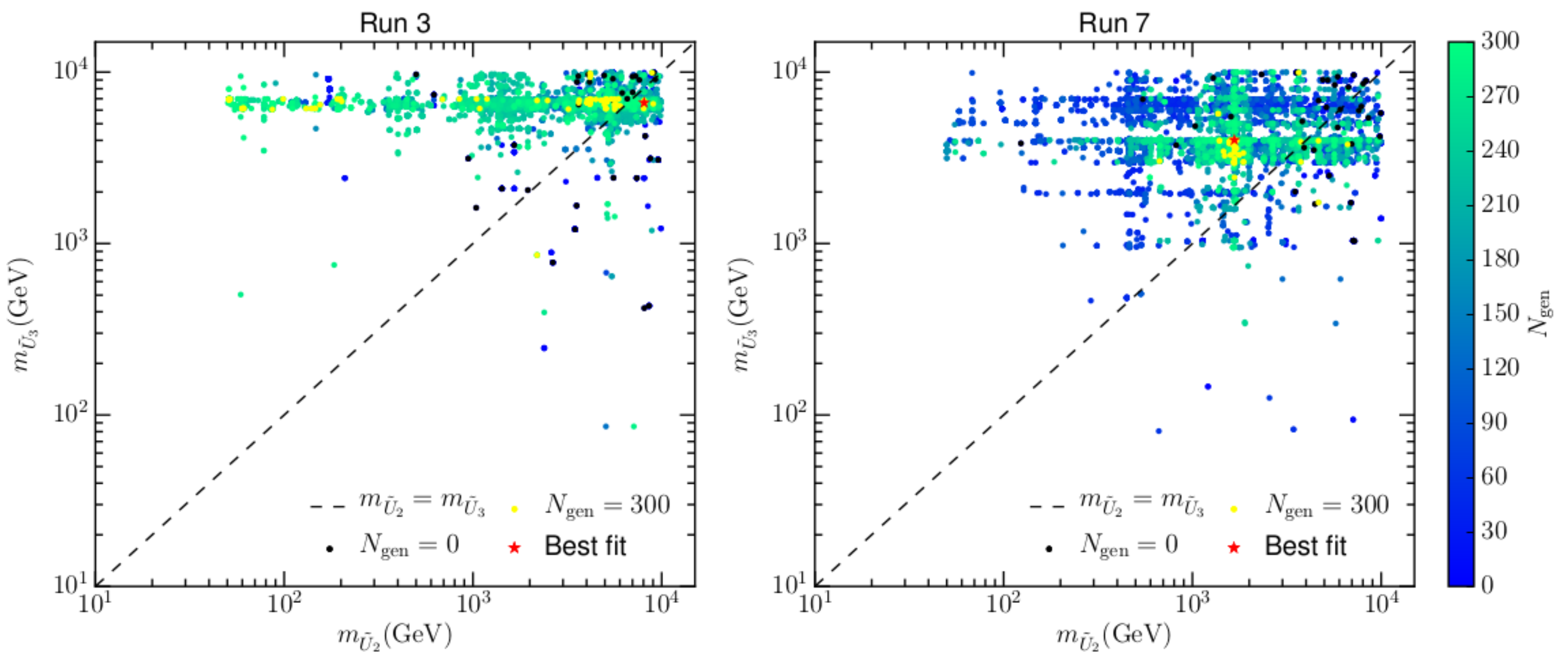} 
\caption{$\msu{3}$ vs. $\msu{2}$, top left (run 1), top right (run 2), bottom left (run 3), bottom right (run 7).}
\label{fig:runs1237_5}
\end{figure}

\begin{figure}[h!]
\centering
  \includegraphics[width=\linewidth]{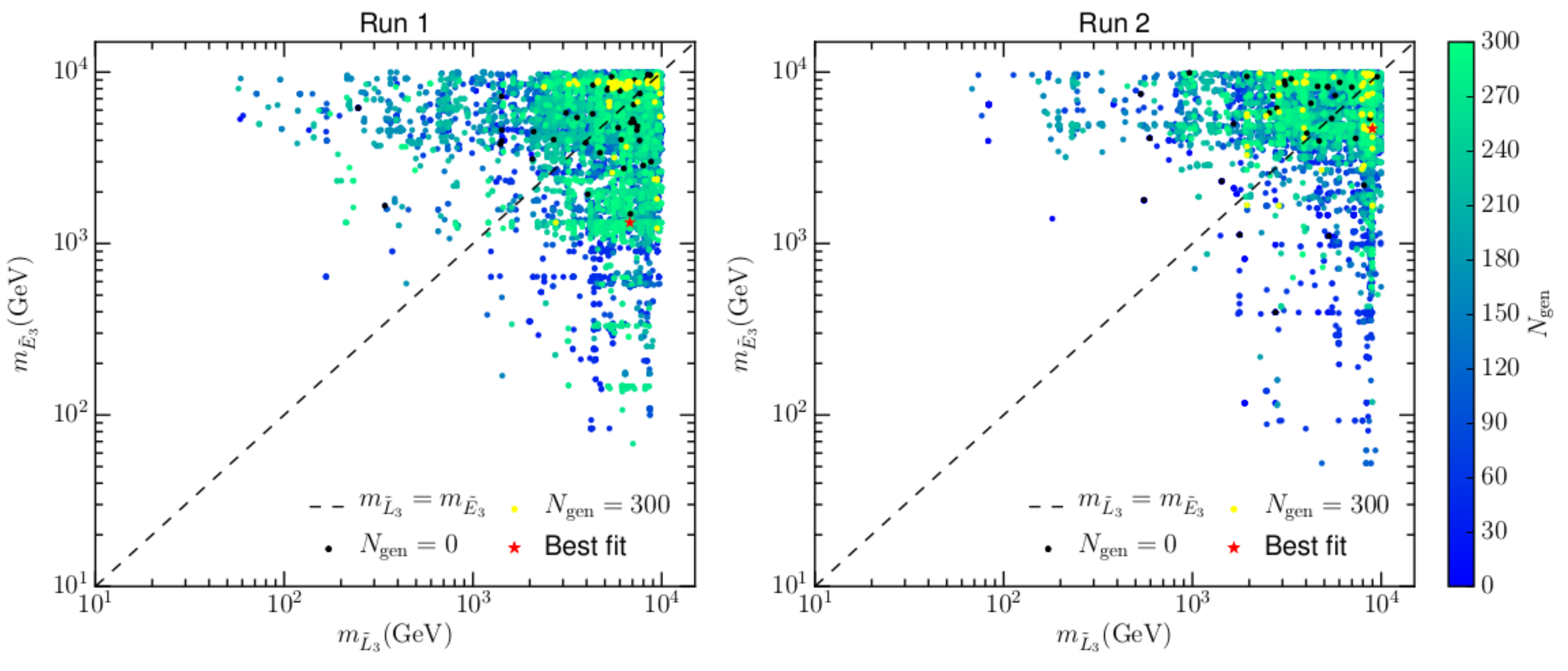}
  \includegraphics[width=\linewidth]{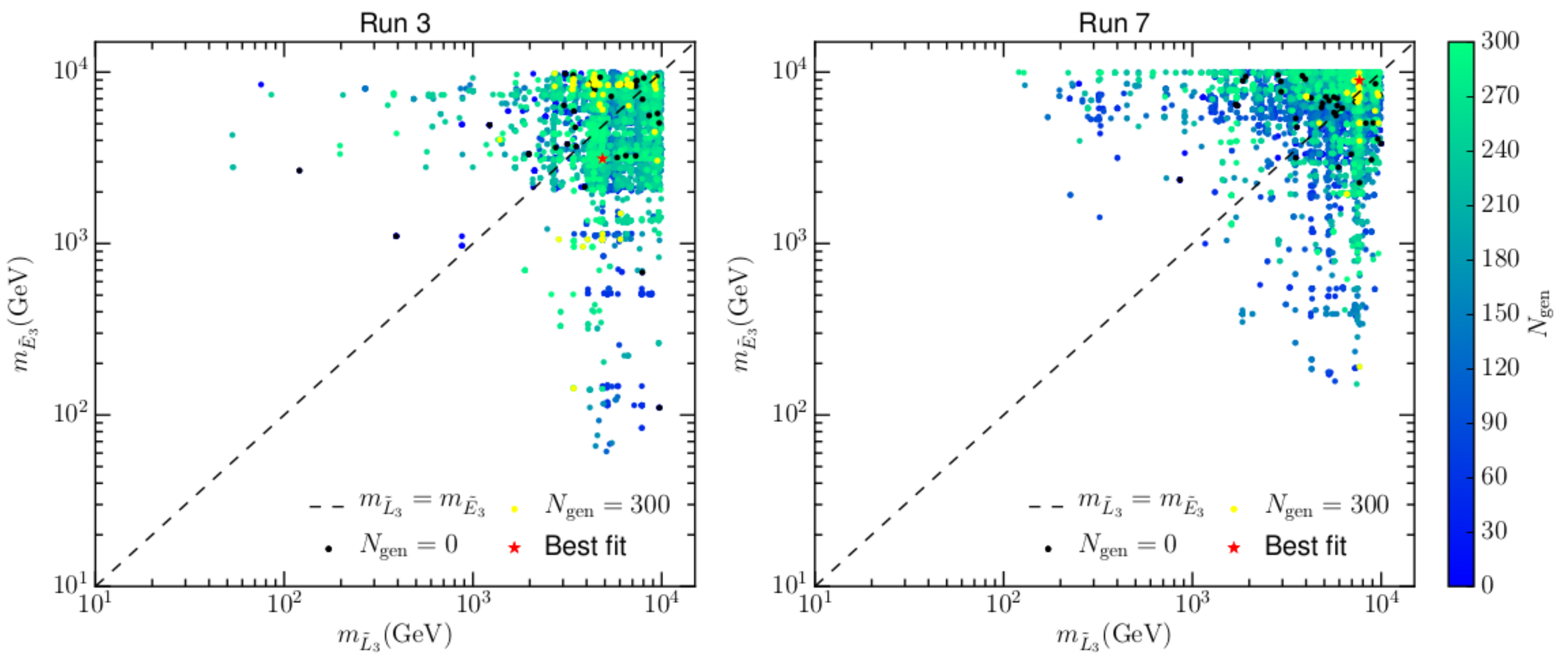}  
\caption{$\mse{3}$ vs. $\msl{3}$, top left (run 1), top right (run 2), bottom left (run 3), bottom right (run 7).}
\label{fig:runs1237_6}
\end{figure}

\clearpage
\bibliographystyle{JHEP-cerdeno}  
\bibliography{references,smodels} 

\end{document}